\documentclass[twocolumn, tighten, trackchanges, resetfootnote]{aastex7}

\usepackage{amsmath}
\usepackage[shortlabels]{enumitem}
\usepackage{longtable}
\usepackage{booktabs}
\usepackage{caption}
\usepackage{amssymb}
\usepackage{subcaption} % The standard replacement

\shorttitle{Gravitational Anomaly from Accurate 3D Velocities of Wide Binaries}

\shortauthors{Chae et al.}

\begin{document}

\title{Detection of Gravitational Anomaly at Low Acceleration from a Highest-quality Sample of 36 Wide Binaries with Accurate 3D Velocities}

\author[orcid=0000-0002-6016-2736, gname=K.-H., sname=Chae]{K.-H. Chae}
\email{chae@sejong.ac.kr}\thanks{corresponding author: chae@sejong.ac.kr \\ kyuhyunchae@gmail.com}
\affiliation{Department of Physics and Astronomy, Sejong University, 209 Neungdong-ro Gwangjin-gu, Seoul 05006, Republic of Korea}
\author[gname=B.-C., sname=Lee]{B.-C. Lee}
\email{bclee@kasi.re.kr}
\affiliation{Korea Astronomy and Space Science Institute, 776 Daedeokdae-ro, Yuseong-gu, Daejeon 34055, Republic of Korea}
\author[gname=X., sname=Hernandez]{X. Hernandez}
\email{xavier@astro.unam.mx}
\affiliation{Universidad Nacional Aut\'{o}noma de M\'{e}xico, Instituto de Astronom\'{i}a, A. P. 70-264, 04510, CDMX, M\'{e}xico}
\author[gname=V., sname=Orlov]{V. G. Orlov}
\email{orlov@astro.unam.mx}
\affiliation{Universidad Nacional Aut\'{o}noma de M\'{e}xico, Instituto de Astronom\'{i}a, A. P. 70-264, 04510, CDMX, M\'{e}xico}
\author[orcid=0000-0001-7277-7175, gname=D., sname=Lim]{D. Lim}
\email{dwlim@yonsei.ac.kr}
\affiliation{Center for Galaxy Evolution Research, Yonsei University, 50 Yonsei-ro, Seodaemun-gu, Seoul 03722, Republic of Korea}
\author[gname=D., sname=Turnshek]{D. A. Turnshek}
\email{turnshek@pitt.edu}
\affiliation{Department of Physics and Astronomy, University of Pittsburgh, Pittsburgh, PA 15260, USA}
\affiliation{Pittsburgh Particle Physics, Astrophysics, and Cosmology Center (PITT PACC), Pittsburgh, PA 15260, USA}
\author[gname=Y.-W., sname=Lee]{Y.-W. Lee}
\email{ywlee2@yonsei.ac.kr}
\affiliation{Center for Galaxy Evolution Research, Yonsei University, 50 Yonsei-ro, Seodaemun-gu, Seoul 03722, Republic of Korea}
\affiliation{Department of Astronomy, Yonsei University, 50 Yonsei-ro, Seodaemun-gu, Seoul 03722, Republic of Korea}

\begin{abstract}
We set out to accurately measure gravity in the low-acceleration range $(10^{-11},10^{-9})\,{\rm m}\,{\rm s}^{-2}$ from 3D motions of isolated wide binary stars. Gaia DR3 provides precise measurements of the four sky-plane components of the 3D relative displacement and velocity ($\mathbf{r}, \mathbf{v}$) for a system, but not comparably precise line-of-sight (radial) separation and velocity $v_{r}$. Based on our new observations and the public databases/publications, we assemble a sample of 36 nearby (distance $<150$pc) wide binaries in the low-acceleration regime with relatively accurate values of $v_r$. Kinematic contaminants such as undetected stellar companions are well under control using various observational diagnostics such as Gaia's {\tt ruwe} parameter, the color-magnitude diagram, multi-epoch observations of radial velocities, Speckle interferometric follow-up observations, and requiring Hipparcos-Gaia proper motion consistency. For $\Gamma \equiv \log_{10}\sqrt{\gamma}$ with $\gamma \equiv G/G_{\rm N}$ (where $G$ is a parameter generalizing Newton's constant $G_{\rm N}$), we find $\Gamma=0.102_{-0.021}^{+0.028}$ for $<10^{-9.1}\,{\rm m}\,{\rm s}^{-2}$ ($0.132_{-0.027}^{+0.035}$ for $<10^{-9.5}\,{\rm m}\,{\rm s}^{-2}$) giving a gravity boost factor of $\gamma=1.60_{-0.14}^{+0.24}$, which rules out Newton. This strong anomaly is dominated by two systems (the pairs of HD 189739, HD 189760, and TYC 259-236-1, TYC 259-906-1) that have 3D relative velocities exceeding their estimated Newtonian escape velocities. Without them, we have $\Gamma=0.042_{-0.024}^{+0.027}$ for $<10^{-9.1}\,{\rm m}\,{\rm s}^{-2}$, $0.057_{-0.031}^{+0.035}$ for $<10^{-9.5}\,{\rm m}\,{\rm s}^{-2}$, and $0.085_{-0.045}^{+0.059}$ for $<10^{-10.1}\,{\rm m}\,{\rm s}^{-2}$. Future radial velocity monitoring and Speckle interferometric imaging for larger samples will be useful to refine the present result.
\end{abstract}


\section{Introduction}\label{sec:intro}

Einstein \citep{Einstein:1916} invented general relativity as a relativistic theory of gravity whose nonrelativistic limit is Poisson's equation, Newtonian gravitation. For the past century this Newton-Einstein standard gravity has been the basis for the modern cosmological paradigm and the requirement of dark matter from astronomical observations. However, Newtonian gravity applicable in the nonrelativistic regime was based on the empirical laws of planetary motions (discovered by Kepler four centuries ago) in the high-acceleration $g_{\rm N} > 10^{-7}$m\,s$^{-2}$ regime. Hereafter, $g_{\rm N}$ refers to Newton's constant ($G_{\rm N}$) times mass over distance squared for the system under consideration, i.e., $g_{\rm N}=G_{\rm N}M_{\rm tot}/r^2$ for a binary with total mass $M_{\rm tot}$ and 3D separation $r$.

From an empirical point of view, Poisson's equation was shown to be broken by Le Verrier's work as gravity gets sufficiently strong, already in the 19th century even before any relativistic theory existed. While a relativistic theory is successful in the strong-gravity regime, the hypothesis that Poisson's equation can be extrapolated indefinitely into the low-acceleration limit requires a direct empirical verification. General relativity's success in strong-gravity regimes does not guarantee its correctness in the low-acceleration limit (perhaps, as the classically perfect theory of Maxwell's electrodynamics has fundamental limitations). Any piece of scientific truth is eventually determined by experimental/observational facts. The same holds for any hypothetically dominant dark matter component as introduced to force agreement with standard gravity, hence the ongoing worldwide campaigns attempting a direct detection of any such component (see, e.g., the report by \citealt{DMreview:2022} and Section~27 of \citealt{ParticlePhysics:2024}), without any positive signal up to now (e.g., \citealt{COSINE-100:2025}). 

Milgrom \citep{Milgrom:1983} first noticed/suggested that galactic rotation curves start to deviate significantly from Newton's $r^{-2}$ law only around a critical acceleration $a_0$ near $10^{-10}$~m\,s$^{-2}$. Milgrom's proposal, known as modified Newtonian dynamics (MOND), posits a modification of standard gravitational dynamics in the low acceleration regime to break the strong equivalence principle (while keeping Einstein's equivalence principle and Galileo's universality of free fall) and makes a number of salient predictions for gravitational dynamics in astrophysical systems \citep{Sanders:2002,Famaey:2012,BanikZhao:2022,Merritt:2020}, including the idiosyncratic external field effects \citep[EFE; ][]{Milgrom:1983,ChaeMilgrom:2022} in the internal dynamics of freely falling systems under external fields.

 Wide binary stars provide a unique probe to test standard gravity and MOND, as first proposed in \cite{Hernandez:2012}, with more recent examples such as \cite{Scarpa:2017}, \cite{PittordisSutherland:2018}, and \cite{BanikZhao:2018}. From a theoretical point of view, binaries of point-like masses are the simplest possible systems to test gravity, and wide binaries can have sufficiently low internal acceleration to probe the `dark matter' regime, while any conceivable effect of hypothetical dark matter is negligible, given the stringent limits on this component locally, coming from observed vertical kinematics of disk stars (e.g., \citealt{Read:2014}). From an observational point of view, these binaries can be found in the solar neighborhood, so that their distances are directly measurable (an unusual advantage in astronomy) and their velocities can be measured with a sufficiently good precision.

 The advent of the Gaia Data Release 3 \citep[DR3; ][]{Gaia:2023} database ignited intensive wide binary gravity tests in recent years. Because Gaia DR3 does not generally provide sufficiently precise line-of-sight (radial) velocities, several statistical methods have been developed based only on sky-plane (tangential) velocities $v_p$ (for a summary of the methods, see Table~3 of \citealt{Chae:2025}). Because gravity is sensitive to the 3D relative velocity $\mathbf{v}$ between the pair at the 3D displacement $\mathbf{r}$, use of $v_p$ does not allow any meaningful gravitational inference from individual wide binaries whose very long orbital periods preclude anything but essentially instantaneous observations. Thus, statistical methods requiring large numbers of wide binaries have been introduced to deal with the issues of projection effects (of $\mathbf{v}$ onto $v_p$), orbital orientation, phase occupancy, and false binaries (i.e.\ gravitationally-unbound fly-bys or chance associations).
 
 All statistical analyses of binaries \citep{Chae:2023,Chae:2024a,Chae:2024b,Hernandez:2023,Hernandez:2024a,HernandezKroupa:2025,Yoon:2025} covering a sufficiently broad dynamic range to include both high ($g_{\rm N} > 10^{-8}$~m\,s$^{-2}$) and low ($g_{\rm N} < 10^{-9}$~m\,s$^{-2}$) acceleration regimes show that the median value of $v_p$ (or related quantity) is boosted about 20\% in the low-acceleration regime as long as the fraction of hierarchical systems  $f_{\rm multi}$ (with stellar multiplicity $\ge3$) is calibrated/checked using the binaries in the high-acceleration regime, regardless of the sample choices that largely dictate $f_{\rm multi}$. See \cite{Hernandez:2024review} for a critical review of some divergent results that are not based on the calibration of $f_{\rm multi}$ and do not include internal validation checks through the presence of a high acceleration Newtonian region. 
 
 More recently, \cite{Pittordis:2025} introduced an improved triple model for a well-defined sample (satisfying in particular Gaia's ${\tt ruwe}<1.2$) and used it to test relative performances of Newton and an approximate MOND model in various bins from the projected separation ($s$) $1.25\,{\rm kau}$ to $20\,{\rm kau}$. They claimed that Newton was preferred over the control MOND model. However, the control model is not a correct representation of MOND gravity in the transition regime $10^{-9}\ga g_{\rm N}\ga 10^{-10}\,{\rm m}\,{\rm s}^{-2}$ \citep{Pf-A:2025}, making their intended comparison questionable. Moreover, when $f_{\rm trip}$ (triple fraction since they considered only triples as possible multiples) is properly calibrated using binaries in a fully Newtonian regime (e.g., $s<0.7\,{\rm kau}$), their sample actually reveals a gravity boost at low acceleration $g_{\rm N}\la 10^{-10}\,{\rm m}\,{\rm s}^{-2}$ with respect to Newton, as shown by \cite{ChaeYoon:2026}. Studies such as \cite{Pittordis:2025} focused on comparing Newton with a control MOND model at $g_{\rm N}\la 10^{-9}\,{\rm m}\,{\rm s}^{-2}$ while studies mentioned in the previous paragraph focused on detecting and quantifying anomaly with respect to Newton in an absolute sense. 
 
Although statistical methods with $v_p$ have been popular, it is clearly desirable to use directly measured 3D velocity $\mathbf{v}$ for gravity tests because individual systems can be analyzed and understood in greater detail, leading to much more accurate gravity inferences. Recently, \cite{Chae:2025,Chae:2025b} developed a Bayesian 3D modeling methodology to infer probability density functions (PDFs) of gravity in individual systems (individually not highly restrictive broad distributions) and then to  statistically consolidate them to derive the effective strength of gravity in a common acceleration regime. \cite{Chae:2025} first carried out an extensive study of 312 wide binaries using a simplified 3D model based on the Gaia DR3 3D velocities, which include relatively less precise radial velocities (RVs), and found a dichotomy such that 125 wide binaries in the strong acceleration regime $g_{\rm N}\ga 10^{-8}$~m\,s$^{-2}$ agree well with Newton, while 111 wide binaries in the transition and low-acceleration regimes show a $4.2\sigma$ anomaly. \cite{Chae:2025b} carried out a pilot study of 32 wide binaries with a fully general 3D model based on accurate 3D velocities including precise HARPS RVs from \cite{Saglia:2025}, and found a moderate indication of dichotomy that 8 wide binaries with $g_{\rm N}<10^{-9}$~m\,s$^{-2}$ show a gravity boost while 24 wide binaries with $g_{\rm N}>10^{-9}$~m\,s$^{-2}$ agree well with Newton.  

The key to a reliable inference of gravity with 3D motions of wide binaries is the construction of a sample of pure binaries free of undetected kinematic contaminants, such as unresolved companion stars or resolvable but too faint stars. In this work we employ an unprecedented combination of observational diagnostics to have a full control of potential kinematic contaminants. They include not only well-known diagnostics such as imposing a color-magnitude diagram (CMD) exclusion region and Gaia's \texttt{ruwe} limit, but also multi-epoch observations of radial velocities over more than several years and detailed Speckle interferometric imaging of the stars of many wide binaries. These diagnostics work together to flag close/unresolved contaminants (within tens of au from the star) and more distant contaminants.

For the first time we carry out Speckle observations of 390 wide binaries selected from recent samples used by two of us (e.g., \citealt{Hernandez:2024a,Chae:2024a}). These observations form part of an ongoing campaign to image wide binaries at the diffraction limit of the 2.1 m telescope at the Observatorio Astronomico Nacional (SPM) using Speckle interferometry. These speckle observations will not only flag certain individual binaries but also give us the measured probability that the stringently selected samples will have resolvable faint companions. On the other hand, we collect as many high-precision radial velocities measured with different telescopes and, more importantly, at different epochs separated by at least several years. The collection includes our new observations of 60 wide binaries with the  Las Cumbres Observatory (LCO) Network of Robotic Echelle Spectrographs (NRES) and 6 wide binaries with the GEMINI-North Observatory MAROON-X spectrograph (hereafter MAROON-X). It also includes wide binaries with radial velocities selected from HARPS \citep{Saglia:2025}, SDSS4 DR17 APOGEE (hereafter APOGEE),\footnote{https://www.sdss4.org/dr17/irspec/radialvelocities/} and \cite{Scarpa:2017}. In addition, we use the comparison of Hipparcos and Gaia proper motions to have a control in the selection process. 

Since many previous studies (e.g., \citealt{Chae:2024a,Chae:2024b,Yoon:2025,Chae:2025,Chae:2025b}) of wide binaries including the recent 3D analyses already confirm that Newton is verified at least for $g_{\rm N}>10^{-8}$~m\,s$^{-2}$, we focus on the low-acceleration regime only. We construct an extremely curated sample of 36 pure wide binaries in the low-acceleration regime that have accurate and precise relative RVs with measurement error $<100$~m\,s$^{-1}$. Our new sample is $\approx 4$ times as large as the \cite{Saglia:2025} sample of wide binaries having radial velocity measurements in the low-acceleration regime, with comparable data qualities, and thus our inferred constraints on gravity can be expected to have twice the precision than the pilot study by \cite{Chae:2025b}.

In Section~\ref{sec:method} we briefly describe theoretical motivations and the Bayesian 3D modeling methodology. In Section~\ref{sec:data}, we describe the process of selecting the statistical sample of wide binaries to be used for gravity inference. We present the results on inferred gravity and relevant discussions in Section~\ref{sec:result}. We give our thoughts on the meanings and implications of the results in Section~\ref{sec:meaning} and conclude in Section~\ref{sec:conclusion} with a future outlook. The Python codes used in this work and the observational data for our sample are available on Zenodo under an open-source Creative Commons Attribution license \dataset[]{https://doi.org/10.5281/zenodo.22803893} . In Appendices~\ref{sec:LCO}, \ref{sec:MAROONX}, \ref{sec:Scarpa}, and \ref{sec:APOGEE}, we describe the observations or collections of RVs. In Appendix~\ref{sec:Speckle}, we describe the Speckle interferometric observations. 

\section{Theoretical motivation and the 3D modeling methodology} \label{sec:method}

In this work we set out to measure gravity in the low-acceleration regime in the context of distinguishing between standard and nonstandard theories of gravity. Although MOND is not yet an established physical theory of relativistic gravity (not to mention quantum gravity), it is useful to consider nonrelativistic MOND models as modifications of standard gravity in the low-acceleration regime. MOND predicts that the $r^{-2}$ law with Newton's constant $G_{\rm N}$ will no longer hold in the acceleration regime $g_{\rm N} \la 10^{-9}\,{\rm m}\,{\rm s}^{-2}$. According to MOND, how gravity will behave depends on the details of the EFE. For a truly isolated system without an external field, the gravity law will gradually switch to a $r^{-1}$ behavior between $10^{-8} \la g_{\rm N} \la 10^{-10}\,{\rm m}\,{\rm s}^{-2}$. In reality, common dynamical systems such as binary stars and galaxies can be subject to external fields of various strength, and consequently, in the regime of low internal acceleration, gravity is predicted by MOND to become pseudo-Newtonian, showing an $r^{-2}$ behavior with a rescaled gravitational parameter $G(>G_{\rm N})$ that depends on the external field (see, e.g., \citealt{ChaeMilgrom:2022} for specific numerical examples). 

\begin{figure}[!htb]
    \centering
    \includegraphics[width=1.03\linewidth]{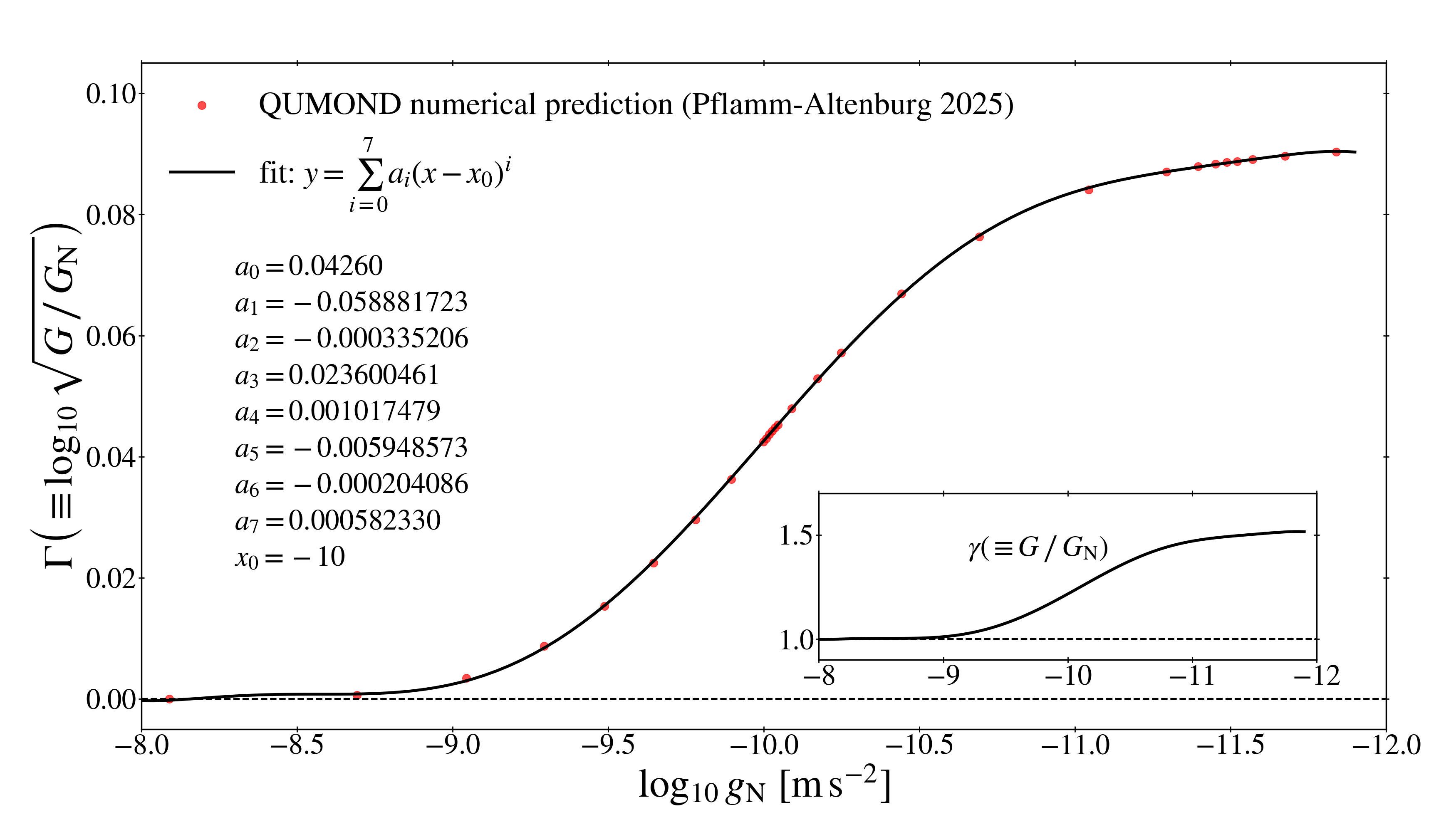}
    \caption{
   This figure shows the QUMOND-predicted behavior of the gravitational anomaly parameter $\Gamma$ (Equation~(\ref{eq:Gamma})) as a function of $g_{\rm N}$ (the internal Newtonian acceleration between the two stars) for wide binaries in the solar neighborhood based on the numerical results by \cite{Pf-A:2025}. The inset shows the behavior of $\gamma$. The curve is for the specific case that two stars have equal masses of $1{\rm M}_\odot$. For realistic cases with mass ratios $\ga 1/3$, similar curves are obtained. But, when the mass ratio $\rightarrow 0$ (the test particle case), resulting curves are different in the transition regime of $10^{-9}\ga g_{\rm N} \ga 10^{-10}\,{\rm m}\,{\rm s}^{-2}$.
    }
    \label{fig:QUMOND}
\end{figure}

In the case of wide binaries in the solar neighborhood, the external field due to the Galaxy is $\approx 1.8 a_0$, which is so strong in the context of MOND that the gravity boost factor is expected to be about $G/G_{\rm N}\approx 1.4$ for a test particle (e.g., \citealt{BanikZhao:2018,ChaeMilgrom:2022}) with $g_{\rm N} \la 10^{-10}\,{\rm m}\,{\rm s}^{-2}$. Complete numerical solutions for a two-body dynamics under an external field (due to a third body) find a lower boost factor in the range $10^{-9}\ga g_{\rm N} \ga 10^{-10.5}\,{\rm m}\,{\rm s}^{-2}$ but a similar boost factor at lower accelerations with a somewhat larger limiting value of $\gamma \approx 1.5$. Figure~\ref{fig:QUMOND} shows the detailed behavior of the QUMOND-predicted boost factor as a function of $g_{\rm N}$ based on the numerical results by \cite{Pf-A:2025}. This curve will be used throughout in this work.  

To test Newtonian and MOND predictions for wide binaries of the solar neighborhood, it is convenient to approximate instantaneous orbits as elliptical even if global orbits deviate from closed ellipses as in the MOND case \citep{Pf-A:2025}. While this assumption is perfectly valid in testing Newtonian gravity, any result deviating from Newton needs to be interpreted correctly in the context of testing MOND models such as AQUAL \citep{Bekenstein:1984} and QUMOND \citep{Milgrom:2010}, or any other alternative model of gravity of interest. Regardless of the details of any modified gravity theory, an approach, in which a $G \to \gamma G_{N}$ model is tested, will reveal the presence of a gravitational anomaly if the inferred value of $\gamma \neq 1$.

\begin{figure}[!htb]
    \centering
    \includegraphics[width=1.\linewidth]{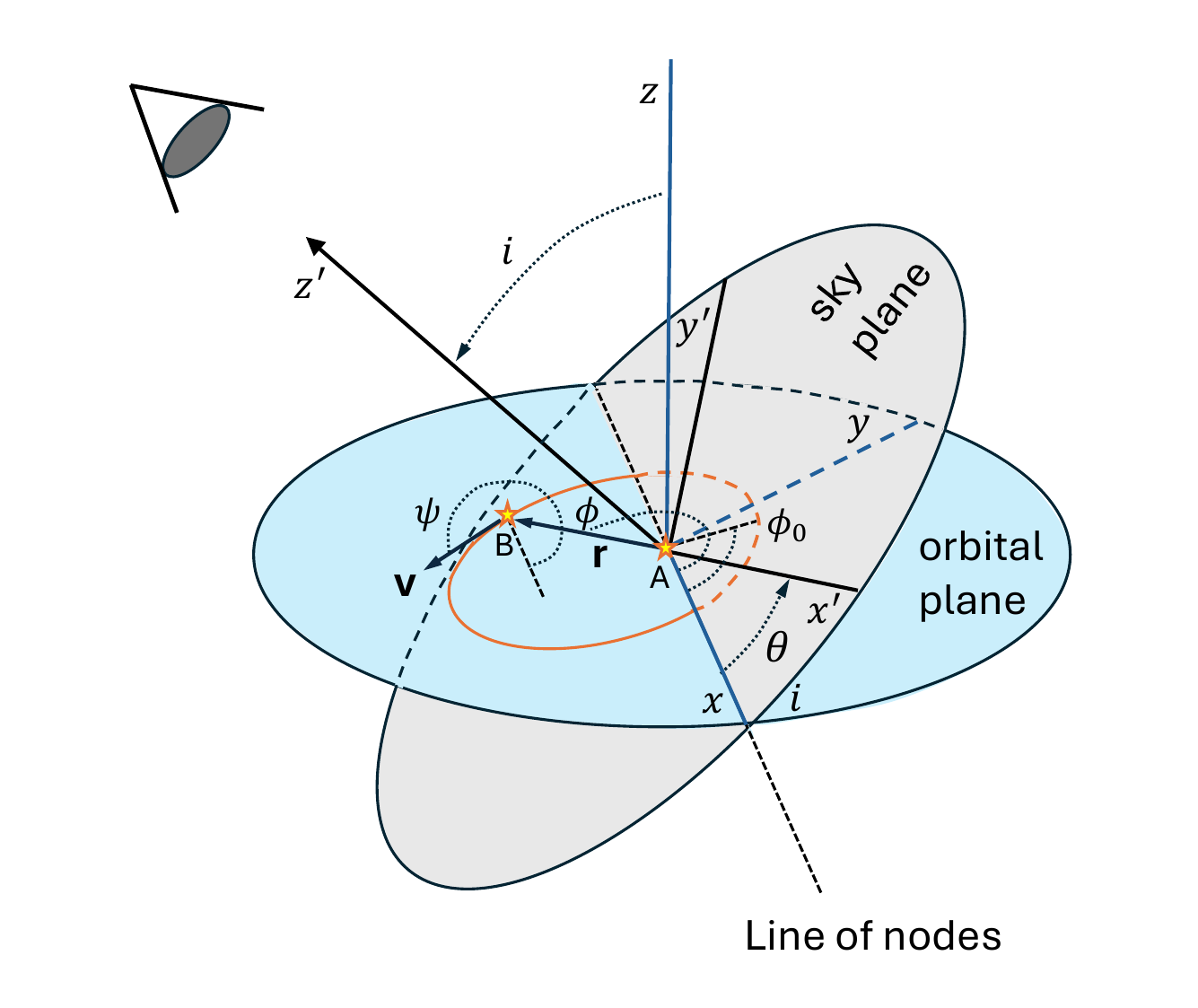}
    \caption{
    (Adapted from \cite{Chae:2025b}) A general 3D geometry of an elliptical orbit. Here $z^\prime$ represents the line-of-sight (radial) direction in observer's frame with the $+$ sign indicating the direction pointing to the observer. The relative radial velocity refers to $v_{z^\prime}(\equiv v_r)$. See \cite{Chae:2025b} for the definition of all the other parameters.
    }
    \label{fig:Euler2angles}
\end{figure}

We use the 3D modeling methodology of \cite{Chae:2025,Chae:2025b} that infers PDFs $p_i(\Gamma)$ ($i=1,\cdots,N_{\rm binary}$) of the parameter
\begin{equation}
    \Gamma \equiv \log_{10}\sqrt{\gamma} \equiv \log_{10}\sqrt{G/G_{\rm N}}
    \label{eq:Gamma}
\end{equation}
for individual systems (along with orbit and orientation parameters: see Figure~\ref{fig:Euler2angles}) and then statistically consolidate them through a normalized product of $p_i$. 

Here we briefly describe the essential points of the methodology and refer the reader to \cite{Chae:2025,Chae:2025b} for the details. Each binary has the six measured quantities of $\mathbf{r}=\{x^\prime,y^\prime,z^\prime\}$ and $\mathbf{v}=\{v_{x^\prime},v_{y^\prime},v_{z^\prime}\}$, where $x^\prime$ and $y^\prime$ are (nearly) exact, $v_{x^\prime}$, $v_{y^\prime}$, and $v_{z^\prime}$ are very precise, and $z^\prime$ is usually not so precise. Taking advantage of the fact that $x^\prime$ and $y^\prime$ are fixed by the data, we have a reduced set of free parameters $\mathbf{\Theta}=\{e,i,\phi_0,\Delta\phi(\equiv\phi-\phi_0),\log_{10}f_M,\Gamma \}$, where $e$ is the eccentricity with the range $(0,1)$, $i$ is the inclination with the range $(0^\circ,180^\circ)$, $\phi_0$ is the phase of the periastron with the range $(0^\circ,360^\circ)$, and $\Delta\phi$ is the phase relative to $\phi_0$ known as the true anomaly with the range $(0^\circ,360^\circ)$.%\footnote{We note that the angle parameter $\theta$ shown in Figure~\ref{fig:Euler2angles}, which is related to the longitude of the ascending node, can be calculated a a function of the free parameters.} 
Here $f_M$ represents the total mass of the binary system normalized by the observational mass. Inclusion of the parameter $f_M$ in the Bayesian modeling means that we are allowing a probability distribution of mass (hence including a confidence interval on this parameter for each binary) with its prior set at the observationally inferred value. 

 {We note that in the algorithm used in this work, the semi-major axis as well as parameter $\theta$ in Figure~\ref{fig:Euler2angles} (which is related to the longitude of the ascending node) are implicitly determined by the two internal self-consistency relations provided by $x^\prime$ and $y^\prime$, or equivalently by $s\left(\equiv\sqrt{{x^\prime}^2+{y^\prime}^2}\right)$ and $b (\equiv y^\prime/x^\prime$). The parameter $b$ is essentially exact because the same distances in $x^\prime$ and $y^\prime$ cancel out so that it is determined by the (essentially) exact Gaia DR3 angular coordinates only. Thus, parameter $\theta$ can always be implicitly determined using $b$ as long as the local Cartesian frame centered on star~A is used (Figure~\ref{fig:Euler2angles}). The uncertainty of $s$ due to the uncertainty of $d_A$ is not essentially zero but comfortably small ($\la 0.1\%$) for our highest-quality wide binaries within $150\,{\rm pc}$. Even if the measured value of $s$ along with its uncertainty is used as an observational constraint, the inference of the parameters is little affected as explicitly verified by \cite{Chae:2026}. Thus, we will use only the algorithm where the uncertainty of $s$ is set to zero and $s$ is used as a fixed input to the internal self-consistency relation. }

The posterior probability of the parameters $p(\mathbf{\Theta})$ is defined by
\begin{equation}
    \ln p(\mathbf{\Theta}) = \ln\mathcal{L} + \sum_l \ln f_{\rm pr}(\Theta_l),
\label{eq:postprob}    
\end{equation}
where $\mathcal{L}$ is the likelihood function (connecting the free parameters with $\mathbf{r}$ and $\mathbf{v}$) whose details can be found in \cite{Chae:2025b}, and $f_{\rm pr}(\Theta_l)$ ($l=1,\cdots,6$) is the prior probability for the parameter $\Theta_l$. The imposed priors are as follows: $f_{\rm pr}(i)$ = $\sin(i)$ (isotropic orientation), $f_{\rm pr}(\phi_0)$ = uniform (random orientation), and 
\begin{equation}
    f_{\rm pr}(\Delta\phi) = \frac{(1-e^2)^{3/2}}{2\pi} \frac{1}{[ 1+ e \cos(\Delta\phi) ]^2}, 
\label{eq:PrDelphi}   
\end{equation}
a phase occupancy probability which is inversely proportional to the scalar velocity at any given phase. Equation~(\ref{eq:PrDelphi}) ensures that each instantaneous motion being observed occurs at a random time during its orbital period. We also impose priors on $e$, $f_M$, and $\Gamma$ as follows: $f_{\rm pr}(e) = (1+\alpha)e^\alpha$ taking $\alpha=1$ (the thermal probability distribution) as the nominal choice but considering also a full range of possibilities given by $0\leq\alpha\leq1.3$ (as inferred by \cite{Hwang:2022} for wide binaries of the solar neighborhood), a normal probability distribution of $\log_{10}f_M$ with $(\mu,\sigma)=(0,0.021)$ (i.e.\ 5\% scatter in the observational total mass), and a uniform distribution of $\Gamma$ in the range $-1<\Gamma<1$. We note that only the flat prior is considered for $\Gamma$ when we seek to measure it from the binary dynamics data.

We will mainly focus on measuring gravity at low acceleration by deriving the PDF of $\Gamma$ while treating the rest of the parameters as essentially nuisance parameters. To ensure an unbiased inference of $\Gamma$, it is necessary to use proper priors on the nuisance parameters as given above (in particular, the priors on inclination $i$ and orbit true anomaly $\Delta\phi$). For example, observational identification of a wide binary is blind to inclination or orbit true anomaly, so a sufficiently large sample of wide binaries should follow the distributions expected from randomness. Thus, the posterior PDFs of the nuisance parameters will provide internal self-consistency checks of the results.

As an auxiliary analysis, we will also carry out Bayesian modeling of the observed 3D motions at fixed gravity models with fixed values of $\Gamma$ including $\Gamma=0$ (i.e., the Newtonian case). The primary purpose of this auxiliary analysis will be to estimate the Newtonian escape velocity for each binary at a fixed Newtonian or pseudo-Newtonian gravity model. We will use the estimated Newtonian escape velocities in conjunction with various observational diagnostics to assemble gravitationally-bound pure binary systems, as will be described in the next section. The pilot study by \cite{Chae:2025b} based on the benchmark sample by \cite{Saglia:2025} will provide a guidance in this regard. In particular, the \cite{Saglia:2025} (sub)sample of 8 (or 9) wide binaries in the low-acceleration regime ($<10^{-9}$~m\,s$^{-2}$) includes one system where the observed 3D velocity exceeds the Newtonian escape velocity by about 12\%. That system is, however, extremely unlikely to be a chance association in the 3D space based on various observational diagnostics and statistical properties of the solar neighborhood. It will be interesting to see how our enlarged sample turns out to be. 

\section{Assembling a Carefully Curated Wide Binary Sample with Accurate 3D velocities} \label{sec:data}

To probe gravity in the low-acceleration regime through 3D modeling of wide binaries, we need kinematically uncontaminated pure binaries with accurate and precise values of the 6 components of $\mathbf{r}$ and $\mathbf{v}$ (Figure~\ref{fig:Euler2angles}). Since the currently available observation facilities do not permit a sufficiently precise measurement of radial separation $z^\prime$, it is unavoidable to work with only 5 precise components along with generally imprecise $z^\prime$. Because Gaia DR3 provides sufficiently precise values for the 4 sky-projected quantities $x^\prime$, $y^\prime$, $v_{x^\prime}$, and $v_{y^\prime}$, we focus on collecting wide binaries with precise $v_r(\equiv v_{z^\prime})=-({\rm RV}_B-{\rm RV}_A)$ where ${\rm RV}_A$ and ${\rm RV}_B$ are, respectively, the RVs of the brighter and fainter components relative to the Sun in the usual sense of the sign (i.e., the positive sign representing the direction of moving away from the Sun). Throughout, the brighter and fainter components are denoted by $A$ and $B$, respectively.  

To collect \emph{pure} wide binaries with accurate and precise $v_r$ in the low-acceleration regime, we follow three steps. In the first step, we collect as many as possible wide binaries with relatively precise RVs for both components and carry out Bayesian modeling for all the systems. In the second step, we select wide binaries that have low internal acceleration ($<10^{-9}$~m\,s$^{-2}$ based on the Bayesian modeling results) and pass the basic observational criteria required for pure binaries. In the final step, we select the clean sample that is most likely to be free from any kinematic contamination based on various observational diagnostics and the Bayesian modeling results.

\subsection{Step 1: Construction of a large sample of wide binaries with relatively precise radial velocities} \label{sec:raw_sample}

The collection includes our new observations carried out during 2024-2025 as well as public database/publications. New RVs are available for 60 wide binaries from LCO (Appendix~\ref{sec:LCO}) and for 6 wide binaries from MAROON-X (Appendix~\ref{sec:MAROONX}). For the case of LCO observations, RVs were measured independently at two epochs separated by a few months for 18 of them. We also collect wide binaries with precise RVs from public databases and individually published results: 195 wide binaries within 300~pc from APOGEE (Appendix~\ref{sec:APOGEE}), 32 wide binaries with HARPS RVs from \cite{Saglia:2025}, and 24 wide binaries from \cite{Scarpa:2017} after excluding obvious chance-alignment and kinematically contaminated cases (see Appendix~\ref{sec:Scarpa}). Because some wide binaries are included more than once in various samples, we have 306 unique wide binaries with relatively precise RVs (compared to Gaia DR3 RVs). This combined sample will be our raw or scratch sample.

\begin{figure}[!htb]
    \centering
    \includegraphics[width=1.\linewidth]{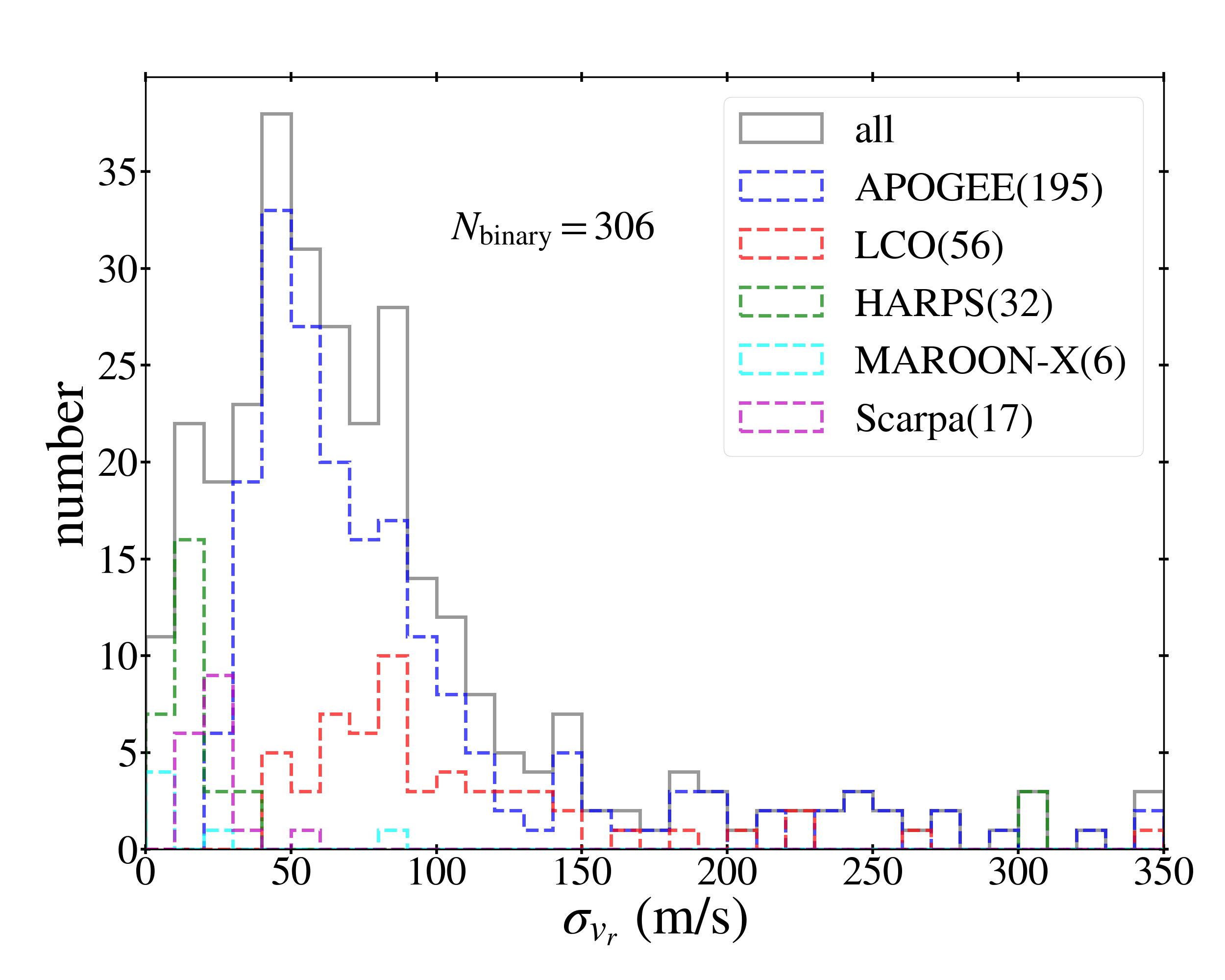}
    \caption{
   The distribution of nominal uncertainties of $v_r(\equiv {\rm RV}_A - {\rm RV}_B)$ for the raw sample of 306 unique wide binaries assembled from various observations and database/publications.
    }
    \label{fig:vr_err}
\end{figure}

\begin{figure*}[!htb]
    \centering
    \includegraphics[width=0.9\linewidth]{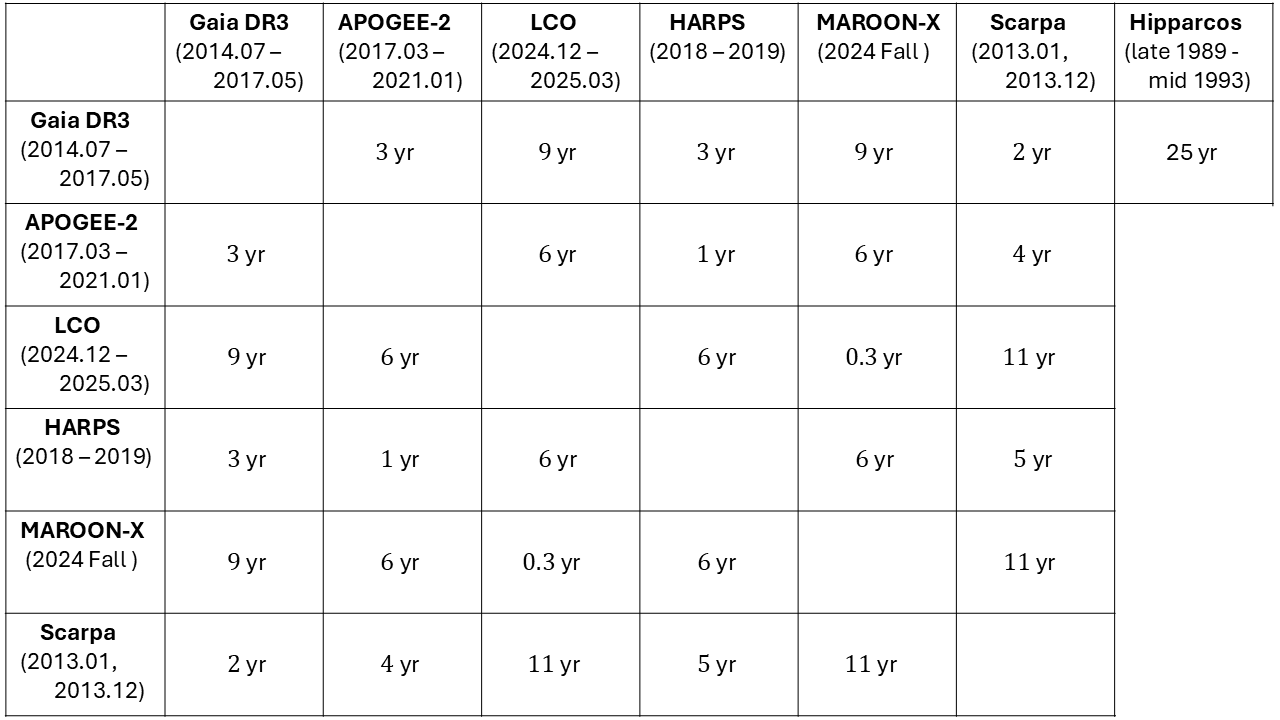}
    \caption{
    Summary of individual (given in parentheses at the top row and the left column) and pairwise time baselines for the various observations of radial velocities used. All given entries represent approximate median values. Hipparcos epoch is for observations of proper motions.
    }
    \label{fig:time}
\end{figure*}

Figure~\ref{fig:vr_err} shows the distribution of the reported nominal uncertainties of $v_r$ for the systems in the raw sample. All systems satisfy $\sigma_{v_r} < 350$~m\,s$^{-1}$ by selection, and the majority have $\sigma_{v_r} < 100$~m\,s$^{-1}$. Table~\ref{tab:WB_overlap} lists wide binaries for which independent measurements of $v_r$ were made with different instruments at different epochs. Time baselines between different observations considered in this work are summarized in Figure~\ref{fig:time}. For all but one system (the fourth) two independent values are consistent with each other within 2.5 times the combined error. The first (obs1) values may be more accurate (and are more precise in most cases), and so only the first values are included in the raw sample. The fourth system (Gaia DR3 5607190344506642432 \& 5607189485513198208) with the HARPS value of $v_r=-0.552\pm0.003$~km\,s$^{-1}$ from \cite{Saglia:2025} will be used despite a $3.5\sigma$ inconsistency with the Scarpa value of $v_r=-0.441\pm0.032$~km\,s$^{-1}$ from \cite{Scarpa:2017}, because the HARPS measurements are generally reliable and the magnitude of the difference ($111\,{\rm m}\,{\rm s}^{-1}$) is relatively small.  

\startlongtable
\begin{deluxetable*}{lcclc}
\tablecaption{\textbf{Wide binaries with two measurements of $v_r$ from independent observations with different instruments}}\label{tab:WB_overlap} 
\centerwidetable
\tabletypesize{\footnotesize}
\startdata
Gaia DR3 identifier & $v_r$\tablenotemark{a}  & $v_r$\tablenotemark{a}  & obs1/obs2\tablenotemark{b} & time baseline\tablenotemark{c} \\
Star A,B &   obs1  &   obs2   &    &    \\
     &  (km\,s$^{-1}$) & (km\,s$^{-1}$) &   & (yr) \\
\hline
2776055105362407680,2776054899203977728 & $-0.458 \pm 0.017$ & $-0.605 \pm 0.150$ & HARPS/LCO  & 3 \\
4940794866807373952,4940794488850252928 &  $0.369 \pm 0.004$ &  $0.387 \pm 0.069$ & HARPS/LCO  & 3 \\
5060104351007433472,5060105897197110144 & $-0.404 \pm 0.035$ & $-0.377 \pm 0.024$ & HARPS/Scarpa  & 5 \\
5607190344506642432,5607189485513198208 & $-0.552 \pm 0.003$ & $-0.441 \pm 0.032$ & HARPS/Scarpa  & 5 \\
3285218186904332288,3285218255623808640 &  $0.510 \pm 0.017$ &  $0.572 \pm 0.020$ & HARPS/Scarpa  & 5 \\
4249652990144051840,4249652783985617920 &  $0.070 \pm 0.004$ &  $0.061 \pm 0.026$ & HARPS/Scarpa  & 5 \\
2201661297490051968,2201661091331626752 & $-0.018 \pm 0.006$ &  $0.002 \pm 0.110$ & MAROON-X/LCO  & 0.3 \\
1172915990414659328,1172920487244742912 &  $0.043 \pm 0.033$ & $-0.027 \pm 0.077$ & APOGEE/LCO   & 6 \\
1282815063829295360,1282817022334383232 & $-0.283 \pm 0.041$ & $-0.170 \pm 0.021$ & LCO/Scarpa   & 11 \\
3230677565443833088,3230677874682668672 & $-0.076 \pm 0.057$ & $-0.135 \pm 0.022$ & LCO/Scarpa   & 11 \\
3550081879381593728,3550084490721711872 & $-0.595 \pm 0.082$ & $-0.410 \pm 0.029$ & LCO/Scarpa   & 11 \\
\enddata
\tablenotetext{a}{Relative radial velocity between the pair $v_r\equiv{\rm RV}_A-{\rm RV}_B$ without any correction for GR and CB.} \tablenotetext{b}{Two independent observations at different epochs. Only the value from obs1 will be used in this work.} \tablenotetext{c}{Approximate median time baseline between obs1 and obs2.}
\end{deluxetable*}

We note that the acquired values of $v_r$, except for those from \cite{Saglia:2025}, do not include corrections for gravitational redshifts (GR) and convective blueshifts (CB) in stellar atmospheres. In our preliminary modeling, we will not consider GR+CB corrections in $v_r$ not from the HARPS sample but quadratically add $40\,{\rm m}\,{\rm s}^{-1}$ to all nominal uncertainties of $v_r$ when using them for modeling as suggested for the HARPS $v_r$. However, we will consider GR+CB corrections in our refined modeling to infer gravity. We also note that the measured 3D velocity components suffer from minor geometric effects known as perspective effects \citep{Shaya:2011,Yoon:2025}, and thus we actually use velocity components corrected for the perspective effects.

Individual Bayesian 3D modeling is carried out for each binary from the raw sample either with a fixed value of $\Gamma$ (including the Newtonian case) or allowing it to vary within the range $-1<\Gamma<1$. These Bayesian outputs will be used in the selection of wide binaries in the low-acceleration regime and in the derivation of the value of $\Gamma$ through statistical consolidation. Since the Bayesian outputs provide orbit solutions, we can estimate various quantities for each system, including the Newtonian gravitational acceleration defined above and the Newtonian escape velocity 
\begin{equation}
v_{\rm escN}(r)\equiv \sqrt{\frac{2 G_{\rm N} M_{\rm tot}}{r}}     
    \label{eq:escN}
\end{equation}
where $r$ is estimated from the Bayesian outputs.

We note that the $\alpha$-Cen AB - Proxima system \citep{Kervella:2017} may satisfy our selection (as the AB system can be treated as a single object) but is not formally included in our sample. There are a few reasons for this. First, as \cite{Kervella:2017} showed assuming Newtonian gravity, Proxima's orbit is likely to be near its apastron phase meaning that it is one of common systems to be found in the Galaxy. \cite{Kervella:2017} obtained $e = 0.50_{-0.09}^{+0.08}$,  {$a=8.7_{-0.4}^{+0.7}\,{\rm kau}$} with $G=G_{\rm N}$ based on 3D velocities transformed to the Galactic coordinate system.  {These results are confirmed in Newtonian gravity as described in \cite{Chae:2026}. However,  \cite{Chae:2026} also obtains $e = 0.60_{-0.09}^{+0.08}$, $a=8.1_{-0.4}^{+0.6}\,{\rm kau}$ with $\gamma=1.255$ (the QUMOND prediction based on Figure~\ref{fig:QUMOND}), and $e = 0.64_{-0.08}^{+0.07}$, $a=7.9_{-0.5}^{+0.6}\,{\rm kau}$ with $\gamma=1.4$.} Because fit qualities in the three cases  {($\gamma=1.0$, $1.255$, and $1.4$)} are indistinguishable, these results cannot distinguish the assumed gravity models other than that for the 3D separation of $\approx 12.9\,{\rm kau}$, $e\ga 0.6$ would be more common than $e\la 0.5$. Second, since this is an exceptional system with precise values of all six components of $(\mathbf{r},\mathbf{v})$, it is appropriate for an individual test of gravity in the future based on long-term observational monitoring of the orbit (see, e.g., \citealt{BanikZhao:2018}). Also, the measurements used by \cite{Kervella:2017} and us here need careful reevaluations (see \citealt{Akeson:2021}).

\subsection{Step 2: Selection of the basic-cut sample of wide binaries in the low-acceleration regime} \label{sec:basic_sample}

\begin{figure}[!htb]
    \centering
    \includegraphics[width=1.\linewidth]{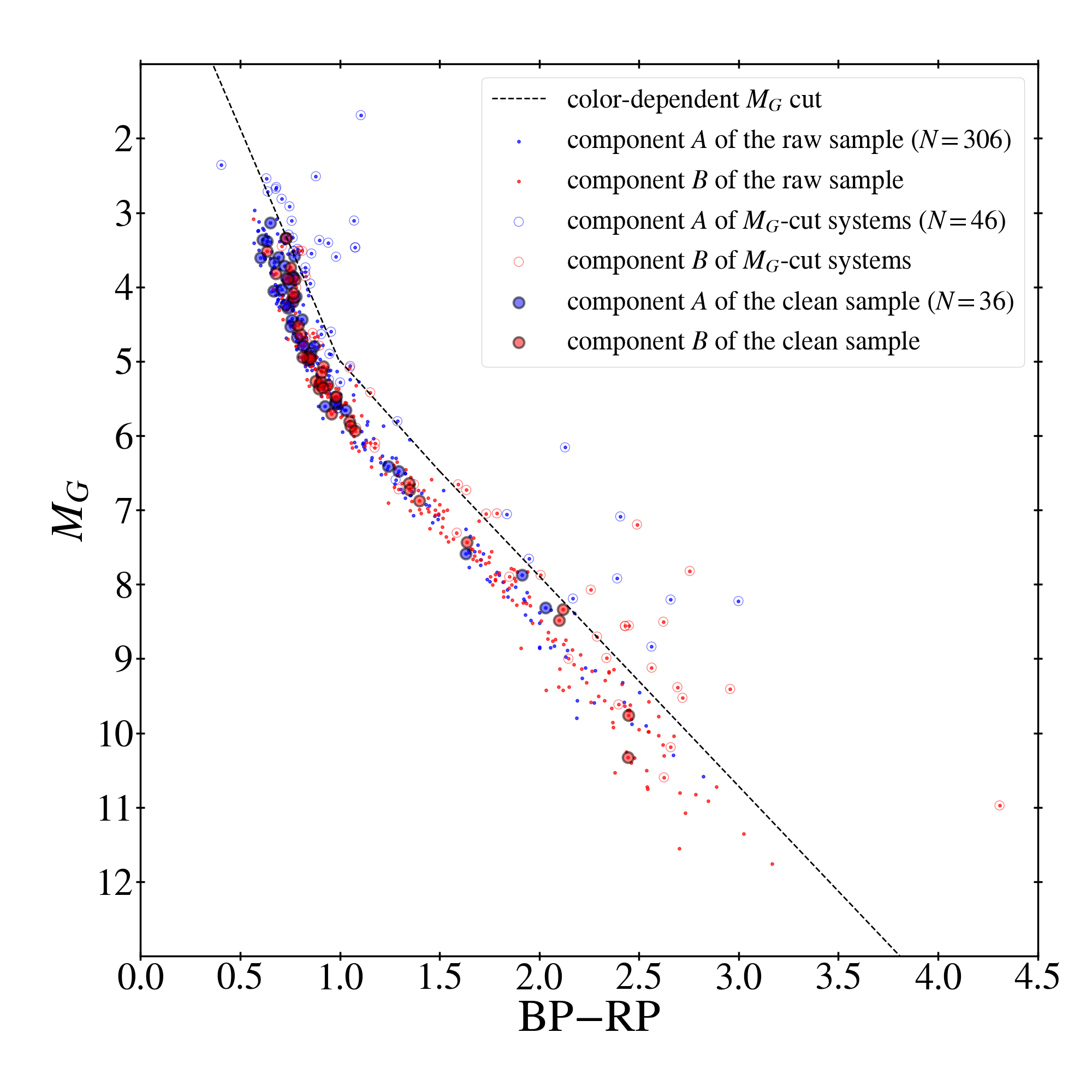}
    \caption{
   This figure shows the color-magnitude diagram of stars in the raw sample of 306 wide binaries using Gaia's BP$-$RP color and absolute magnitude $M_G$ in Gaia's $G$ band. All stars in the clean sample are fainter than the color-dependent $M_G$ cut line. 
    }
    \label{fig:CM}
\end{figure}

In the second step, we select a statistical sample of wide binaries in the low-acceleration regime that have sufficiently precise $v_r$ and are likely to be relatively free of kinematic contaminants based on various observational diagnostics available at present. We first require basic selection and quality cuts of the following:
\begin{itemize}
    \item Newtonian gravitational acceleration $g_{\rm N}<10^{-9}$m\,s$^{-2}$ where $g_{\rm N}$ is derived from Bayesian reconstructed orbits in generalized gravity.
    \item Distance of the binary system from the Sun is less than 150~pc.
    \item Relative RV $v_r$ and relative 3D velocity $v_{\rm obs}=|\mathbf{v}_{\rm obs}|$ have measurement errors less than $100$~m\,s$^{-1}$.
    \item Both stars have Gaia's {\tt ruwe} $<1.25$. (This threshold follows \cite{Saglia:2025} but a slightly different threshold such as $<1.20$ or $<1.30$ yields consistent results.)
    \item Both stars are below a color-dependent luminosity cut line in the main sequence of the color-magnitude diagram as shown in Figure~\ref{fig:CM}. This threshold helps remove unresolved close binaries, i.e., photometric binaries.
\end{itemize}

The above selection criteria retain only wide binaries of low internal acceleration that are relatively nearby and thus have highest data qualities. The criteria also help remove cases with potential kinematic contaminants towards the goal of selecting pure binary individuals. Only 75 systems (about 25\%) of the raw sample pass the above cuts. This is our basic-cut sample. Since GR and CB corrections are available only for 8 wide binaries from the HARPS sample, we now consider GR and CB effects for the rest of the sample. When the two stars of a binary are not of the same type, the GR+CB effects for the two stars are different and the relative effect introduces a bias in the observed relative radial velocity $v_r$. This bias can either increase or decrease the magnitude of $v_r$ for an individual system, but the mean/median change in a population is expected to be close to zero as we will show below.

For 24 wide binaries from the APOGEE sample, we use the measured values of the surface gravity and the effective temperature provided by the SDSS DR17 APOGEE database to estimate the GR and CB effects. For the rest of 43 wide binaries without either HARPS or APOGEE data, we estimate realistic uncertainties of the uncorrected $v_r$ as a function of mass ratio using the statistics of the GR+CB corrections of $v_r$.

With {\tt logg} (which is the logarithm of the surface gravity in units of ${\rm cm}\,{\rm s}^{-2}$) and mass $M$ of a star, the GR effect is given by 
\begin{equation}
    V_{\rm GR} = 636 \left(\frac{(M/{\rm M}_\odot)10^{\tt logg}}{2.742\times 10^4} \right)^{1/2} {\rm m}\,{\rm s}^{-1}.
    \label{eq:GR}
\end{equation}
The uncertainty of $V_{\rm GR}$ is estimated to be $15\,{\rm m}\,{\rm s}^{-1}$ considering the uncertainties of $M$ and {\tt logg}.

We estimate the CB effect using the empirical relation derived by \cite{Liebing:2021} based on the HARPS spectra of 810 F and G stars. With the effective temperature $T_{\rm eff}$, the CB effect is given by 
\begin{equation}
    V_{\rm CB} = -350\,{\rm m}\,{\rm s}^{-1}\left[0.258\left(\frac{T_{\rm eff}-4400\,{\rm K}}{1000\,{\rm K}} \right)^3 +0.233\right],
    \label{eq:CB}
\end{equation}
for $4094\,{\rm K}\le T_{\rm eff}\le 5970\,{\rm K}$. For $T_{\rm eff}$ outside this range, we use the value at the corresponding limit of the range based on Figure~6 of \cite{Liebing:2021}. We estimate the scatter around the median relation given by Equation~(\ref{eq:CB}) as follows: $\sigma_{\rm CB}=45.1+0.0192(T_{\rm eff}-T_0)+0.0000174(T_{\rm eff}-T_0)^2\,{\rm m}\,{\rm s}^{-1}$ with $T_0=5265\,{\rm K}$ for $4200\,{\rm K}\le T_{\rm eff} \le 6101\,{\rm K}$, and the corresponding limiting value for $T_{\rm eff}$ outside the range. This scatter (of $\approx 45\,{\rm m}\,{\rm s}^{-1}$ typically) will be used as the error of $V_{\rm CB}$.

\begin{figure}[!htb]
    \centering
    \includegraphics[width=1.03\linewidth]{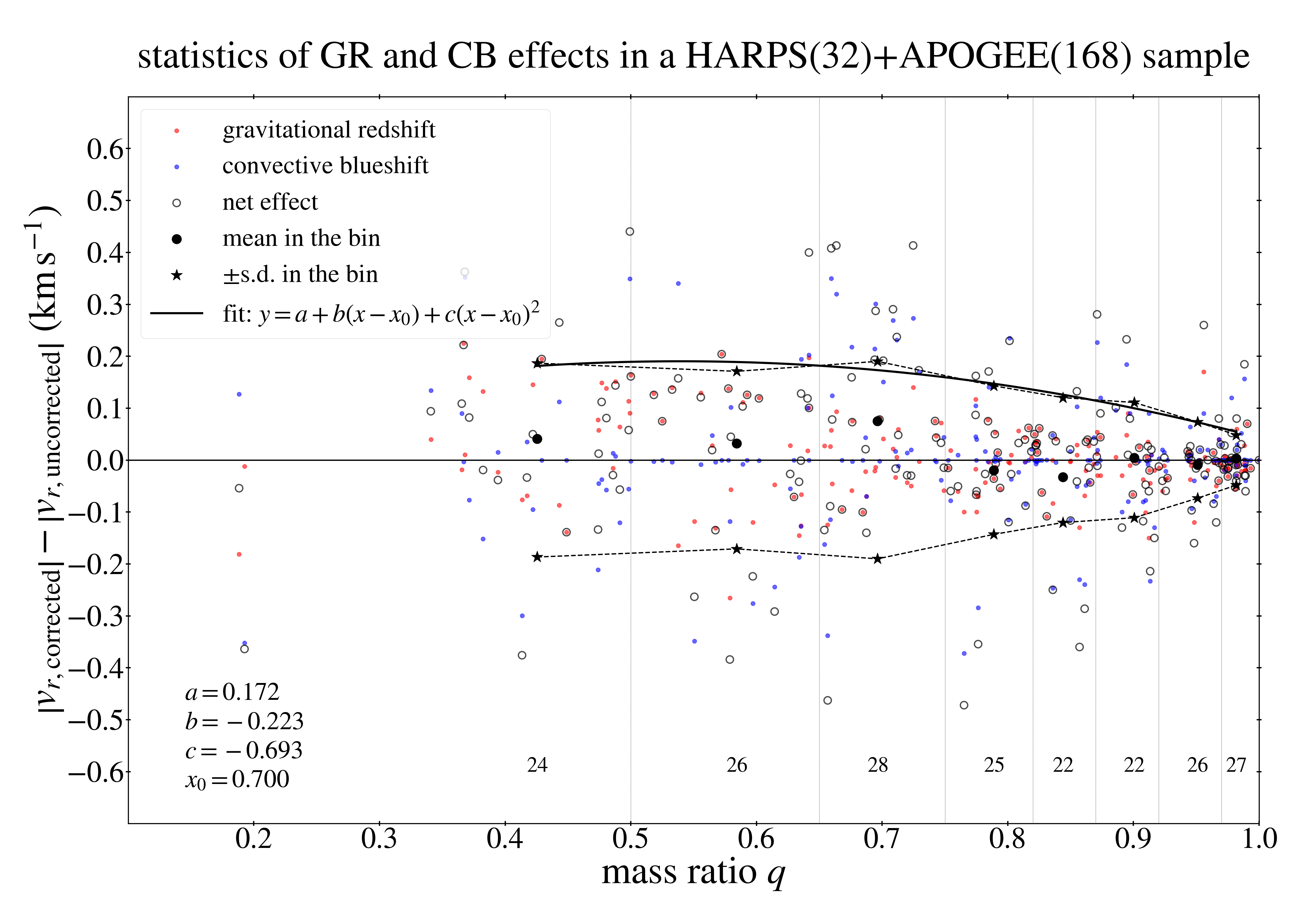}
    \caption{
   This figure shows the the effects of GR and CB and the combined effect for HARPS and APOGEE wide binaries (taken from the raw sample) whose GR and CB corrections are available. Here the change in the \emph{magnitude} of the relative radial velocity between the two stars is displayed.
    }
    \label{fig:RVcorrections}
\end{figure}

An observed RV is corrected to ${\rm RV}-V_{\rm GR}-V_{\rm CB}$. Then, the relative velocity $v_r = {\rm RV}_A - {\rm RV}_B$ is corrected to $v_r +\Delta v_r$ with $\Delta v_r = (V_{{\rm GR},B}- V_{{\rm GR},A})+(V_{{\rm CB},B}- V_{{\rm CB},A})$. Both $v_r$ and $\Delta v_r$ can be either positive or negative and what matters most in gravity tests is whether the GR and CB effects increase or decrease the \emph{magnitude} of $v_r$. Figure~\ref{fig:RVcorrections} shows the effects of GR and CB for 200 wide binaries whose GR and CB effects are available from \cite{Saglia:2025} or estimated here using Equations~(\ref{eq:GR}) and (\ref{eq:CB}) for the available APOGEE values of {\tt logg} and $T_{\rm eff}$ from the raw sample. The mean and median of the change in the magnitude are close to zero for the entire sample but they are slightly negative for the subsample with mass ratio $q \ga 0.75$ which is more relevant for the sample to be used for gravity tests. As expected, the scatter of the change in the $v_r$ magnitude increases as $q$ decreases. We will use this $q$-dependent scatter as the added uncertainty of $v_r$ when the GR+CB correction is not available for a specific system. 

The basic-cut sample is listed in Table~\ref{tab:sample}. In the table, we provide the measured relative radial velocity $v_r$ and its GR+CB correction along with its estimated error when the correction is available or just the scatter shown in Figure~\ref{fig:RVcorrections} when the correction is not available. In the table, we also provide the Bayesian inferred values of $\log_{10}g_{\rm N}$ and $\Gamma$, the observational source of $v_r$, and the selection merits for the final clean sample to be described below. Table~\ref{tab:physparameters} provides the physical scales, masses, and the ratio
\begin{equation}
    \eta_r \equiv v_{\rm obs}/v_{\rm escN}(r),
    \label{eq:eta_r}
\end{equation}
where $v_{\rm obs}$ is the 3D relative velocity corrected for the perspective effect (and the GR+CB effect when it is available) and $v_{\rm escN}(r)$ is the Newtonian escape velocity given by Equation~(\ref{eq:escN}). Here we consider Bayesian inferred values of the 3D separation $r$ as well as the sky-projected 2D separation $s$ that provides a model-independent observational lower bound on $r$ and thus an upper bound on $v_{\rm escN}(r)$ and a lower bound on $\eta_r$. In Table~\ref{tab:physparameters}, we provide the values of $\eta_s$ and $\eta_r$ along with the values of $s$ and other measured physical parameters. The values of $r$ as well as the fitted parameters can be found in Table~\ref{tab:fitparameters}. These tables give results not only for the general gravity model but also for two fixed gravity models of $G=G_{\rm N}$ and $G=1.4G_{\rm N}$.

\begin{figure}[!htb]
    \centering
    \includegraphics[width=1.0\linewidth]{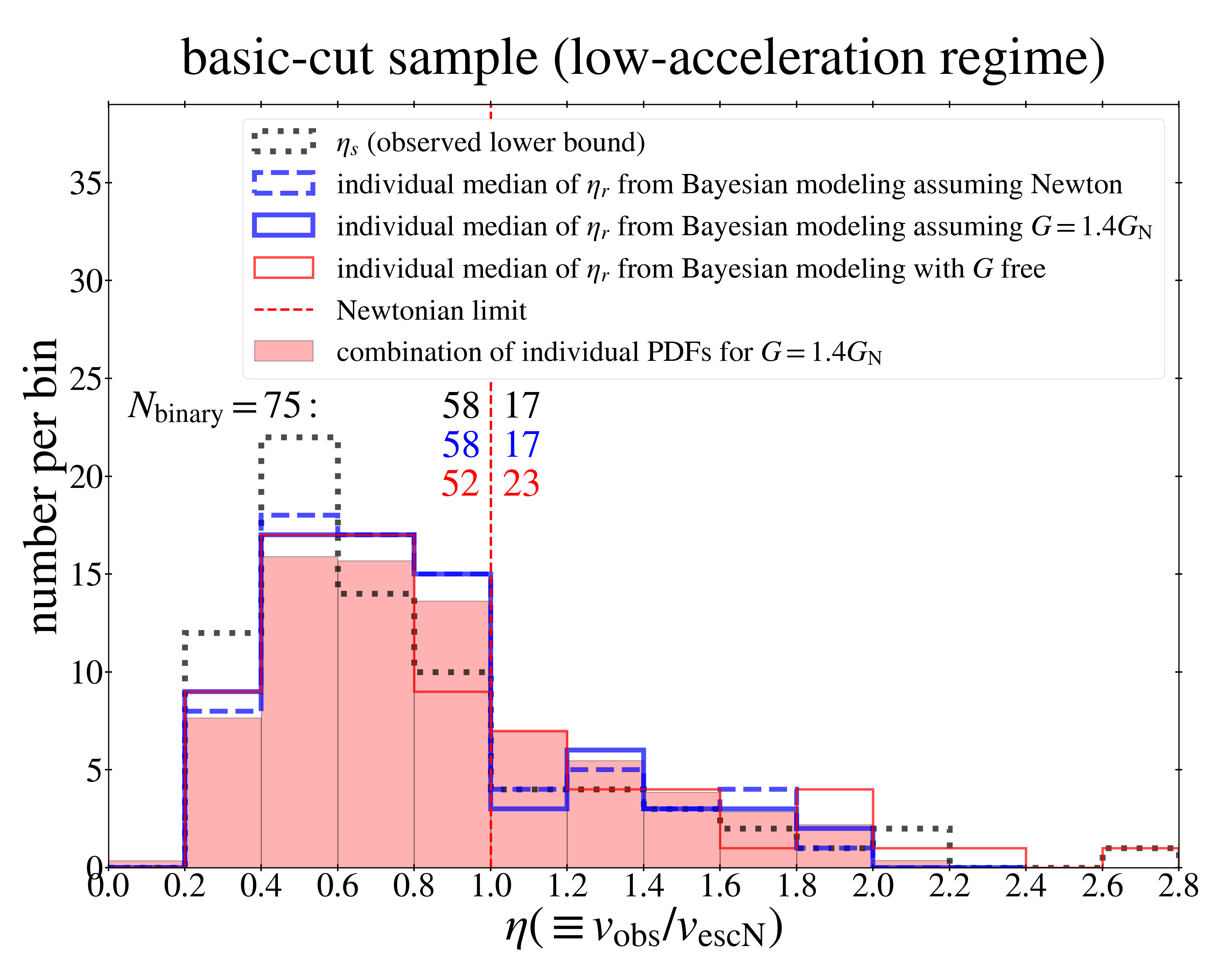}
    \caption{
    The distribution of $\eta(\equiv v_{\rm obs}/v_{\rm escN})$ is shown for the basic-cut sample of 75 wide binaries that may include kinematically contaminated cases. Here $v_{\rm obs}$ is the magnitude of the observed relative 3D velocity between the pair while $v_{\rm escN}$ (Equation~(\ref{eq:escN})) is the Newtonian escape velocity that depends on the mass of the two stars and their physical separation, for which we consider the observed sky-projected separation $s$ as well as the 3D separation $r$ predicted by 3D elliptical orbit modeling. See the text for the discussion of this figure.
    }
    \label{fig:v_over_vescN_basic}
\end{figure}

The distributions of $\eta_r$ and $\eta_s$ for the basic-cut sample can be found in Figure~\ref{fig:v_over_vescN_basic}. It shows that this sample includes cases violating the Newtonian limit ($>1$). The fraction $f(\eta_s>1)=f(\eta_r>1)=17/75=0.23$ (for fixed gravity models) or $f(\eta_r>1)=23/75=0.31$ (with $G$ free) is higher than expectations from empirical statistics. For a sample within 300\,pc satisfying ${\tt ruwe}<1.2$ and a CMD cut, \cite{Pittordis:2025} estimate a triple fraction of $\approx 0.19$ using a low-acceleration subsample by simultaneously fitting triple and flyby fractions, while \cite{ChaeYoon:2026} estimate $0.10\pm0.03$ using a Newtonian-regime subsample that is free of flybys. Moreover, for a higher-quality sample within 150\,pc which is more relevant to our basic-cut sample, \cite{ChaeYoon:2026} estimate $0.03\pm0.03$. These results indicate that the expected contamination fraction is $\la 0.1$ and thus at least one half (or $\ga 8$) of the $\eta_r>1$ cases are likely to be gravitationally-bound pure binaries.

\begin{figure}[!htb]
    \centering
    \includegraphics[width=1.0\linewidth]{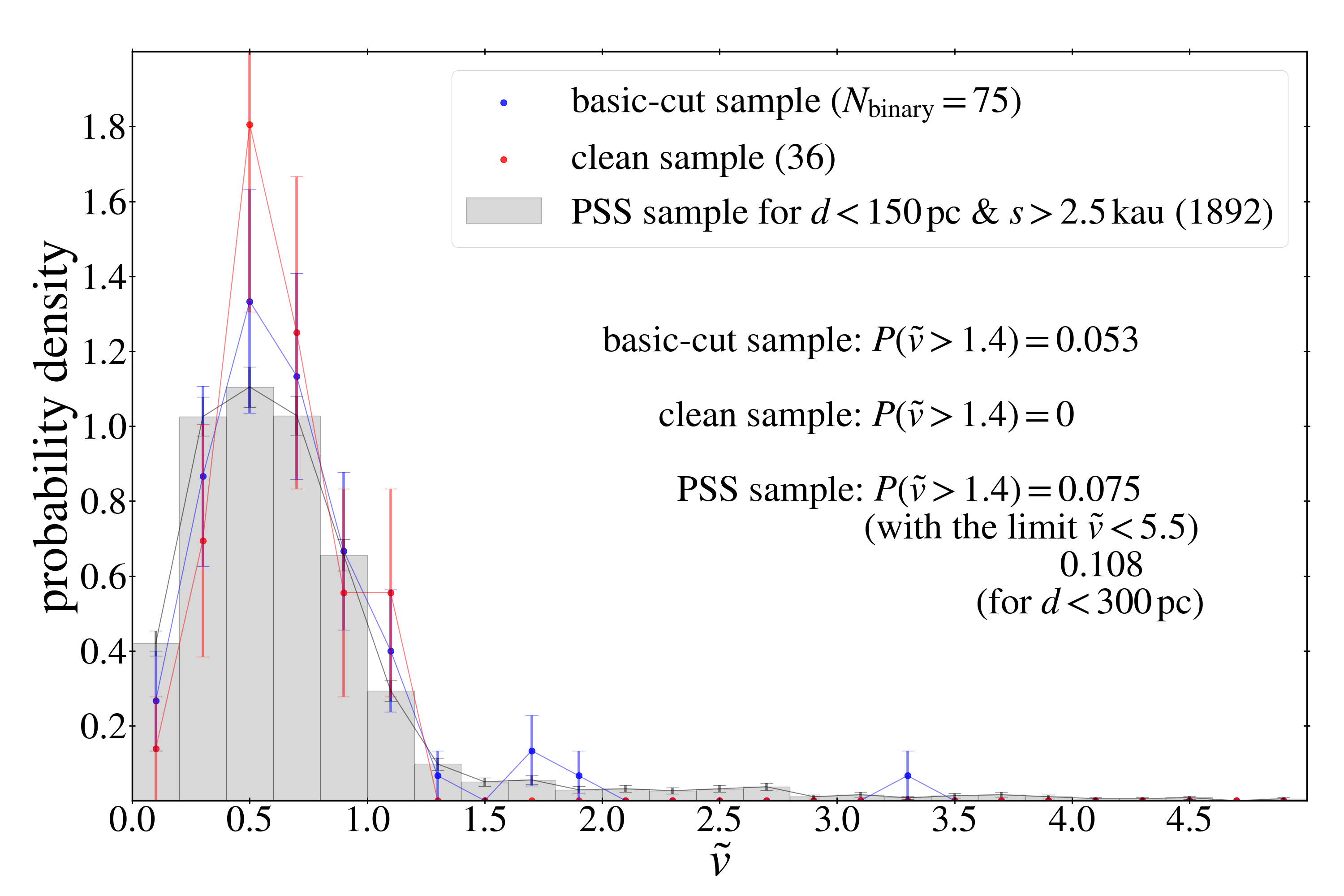}
    \caption{
    Distribution of $\tilde v$ ($\equiv v_p/v_c$, i.e.\ the 2D sky-plane velocity $v_p\equiv\sqrt{v_{x^\prime}^2 + v_{y^\prime}^2})$ over the Newtonian circular velocity $v_c\equiv\sqrt{G_{\rm N}M_{\rm tot}/s}$) calculated at $s$ in our wide binaries is compared with that of a "PSS" sample satisfying the same distance limit $d<150\,{\rm pc}$ and a similar sky-plane separation limit $s>2.5\,{\rm kau}$ taken from \cite{Pittordis:2025} sample. Probability of $P(\tilde v > 1.4)=0.053$ in our basic-cut sample is lower than $0.075$ in the PSS sample (after removing $\tilde v > 5.5$ which are mostly flybys: see \cite{ChaeYoon:2026}). No cases with $>1.4$ (or even $>1.2$) are found in the clean sample of 36 wide binaries. The PSS sample for the whole distance range $d<300\,{\rm pc}$ has a higher value of $P(\tilde v > 1.4)=0.108$ consistent with a higher fraction of triples as noticed by \cite{ChaeYoon:2026}. All error bars are based on Poisson statistics in the bins. 
    }
    \label{fig:vtilde}
\end{figure}

From Figure~\ref{fig:v_over_vescN_basic} and Table~\ref{tab:physparameters}, we note that $\eta_r > \eta_s$ is satisfied for all systems when $G$ is free, but some systems have $\eta_r < \eta_s$ in fixed gravity models. Since $\eta_r > \eta_s$ must be satisfied for fixed masses, cases of $\eta_r < \eta_s$ in fixed gravity models mean that masses are forced to change in trying for the assumed gravity models to fit the observed 3D velocities. For $G=1.4G_{\rm N}$ ($G_{\rm N}$), there are five (eight) cases with $\eta_r/\eta_s\le0.98$. The five cases for $G=1.4G_{\rm N}$ are Binary {\#}18, {\#}68, {\#}71, {\#}73, and {\#}74. All but one system are from the Scarpa sample. These systems may be suspicious as they have extraordinarily large posterior values of $\log_{10}f_M$ when gravity is fixed ($G=G_{\rm N}$ or $1.4G_{\rm N}$) and $\Gamma$ ($>0.5$ in all cases) when it is free. None of the five cases satisfy any of the selection merits for the clean sample to be presented in the following subsection. Binary {\#}59 from the HARPS sample and included in the clean sample has $\eta_r/\eta_s=0.97$ for $G=G_{\rm N}$ but $\eta_r/\eta_s=1.00$ for $G=1.4G_{\rm N}$. 

Further insight on our sample can be gained by checking the distribution of the parameter $\tilde v$ that is based only on accurate sky-plane parameters (Figure~\ref{fig:vtilde}) and widely used in the wide binary community. Since Newtonian pure binaries satisfy $\tilde v < \sqrt{2}$, occurrence rate of $\tilde v >1.4$ (or a similar threshold) in a sample provides a useful statistical information. A system with $\tilde v >1.4$ indicates a boosted velocity (due to a hidden tertiary or modified gravity) or a state of not being gravitationally bound. The occurrence rate of $\tilde v >1.4$ in our basic-cut sample is $P(\tilde v > 1.4)=0.053$, which is lower than $0.075$ for a sample taken from \cite{Pittordis:2025} that satisfy similar selection criteria including $d<150\,{\rm pc}$ (we note that flyby fraction is corrected for in this estimate). Moreover, a \cite{Pittordis:2025} sample with their full distance range $d<300\,{\rm pc}$ has a higher $P(\tilde v > 1.4)=0.108$, which is twice that in the basic-cut sample. These statistics indicate that our basic-cut sample contains a smaller fraction of systems with hidden tertiaries than the PSS sample. Even if we consider the relatively high triple fraction of $\approx 0.19$ estimated by \cite{Pittordis:2025}, these statistics indicate that the fraction in our basic cut sample cannot be larger than $\approx 0.1$ that is one half that by \cite{Pittordis:2025}. Thus, consistent with our arguments above, at least one half of the $\eta_r > 1$ cases are likely to be pure binaries. These statistics are in line with the fitted/inferred values of the triple fraction by \cite{ChaeYoon:2026}. 

The aforementioned five suspicious cases based on $\eta_r/\eta_s\le0.98$ with $G=1.4G_{\rm N}$ have $\tilde v >1.4$ except for one. Our clean sample to be selected independently of $\tilde v$ has no cases with $\tilde v>1.4$ (the maximum value is $1.17$). This indicates that uncontaminated pure binaries (whatever gravity they obey) may have low (or zero) probability of $\tilde v >1.4$. If we consider both $\eta_r/\eta_s$ and $\tilde v$, our basic-cut sample may contain six suspicious systems. 

\startlongtable
\begin{deluxetable*}{rlcccccl}
\tablecaption{\textbf{Measured $v_r$ and Bayesian inferred $\Gamma$ in the basic-cut sample}\label{tab:sample}} 
\centerwidetable
\tabletypesize{\scriptsize}
\startdata
ID & Gaia DR3 identifier  &  $v_{r,\rm{DR3}}$\tablenotemark{a}   &  $v_r$\tablenotemark{b}(correction\tablenotemark{c}) & $\log_{10}g_{\rm N}$\tablenotemark{d} & $\Gamma$\tablenotemark{e} & obs\tablenotemark{f} & merits\tablenotemark{g} \\
  & Star A,B  & (km\,s$^{-1}$)  & (km\,s$^{-1}$) &  (m\,s$^{-2}$)  &   &   &    \\
 \hline
1 & 2642922251741817216,2642922286101533952 & $0.503\pm0.657$ & $-0.156\pm0.040$($-0.015\pm0.066$) & $-9.63_{-0.34}^{+0.10}$ & $-0.367_{-0.165}^{+0.263}$ & A &  4 \\  
2 & 3840226230398314368,3840219358450646656 & $4.099\pm3.087$ & $-0.140\pm0.085$($-0.012\pm0.066$) & $-11.70_{-0.30}^{+0.10}$ & $-0.699_{-0.144}^{+0.390}$ & A &  None \\  
3 & 3861880665230888960,3858878375716469120 & $0.527\pm0.327$ & $0.386\pm0.051$($0.291\pm0.090$) & $-10.58_{-0.30}^{+0.11}$ & $0.491_{-0.127}^{+0.207}$ & A &  4 \\  
4 & 315289980082496000,315289980082496256 & $-1.707\pm2.100$ & $-0.869\pm0.061$($-0.136\pm0.063$) & $-9.34_{-0.36}^{+0.10}$ & $0.400_{-0.119}^{+0.215}$ & A &  None \\  
5 & 2627346157705716352,2627346535662838272 & $-0.929\pm1.605$ & $0.145\pm0.057$($-0.015\pm0.065$) & $-9.12_{-0.41}^{+0.11}$ & $-0.030_{-0.117}^{+0.252}$ & A &  4 \\  
6 & 2581807955200716800,2581806619466249728 & $-3.116\pm99.002$ & $-0.011\pm0.072$($-0.103\pm0.063$) & $-10.89_{-0.40}^{+0.13}$ & $0.441_{-0.131}^{+0.220}$ & A &  None \\
\enddata
\end{deluxetable*}

\startlongtable
\begin{deluxetable*}{rccccccccccc}
\tablecaption{\textbf{Physical scale, mass, relative 3D velocity, and Newtonian escape velocity} \label{tab:physparameters}} 
\centerwidetable
\tabletypesize{\scriptsize}
\startdata
ID  & $s$\tablenotemark{a} &  $d_A$\tablenotemark{b} & $d_B$\tablenotemark{b}  &  $M_A$\tablenotemark{c} & $M_B$\tablenotemark{c} &  $v_{\rm obs}$\tablenotemark{d} & $v_{\rm escN}(s)$\tablenotemark{e} &  $\eta_s$\tablenotemark{f} & $\eta_r(G_{\rm N})$\tablenotemark{g} & $\eta_r(1.4G_{\rm N})$\tablenotemark{g} & $\eta_r(\Gamma)$\tablenotemark{g} \\
   & (kau) & (pc) & (pc) & ($M_\odot$)  & ($M_\odot$) & (km\,s$^{-1}$) & (km\,s$^{-1}$) &   &   &   &  \\
 \hline
1 & $4.581\pm0.005$ & $56.03\pm0.06$ & $55.99\pm0.06$ & $0.548$ & $0.528$ & $0.197\pm0.063$ & $0.646_{-0.017}^{+0.016}$ & $0.306_{-0.098}^{+0.099}$ & $0.366_{-0.119}^{+0.214}$ & $0.371_{-0.121}^{+0.227}$ & $0.341_{-0.110}^{+0.159}$ \\ 
2 & $48.488\pm0.111$ & $118.10\pm0.27$ & $117.89\pm0.29$ & $0.523$ & $0.494$ & $0.151\pm0.094$ & $0.193_{-0.010}^{+0.014}$ & $0.781_{-0.488}^{+0.498}$ & $0.885_{-0.553}^{+0.648}$ & $0.886_{-0.554}^{+0.649}$ & $0.869_{-0.544}^{+0.614}$ \\ 
3 & $19.322\pm0.046$ & $131.48\pm0.32$ & $131.12\pm0.33$ & $1.301$ & $0.922$ & $0.755\pm0.090$ & $0.452_{-0.015}^{+0.016}$ & $1.670_{-0.205}^{+0.212}$ & $1.667_{-0.203}^{+0.209}$ & $1.682_{-0.205}^{+0.216}$ & $1.843_{-0.249}^{+0.517}$ \\ 
4 & $3.346\pm0.006$ & $82.69\pm0.16$ & $83.07\pm0.16$ & $0.730$ & $0.389$ & $1.032\pm0.085$ & $0.770_{-0.020}^{+0.019}$ & $1.340_{-0.114}^{+0.118}$ & $1.334_{-0.115}^{+0.121}$ & $1.344_{-0.116}^{+0.124}$ & $1.467_{-0.151}^{+0.469}$ \\ 
5 & $2.975\pm0.008$ & $148.36\pm0.37$ & $147.12\pm0.35$ & $0.873$ & $0.654$ & $0.459\pm0.031$ & $0.954_{-0.025}^{+0.024}$ & $0.481_{-0.034}^{+0.035}$ & $0.534_{-0.050}^{+0.155}$ & $0.547_{-0.056}^{+0.198}$ & $0.527_{-0.048}^{+0.280}$ \\ 
6 & $19.842\pm0.058$ & $146.05\pm0.43$ & $145.38\pm0.71$ & $0.764$ & $0.450$ & $0.440\pm0.040$ & $0.329_{-0.012}^{+0.013}$ & $1.334_{-0.129}^{+0.136}$ & $1.324_{-0.124}^{+0.130}$ & $1.339_{-0.126}^{+0.134}$ & $1.492_{-0.171}^{+0.671}$ \\
\enddata
\tablenotetext{a}{The sky-projected separation of component B from A given by $s\equiv \Delta\theta\times d_A$ where $\Delta\theta$ is the angular separation between the pair and $d_A$ is the distance of component A} \tablenotetext{b}{Distances of the two components from the Sun}  \tablenotetext{c}{Masses of the two components} \tablenotetext{d}{The magnitude of the measured 3D velocity including the correction of the perspective effect and the GR+CB effect of $v_r$ (if available)} \tablenotetext{e}{The Newtonian escape velocity set by the observed masses and the separation $s$. This is the upper limit of the Newtonian escape velocity for the system. Here an uncertainty of 5\% is assumed for the total mass of the system. \tablenotetext{f}{$\eta_s\equiv v_{\rm obs}/v_{\rm escN}(s)$} \tablenotetext{g}{Bayesian inferred value of $\eta_r$ (Equation~(\ref{eq:eta_r})) in three gravity models. Here masses are allowed to vary with a Gaussian constraint.} } 
\end{deluxetable*}

\startlongtable
\begin{deluxetable*}{rccccccccc}
\tablecaption{\textbf{Fitted and derived parameters in three gravity models}\label{tab:fitparameters}} 
\centerwidetable
\tabletypesize{\scriptsize}
\startdata
ID & $G$\tablenotemark{a}   &  $r$\tablenotemark{b}   &  $a$\tablenotemark{c} &  $e$\tablenotemark{d} & $i$\tablenotemark{d} & $\phi_0$\tablenotemark{d} & $\Delta\phi$\tablenotemark{d} &  $\log_{10}f_M$\tablenotemark{d} & $v_{\rm mod}$\tablenotemark{e} \\
  &    &  (kau) & (kau)  &    &  (deg) &  (deg)  &  (deg)  &    &  (${\rm km}\,{\rm s}^{-1}$)  \\
 \hline
 \hline
1 & $G_{\rm N}$ & $5.97_{-1.28}^{+5.08}$ & $3.45_{-0.84}^{+3.66}$ & $0.857_{-0.106}^{+0.065}$ & $99.5_{-39.3}^{+20.1}$ & $164.2_{-68.5}^{+83.5}$ & $181.1_{-9.3}^{+10.8}$ & $-0.001_{-0.021}^{+0.021}$ & $0.184_{-0.055}^{+0.067}$ \\ 
   & $1.4G_{\rm N}$ & $6.12_{-1.40}^{+5.60}$ & $3.39_{-0.84}^{+3.70}$ & $0.893_{-0.080}^{+0.048}$ & $83.5_{-22.9}^{+35.2}$ & $164.8_{-67.4}^{+81.7}$ & $180.8_{-6.8}^{+7.8}$ & $-0.001_{-0.021}^{+0.021}$ & $0.187_{-0.056}^{+0.067}$ \\ 
  & $10^{2\Gamma}$ & $5.21_{-0.57}^{+2.52}$ & $4.61_{-1.51}^{+4.17}$ & $0.603_{-0.255}^{+0.220}$ & $80.6_{-34.5}^{+52.7}$ & $173.7_{-101.5}^{+109.3}$ & $-5.0_{-190.2}^{+202.9}$ & $-0.000_{-0.021}^{+0.021}$ & $0.138_{-0.032}^{+0.060}$ \\ 
\hline 
2 & $G_{\rm N}$ & $56.52_{-7.33}^{+28.23}$ & $36.94_{-9.39}^{+26.22}$ & $0.849_{-0.282}^{+0.116}$ & $90.2_{-20.4}^{+20.9}$ & $149.2_{-71.1}^{+101.5}$ & $101.5_{-283.2}^{+83.8}$ & $-0.001_{-0.021}^{+0.021}$ & $0.062_{-0.031}^{+0.047}$ \\ 
   & $1.4G_{\rm N}$ & $56.67_{-7.47}^{+28.53}$ & $36.61_{-9.21}^{+25.86}$ & $0.860_{-0.277}^{+0.109}$ & $90.1_{-18.1}^{+18.6}$ & $150.4_{-67.8}^{+95.5}$ & $118.9_{-300.1}^{+66.3}$ & $-0.001_{-0.021}^{+0.021}$ & $0.072_{-0.038}^{+0.055}$ \\ 
  & $10^{2\Gamma}$ & $54.81_{-5.78}^{+22.99}$ & $42.10_{-12.02}^{+32.04}$ & $0.736_{-0.309}^{+0.194}$ & $89.9_{-40.3}^{+40.3}$ & $153.2_{-96.5}^{+133.5}$ & $-30.8_{-157.6}^{+220.3}$ & $-0.000_{-0.021}^{+0.021}$ & $0.018_{-0.008}^{+0.018}$ \\
\enddata
\tablenotetext{a}{Gravitational parameter: the value of $\Gamma$ in the general model is given in Table~\ref{tab:sample}.} \tablenotetext{b}{Derived (model-predicted) 3D separation between the two stars}  \tablenotetext{c}{Derived semi-major axis of the elliptical orbit} \tablenotetext{d}{Fitted parameters. We note that the argument of ascending node is not given here because it is related to the parameter $\theta$ introduced in \cite{Chae:2025b} that can be obtained by Equation~(5) of that paper from the other parameters given here.} \tablenotetext{e}{Derived/predicted 3D velocity} 
\end{deluxetable*}

\subsection{Step 3: Selection of the \emph{clean} sample of wide binaries in the low-acceleration regime} \label{sec:clean_sample}

The driving philosophy of our sample selection is to maximally control systematic uncertainties even if statistical power is significantly sacrificed. Therefore, we require that each binary satisfy at least one of the following additional criteria to remove potentially contaminated or unreliable systems:
\begin{enumerate}[(1)]
    \item Speckle interferometric observations of the fields around both stars were carried out and no objects were detected. In addition, the measured value of $v_r$ (whatever its source is) is consistent with the Gaia DR3 value within $2\sigma$. 
    \item Both stars were observed at two epochs with the LCO and two independent values of $v_r$ are consistent with each other within $2\sigma$.
    \item $v_r$ values come from the HARPS RVs of the \cite{Saglia:2025} sample.
    \item $v_r$ values come from APOGEE, and value(s) of \texttt{VSCATTER} (scatter from multiple visits in an observation) is(are) available and its (maximum) value is less than $100$~m\,s$^{-1}$.
    \item Two independent values of $v_r$ are available from two different sources/telescopes and are consistent with each other (see Table~\ref{tab:WB_overlap}).
    \item Unusually precise Gaia DR3 values (with error $<200$~m\,s$^{-1}$) are available and agree with the values from other sources.
\end{enumerate}
Some systems satisfy multiple selection merits. In Table~\ref{tab:sample}, we indicate the selection merit(s) for each binary.

We note that the targets of Speckle observations were selected from the samples recently defined and used by two of us (e.g.\ \citealt{Hernandez:2024a,Chae:2024a}). These samples were selected with a number of indirect criteria to exclude apparent binaries hosting unresolved close companion star(s). The Speckle observations (Appendix~\ref{sec:Speckle}) show that only 5\% of the target systems potentially have additional faint star(s), confirming that the employed criteria were efficient in selecting samples of pure binaries in a \emph{statistical} sense. The fraction of 5\% is similar to the triple fraction of $3\pm3$\% obtained by \cite{ChaeYoon:2026} for a sample within 150\,pc satisfying similar criteria. By comparison, in random samples of local stars of comparable mass, Speckle studies generally identify about 40\% unresolved binaries.

Only 39 systems satisfy at least one of the requirements listed above. We manually check all objects found in the 3D space around each binary using the Gaia DR3 archive, and find that three pairs (Binary {\#}64, {\#}70, and {\#}75) have nearby third objects that may perturb the system. So we exclude these systems and are left with 36 systems, which will define our clean sample that also passes additional tests as will be shown in the following two subsections. Aside from the membership inclusion criteria mentioned above, no system having Speckle observations identifying any extra component (some of which might just be background sources) was included in the final clean sample. The properties of the clean sample will be described in Section 3.4.

\subsubsection{Additional check: Testing Gaia-Hipparcos proper motion consistency over 25 years} \label{sec:Hip_Gaia}

\begin{figure*}[!htb]
    \centering
    \includegraphics[width=0.8\linewidth]{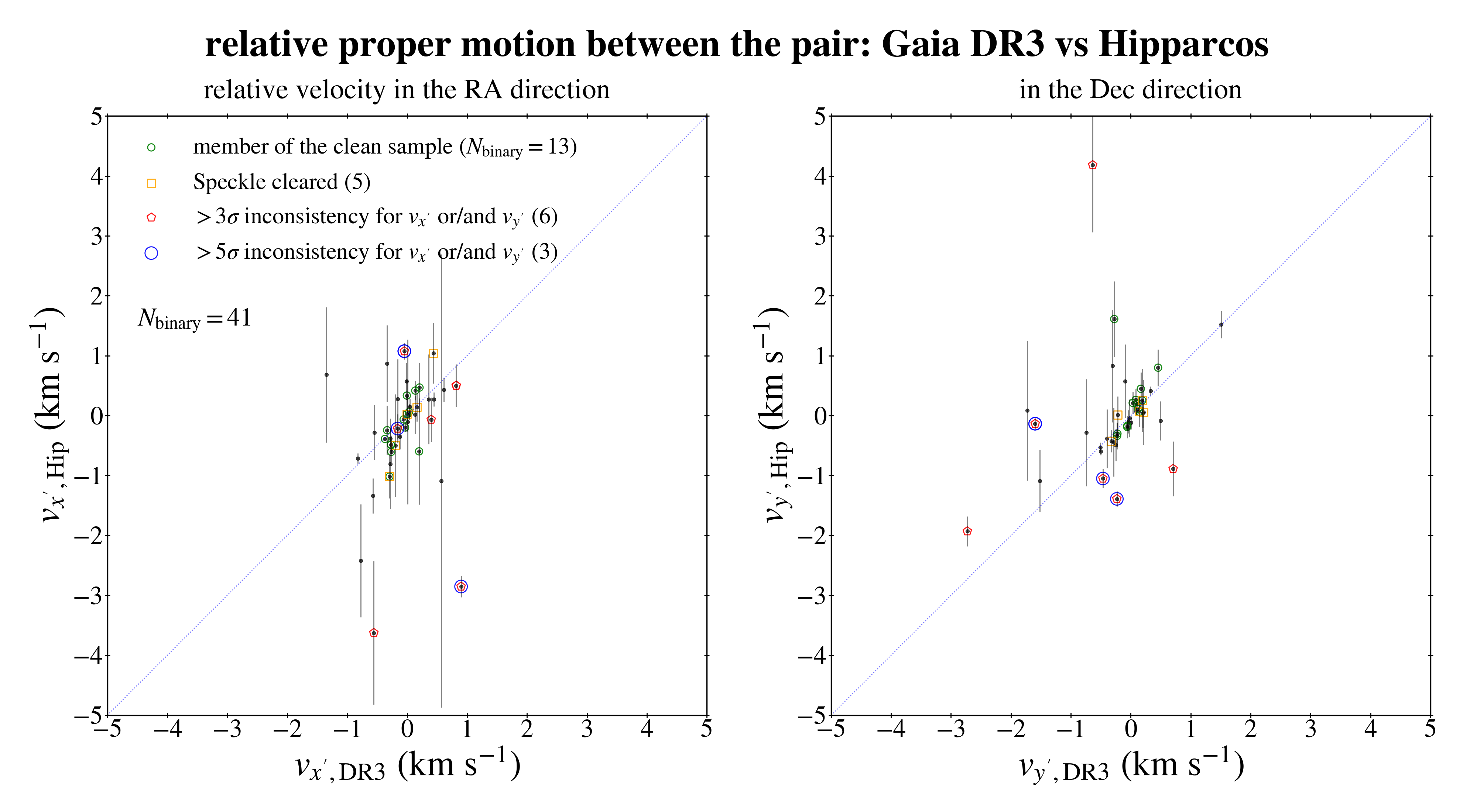}
    \caption{
    This figure compares sky-projected transverse velocities derived from the relative proper motions (PMs) between the pair reported in Gaia DR3 and Hipparcos for 41 wide binaries selected from the raw sample. The three systems with the $>5\sigma$ discrepancies between the two observations show abnormal properties in the posterior PDFs of the parameters from Bayesian modeling (see the text). 
    }
    \label{fig:PM_dr3_hip}
\end{figure*}

We check whether the Gaia DR3 proper motions (PMs) of the stars in the clean sample are consistent with the Hipparcos PMs measured about 25 years earlier than Gaia. Although the Hipparcos precision is poorer by an order-of-magnitude and not all binaries in the Gaia DR3 were observed by Hipparcos, the comparison between the two can provide an additional test. We only consider the relative PM between the pair for this test because it is free from issues of absolute calibrations between the two observations \citep{Brandt:2018} and what is relevant for internal dynamics is the relative velocity. We find that 41 binaries from the raw sample (including 13 that satisfy the above selection merits) have the Hipparcos PMs for both stars.

Figure~\ref{fig:PM_dr3_hip} shows the comparison between the two observations in terms of the sky-projected relative velocity components $v_{x^\prime}$ and $v_{y^\prime}$ derived from the PMs for 41 systems from the raw sample including 13 systems belonging to the clean sample. First of all, all 13 systems from the clean sample satisfy the Gaia-Hipparcos consistency at least within $3\sigma$, while there are cases that do not satisfy such a consistency.

Three binaries show $>5\sigma$ discrepancy between Gaia DR3 and Hipparcos, strongly indicating that these binaries have been undergoing kinematic perturbations during the time baseline of $\approx 25$ years (see Figure~\ref{fig:time}) between the two observations. If this is true, the Bayesian modeling results for these systems are very likely to display abnormal properties in the posterior PDFs of the parameters beyond the ranges predicted by any viable gravity models including Newton and MOND. Indeed, we find that the posterior PDFs for the systems return abnormally large values of $\Gamma > 0.5$ and $v_{\rm obs}/v_{\rm escN}>1.3$. Interestingly, all these systems are from the Scarpa sample (Gaia DR3 777967084390189696 \& 777967702865481344, Gaia DR3 6193279279612173952 \& 6193280031230266752, and Gaia DR3 1754191435419155456 \& 1754191229260708736: the former two are included in the basic-cut sample of Table~\ref{tab:sample} while the latter is not), and none of them satisfy any of the selection merits for the clean sample and were therefore excluded. We note also that two of these systems fail the velocity threshold introduced by \cite{Chae:2024a} (see Appendix~\ref{sec:Scarpa}).

While the $>5\sigma$ discrepancy between Gaia DR3 and Hipparcos PMs appears to be a decisive indicator of kinematic contaminants, the interpretation of the less strong discrepancy $>3\sigma$ (but not $>5\sigma$) is unclear. Three systems exhibit such discrepancies but no abnormalities in the PDF of $\Gamma$ or the inferred value of $v_{\rm obs}/v_{\rm escN}$. This may indicate that a $3\sigma$-threshold in the Gaia-Hipparcos comparison is not effective or reliable. However, the detail on the threshold appears to be irrelevant for the present clean sample because none of the 13 systems from the clean sample show a $>3\sigma$ discrepancy.

The remarkable agreement between the Gaia-Hipparcos $>5\sigma$ discrepancy and the Bayesian-inferred large value of $v_{\rm obs}/v_{\rm escN}\ga 1.3$ in identifying kinematically contaminated systems suggests that the PDF of $\Gamma$ (as was already discussed in \cite{Chae:2025}) or the quantity $v_{\rm obs}/v_{\rm escN}$ may be used when the direct observational test such as the Gaia-Hipparcos PM test is not available. As will be shown below, all systems from the clean sample have $v_{\rm obs}/v_{\rm escN}\la 1.2$ in concordance with the Gaia-Hipparcos test results.

\subsubsection{Additional check: Comparing metallicities of the two components} \label{sec:metal}

\begin{figure*}[!htb]
    \centering
    \includegraphics[width=1.0\linewidth]{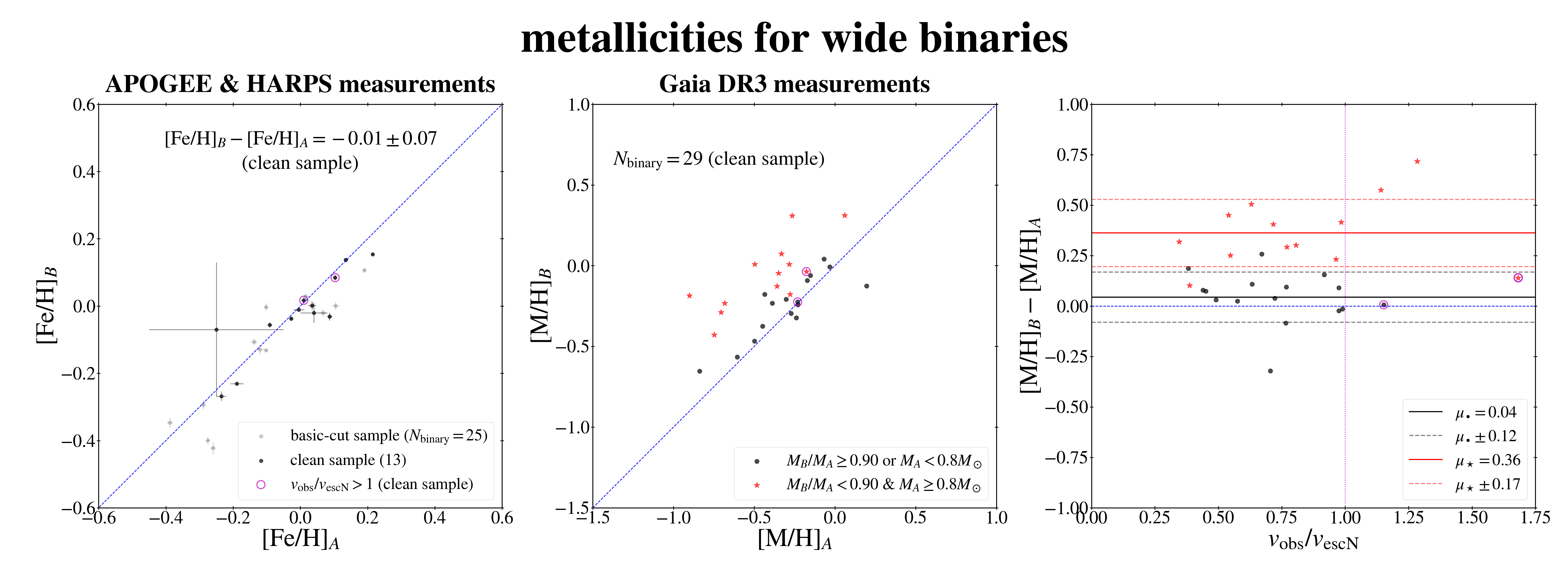}
    \caption{
    The left panel shows APOGEE and HARPS measurements of iron abundances [Fe/H] for the stars of 25 wide binaries from the basic-cut sample including 13 from the clean sample. The brighter ($A$) and the fainter ($B$) components of binaries show a negligible difference: ${\rm[Fe/H]}_B-{\rm[Fe/H]}_A=-0.01\pm 0.07$. Two systems (marked by encircled points) gravitationally unbound by Newtonian gravity ($v_{\rm obs}>v_{\rm escN}$) also have nearly equal values of [Fe/H] for the two components. The middel panel shows Gaia DR3 metallicities [M/H] \citep{Gaia:GSPP} for 29 wide binaries from the clean sample. There is a good correlation between the two components of the binaries. The right panel shows the systematic difference between the two components. The systematic difference is largely due to the binaries marked by red five-pointed stars in which component A is brighter (more massive) than component B, which is consistent with the known systematic for Gaia's [M/H] \citep{Gaia:GSPP}. See the text for the details.
    }
    \label{fig:metallicity}
\end{figure*}

While the formation and long-term survival mechanisms of gravitationally-bound wide binaries are not well understood, wide binary stars are thought to be formed in a common interstellar gas cloud, through mechanisms such as dissolution of star clusters (e.g.,\citealt{Kouwenhoven:2010,Moeckel:2011}, dynamical unfolding of triples \citep{Reipurth:2012} ), formation from adjacent cores \citep{Tokovinin:2017}, gravitational capture \citep{Rozner:2023}), and star formation in the turbulent interstellar medium \citep{Xu:2023}. If the two main-sequence stars of a wide binary are born together from a common cloud, they will have identical or similar chemical compositions regardless of their ages or formation mechanisms. Observational studies generally find that metallicities of the two stars match well (e.g., \citealt{Andrews:2019,Hawkins:2020,Nelson:2021,Lim:2024}).

Thus, we seek to check whether our wide binaries satisfy the expected consistency of metallicities between the two components. In general, high-SNR and high-resolution spectra are needed to measure metallicities of stars, but we collect metallicities from the public databases/publications for as many stars of our sample as possible. Figure~\ref{fig:metallicity} shows the metallicities collected from APOGEE,\footnote{https://www.sdss4.org/dr17/irspec/abundances/} HARPS \citep{Saglia:2025}, and Gaia DR3 \citep{Gaia:GSPP}. The former two provide accurate and precise results based on high-resolution spectra while the latter provides much less precise results.

The left panel of Figure~\ref{fig:metallicity} compares APOGEE or HARPS iron abundances [Fe/H] of the the brighter ($A$) and the fainter ($B$) components of 25 (17 APOGEE + 8 HARPS) wide binaries selected from the basic-cut sample that include 13 wide binaries of the clean sample. The metallicity difference ${\rm[Fe/H]}_B-{\rm[Fe/H]}_A$ is consistent with zero with an rms scatter $<0.1$. In particular, the 13 wide binaries of the clean sample have ${\rm[Fe/H]}_B-{\rm[Fe/H]}_A=-0.01\pm 0.07$. Moreover, the two systems with $v_{\rm obs}/v_{\rm escN}>1$ from the clean sample satisfy well the expected property ${\rm[Fe/H]}_B \simeq {\rm[Fe/H]}_A$ of true binaries.

The middle panel of Figure~\ref{fig:metallicity} shows Gaia DR3 metallicities [M/H] (where M represents all metal elements) for the 29 wide binaries of the clean sample for which this quantity is available for both components. The two metallicities ${\rm[M/H]}_A$ and ${\rm[M/H]}_B$ are well correlated, but there is a clear systematic shift for those with relatively large mass difference ($M_B/M_A < 0.9$) between the two components while there is no tangible shift for those with similar masses or those having relatively small masses regardless of mass ratio. The selective systematic difference is well consistent with the known systematic of the Gaia DR3 general stellar parameterizer for photometric (GSP-Phot) [M/H] as a function of effective temperature $T_{\rm eff}$. Figure~11 of \cite{Gaia:GSPP} shows that [M/H] for stars with $T_{\rm eff}\la 4500$K have no systematic shift while [M/H] has a systematic shift $\approx -0.4$ - $-0.3$ dex for $T_{\rm eff} > 4600$K, which corresponds to $M\approx 0.8 M_\odot$ (Figure~7 of \cite{Eker:2015}). Moreover, the same figure shows that the scatter around the median systematic shift is large $\ga 0.3$ dex. Thus, the measured statistic of ${\rm[M/H]}_B-{\rm[M/H]}_A=0.36\pm 0.17$ dex (i.e., the more massive star has a lower value) is well consistent with random fluctuations.

The above results demonstrate that all available metallicties for our clean sample of wide binaries are consistent with the current understanding of wide binary stars that the two components have identical or similar metallicities.

\subsection{Properties of the clean sample} \label{sec:property}

The clean sample was defined with various selection merits, and additional checks (Sections~\ref{sec:Hip_Gaia} and \ref{sec:metal}) using the comparison of Gaia and Hipparcos relative PMs and the comparison of metallicities of the two components have not found any noticeable issue in the clean sample. Thus, all 36  wide binaries from the clean sample will be used to infer gravity. Here we describe key statistical properties of the clean sample.

\begin{figure}[!htb]
    \centering
    \includegraphics[width=1.0\linewidth]{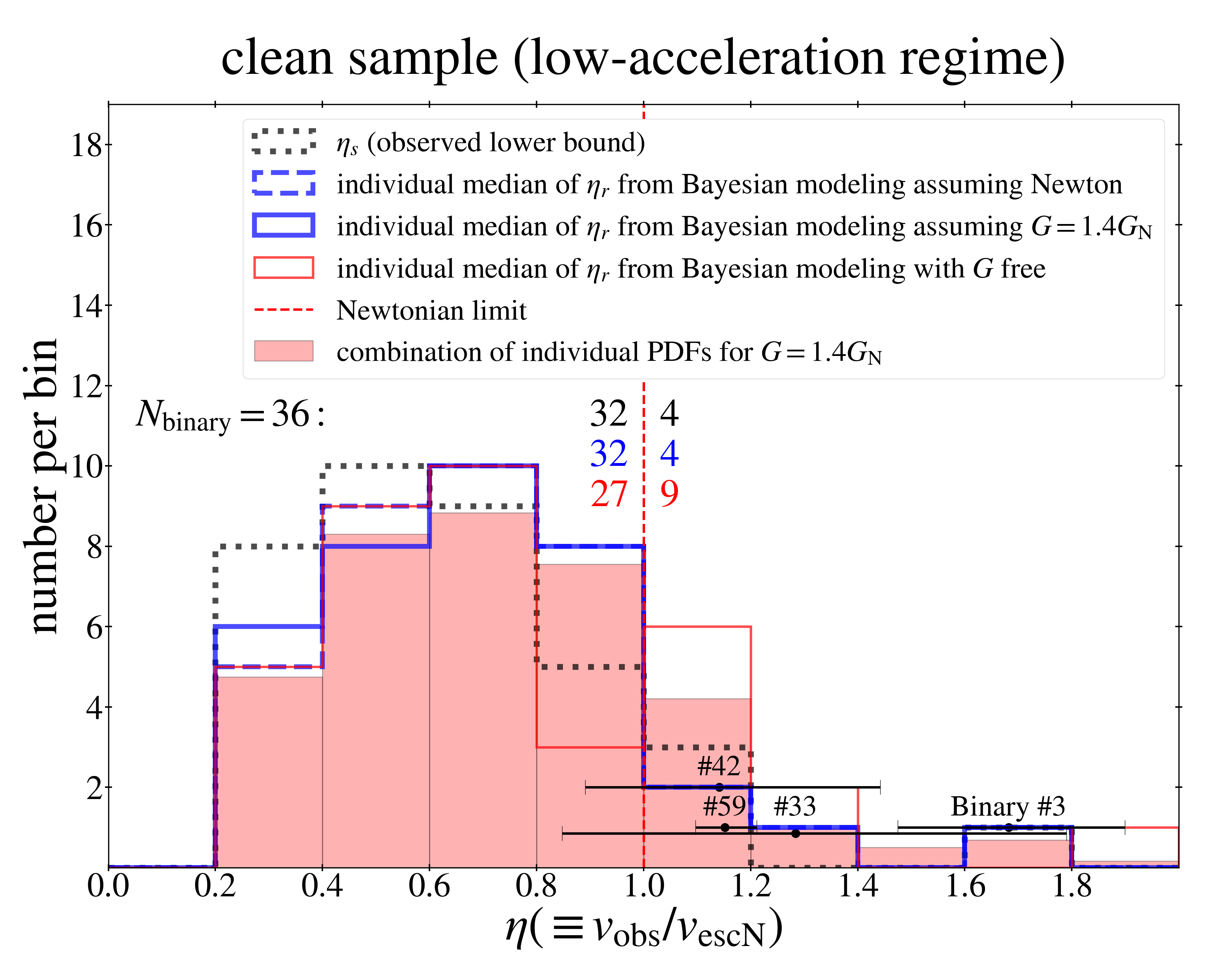}
    \caption{
    Same as Figure~\ref{fig:v_over_vescN_basic} but for the clean sample. Four systems have $\eta_r>1$ from the Bayesian results in fixed gravity models. The displayed individual values with error bars are for $G=1.4G_{\rm N}$. Considering the uncertainties of the individual values, there are only two clear cases of $\eta_r>1$: Binary {\#}3 and {\#}59. We note that several systems have values in the range $0.8<\eta_r<1.2$ containing the Newtonian limit within the uncertainties. The modeling results with free $G$ have twice as many cases of $\eta_r>1$ but most of them are in the range $1.0<\eta_r<1.2$ (containing $\eta=1$ within the uncertainties) and thus there are similar cases of $\eta_r>1.2$.
    }
    \label{fig:v_over_vescN_clean}
\end{figure}

\begin{figure*}[!htb]
    \centering
    \includegraphics[width=0.9\linewidth]{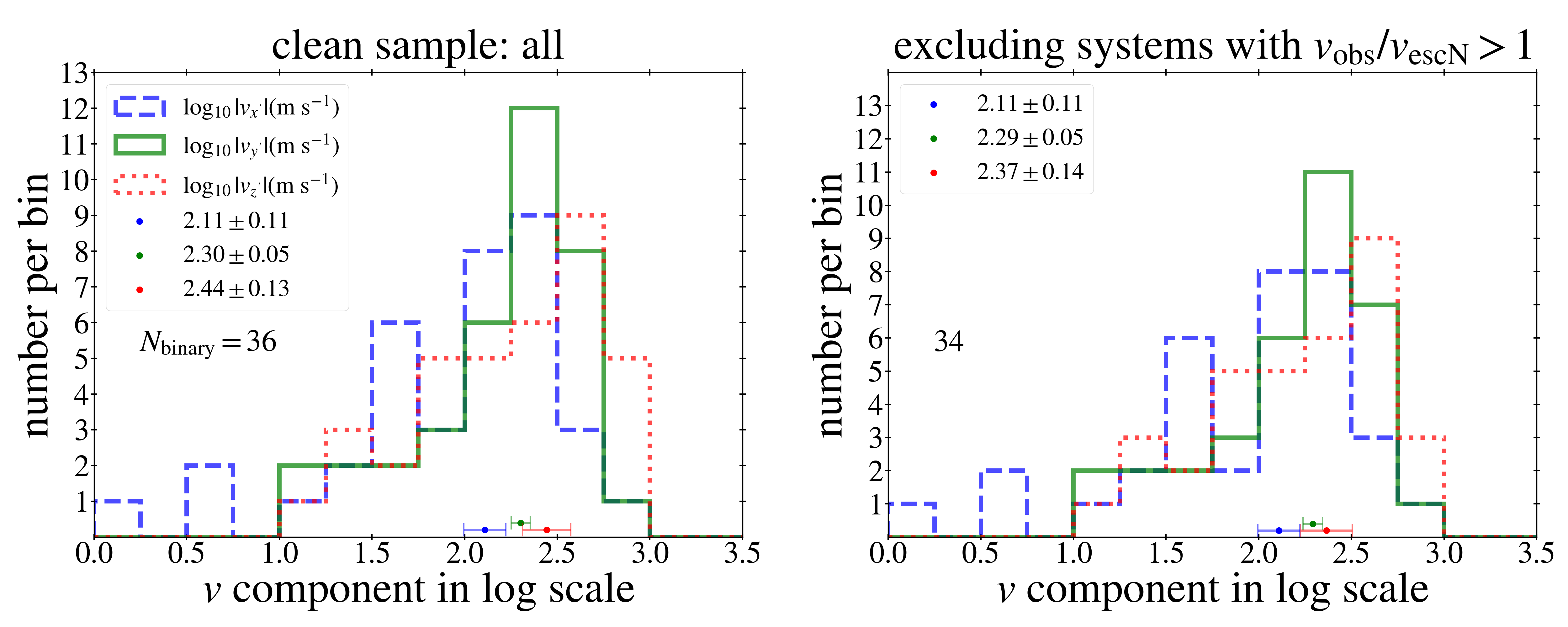}
    \caption{
    The distribution of the magnitude of relative velocity component is shown for the clean sample of wide binaries. The left panel shows the distribution for all wide binaries while the right panel shows the distribution excluding the two systems that are not gravitationally bound by Newtonian gravity. In both panels, all three components are statistically consistent with one another, dots with error bars giving the mean velocity values for the 3 velocity components and their confidence intervals.
    }
    \label{fig:v_components}
\end{figure*}

The distributions of $\eta_r$ (Equation~(\ref{eq:eta_r})) and $\eta_s$ for the clean sample can be found in Figure~\ref{fig:v_over_vescN_clean}. Compared with Figure~\ref{fig:v_over_vescN_basic}, the fraction $f(\eta_s>1)=f(\eta_r>1)=4/36=0.11$ (for fixed gravity models) is smaller by a factor of two. Moreover, considering the uncertainties of the individual values there are only two clear cases with $\eta>1$. Binary~{\#}59 from the HARPS sample \citep{Saglia:2025} has a value $\eta_s=1.156_{-0.055}^{+0.058}$ which is within MOND predictions, while Binary~{\#}3 from this work has a value $\eta_s=1.670_{-0.205}^{+0.213}$ which is $2.2\sigma$ above the limit $\approx \sqrt{1.5}$ predicted by realistic numerical solutions of QUMOND (Figure~\ref{fig:QUMOND}) for systems with large separations. Because the basic-cut sample is expected to have $\la 8$ cases of contamination based on the statistical analyses of large samples in the literature (see Setion~\ref{sec:basic_sample}), Figure~\ref{fig:v_over_vescN_basic} suggest that we expect $\ga 9$ cases of $\eta > 1$ from the remaining $\la 67$ pure binaries. Thus, 4 cases of $\eta>1$ from the clean sample are consistent with the expectation. 

Figure~\ref{fig:v_components} shows the distribution of the magnitude of three relative velocity components for the clean sample. For a randomly selected sample, the three components are expected to be statistically equivalent if they are measured with comparable precision. Indeed, the three components are consistent with one another, in terms of both the overall shape and the median value.

\begin{figure*}[!htb]
    \centering
    \includegraphics[width=1.\linewidth]{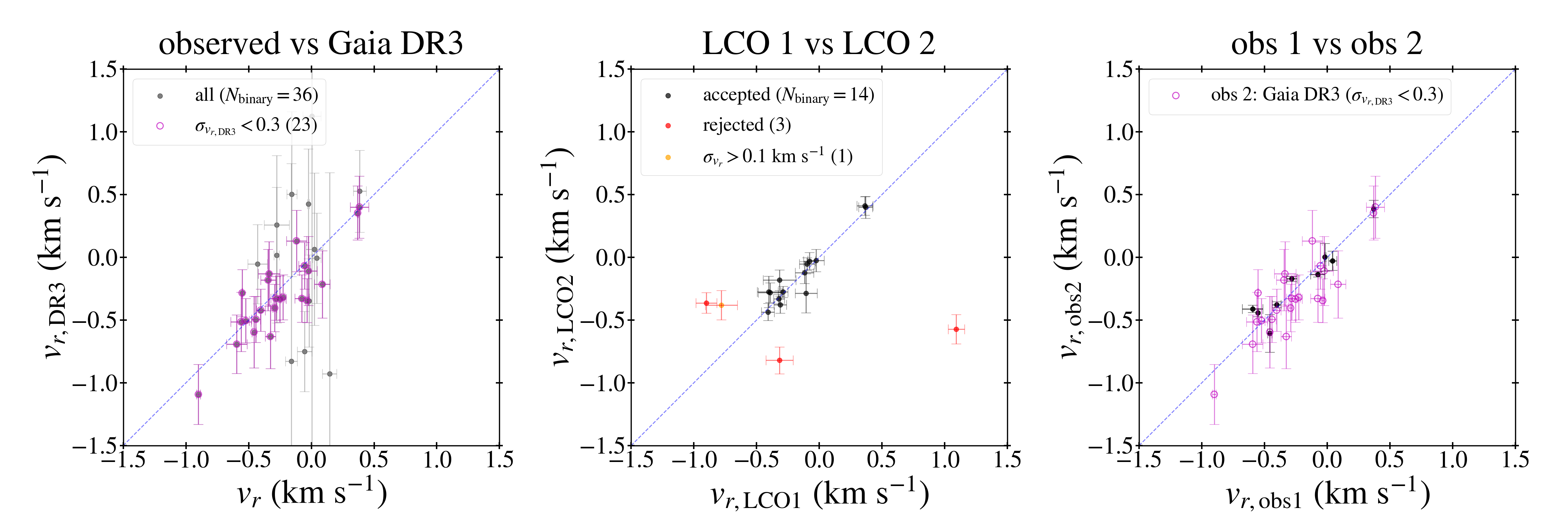}
    \caption{
    The relative radial velocity ($v_r$) measurements from different observations are compared for the wide binaries of the clean sample. The left panel compares the values collected in this work with the values from the Gaia DR3 database. Relatively precise DR3 values represented with magenta color match excellently with $v_r$. The middle panel compares two independent measurements within the LCO network as described in Appendix~\ref{sec:LCOdouble}. Those with $>2\sigma$ discrepancy (marked by red dots) are precluded in this work. The right panel compares two independent measurements from different observations/sources separated by more than several years. Black dots are from Table~\ref{tab:WB_overlap} and magenta circles are from the left panel. Seven systems are represented by both black dots and magenta circles, meaning that this panel includes only 25 unique systems (see the text for the details).
    }
    \label{fig:vr_compare}
\end{figure*}

\begin{figure*}[!htb]
    \centering
    \includegraphics[width=0.8\linewidth]{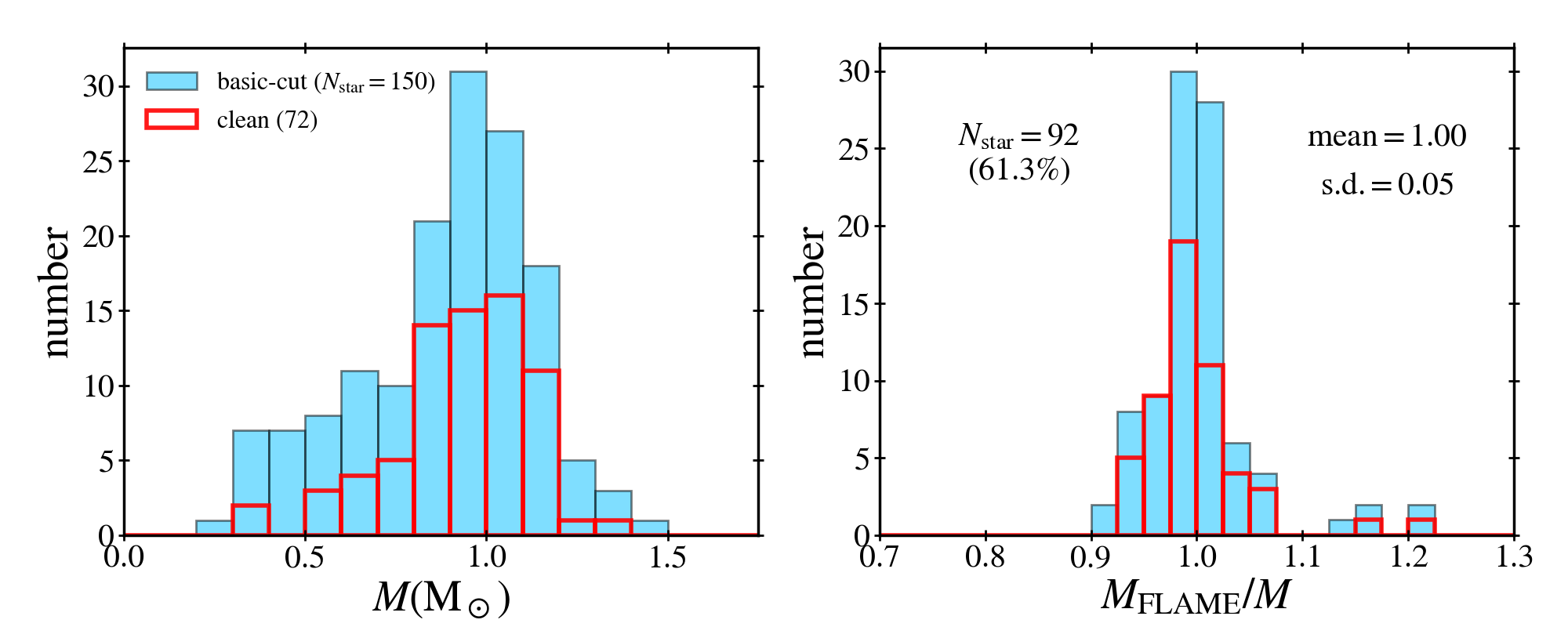}
    \caption{
    The left panel shows the distribution of masses of the stars to be used in this work. The right panel compares the Gaia DR3 FLAME masses with our masses for those stars whose FLAME masses are available. The two masses agree well with a scatter of only 5\%.
    }
    \label{fig:mass}
\end{figure*}
 
 In Figure~\ref{fig:vr_compare}, we compare various independently measured values of $v_r$. Because the major criterion used to define the clean sample is the availability of multiple values of $v_r$ measured at multiple (mostly two) epochs (although the time difference between epochs is quite diverse: see Figure~\ref{fig:time}), it is interesting to check/test how well independent values match each other. The left panel of Figure~\ref{fig:vr_compare} compares the values ($v_r$) of the clean sample listed in Table~\ref{tab:sample} with the Gaia DR3 values ($v_{r,\rm{DR3}}$) that happen to be available for all the binaries of the sample. They are consistent with each other up to their measurement errors. The difference $v_{r,\rm{DR3}}-v_r$ has a mean of $-0.009$~km\,s$^{-1}$ and an rms scatter of $0.385$~km\,s$^{-1}$. We find that $v_{r,\rm{DR3}}$ has a much larger rms scatter of $0.455$~km\,s$^{-1}$ (due to the larger measurement errors) compared with $0.282$~km\,s$^{-1}$ for $v_r$. The latter is close to the intrinsic scatter. For the subsample with relative precise $v_{r,{\rm DR3}}$ (with error $<0.3$~km\,s$^{-1}$), $v_{r,\rm{DR3}}$ has a much smaller scatter of $0.322$~km\,s$^{-1}$ which is similar to $v_r$'s scatter of $0.296$~km\,s$^{-1}$ for the same binaries. For this subsample, there is an excellent match between $v_r$ and $v_{r,{\rm DR3}}$ that were observed at different epochs separated by 2 - 9 years (Figure~\ref{fig:time}).

The middle panel of Figure~\ref{fig:vr_compare} compares two independent values measured by the LCO at two epochs separated by a few months (see Appendix~\ref{sec:LCOdouble}). In 14 of the 18 systems, two values agree well within $2\sigma$. For the rest, one system agrees within $3\sigma$, but the other systems are discrepant by $>3\sigma$. All such cases were excluded from the clean sample. In one system (Stars HD 101574 and BD-01 2557, Table~\ref{tab:LCO}) there is a large discrepancy between two measurements of $v_r$ with different telescopes (within the LCO) due to the discrepant RV values for HD 101574 (see Appendix~\ref{sec:LCOdouble}). It is unclear whether this kind of discrepancy is due to intrinsic variations from kinematic perturbations or issues in measurements. Whatever the case, this possibility of large measurement-to-measurement variation highlights the importance of reproduction and confirmation of $v_r$ values.

In the right panel of Figure~\ref{fig:vr_compare}, black dots compare two independent values of $v_r$ measured with different telescopes at two epochs separated by 3 - 11 years (except for one system which has a 0.3 year baseline) for 9 wide binaries satisfying the selection merit \#(5). The values $v_{r,\rm{obs1}}$ and $v_{r,\rm{obs2}}$, star identifiers, and observation sources of the 9 systems can be found in Table~\ref{tab:WB_overlap} and Table~\ref{tab:sample}. They match well each other within a few times the small nominal errors. We note that nominal measurement errors for some systems shown in the panel are extremely small ($<10$m\,s$^{-1}$), but realistic errors considering the effects of gravitational redshift and convective flow may be at least a few tens of m\,s$^{-1}$ \citep{Saglia:2025}. Magenta circles shown in the left panel are reproduced here to show all possible multi-epoch measurements that are reasonably precise. Because seven systems from the subsample represented by magenta circles are also included in the list of Table~\ref{tab:WB_overlap}, there are only 25 systems with multi-epoch values. For 21 of them, time baselines are in the range of 3 - 11 years (see Figure~\ref{fig:time}).

Regarding masses of the stars to be used as the observational input, individual estimates are available for wide binaries from the \cite{Saglia:2025} sample and the LCO sample (Appendix~\ref{sec:LCO}). For wide binaries for which individual estimates are not available from stellar modeling, we use the mass-magnitude relation derived by \cite{Chae:2023} based on \cite{Pecaut:2013}, which is consistent with the relation derived by \cite{Chevalier:2023} for Gaia binaries.. As explicitly shown in \cite{Chae:2025} and \cite{Yoon:2025}, the \cite{Chae:2023} masses agree well with the Gaia DR3 FLAME masses for stars similar to those used in this work whenever the latter are available from the Gaia archive (note that not for all stars FLAME masses are available). Figure~\ref{fig:mass} shows the distribution of our masses and a comparison with the corresponding FLAME masses for those stars that have the FLAME masses. We note that our masses are statistically consistent with the FLAME masses with an rms scatter of 5\% and our Bayesian methodology allows for the uncertainty of mass through the parameter $\log_{10}f_M$.

\section{Results along with relevant discussions} \label{sec:result}

We now present our inference of gravity at low acceleration ($< 10^{-9}$~m\,s$^{-2}$) through the parameter $\Gamma$ (Equation~(\ref{eq:Gamma})) by applying the Bayesian methodology (Section~\ref{sec:method}) to the clean sample of wide binaries described in \ref{sec:data}. While the priors on inclination ($i$) and orbit true anomaly ($\Delta\phi$) described in Section~\ref{sec:method} are fixed because they are statistical properties that a random sample of orbits must satisfy, we consider a wide range of  possibilities on the prior probability distribution on eccentricity ($e$): $\alpha=0$ (flat), $\alpha=1$ (thermal), and $\alpha=1.3$ (superthermal for Gaia DR3 wide binaries of relevance: \citealt{Hwang:2022}) in $f_{\rm pr}(e)=(1+\alpha)e^{\alpha}$. Our nominal choice will be the thermal distribution and it is implicitly assumed if not stated otherwise.

\begin{figure}[!htb]
    \centering
    \includegraphics[width=1.\linewidth]{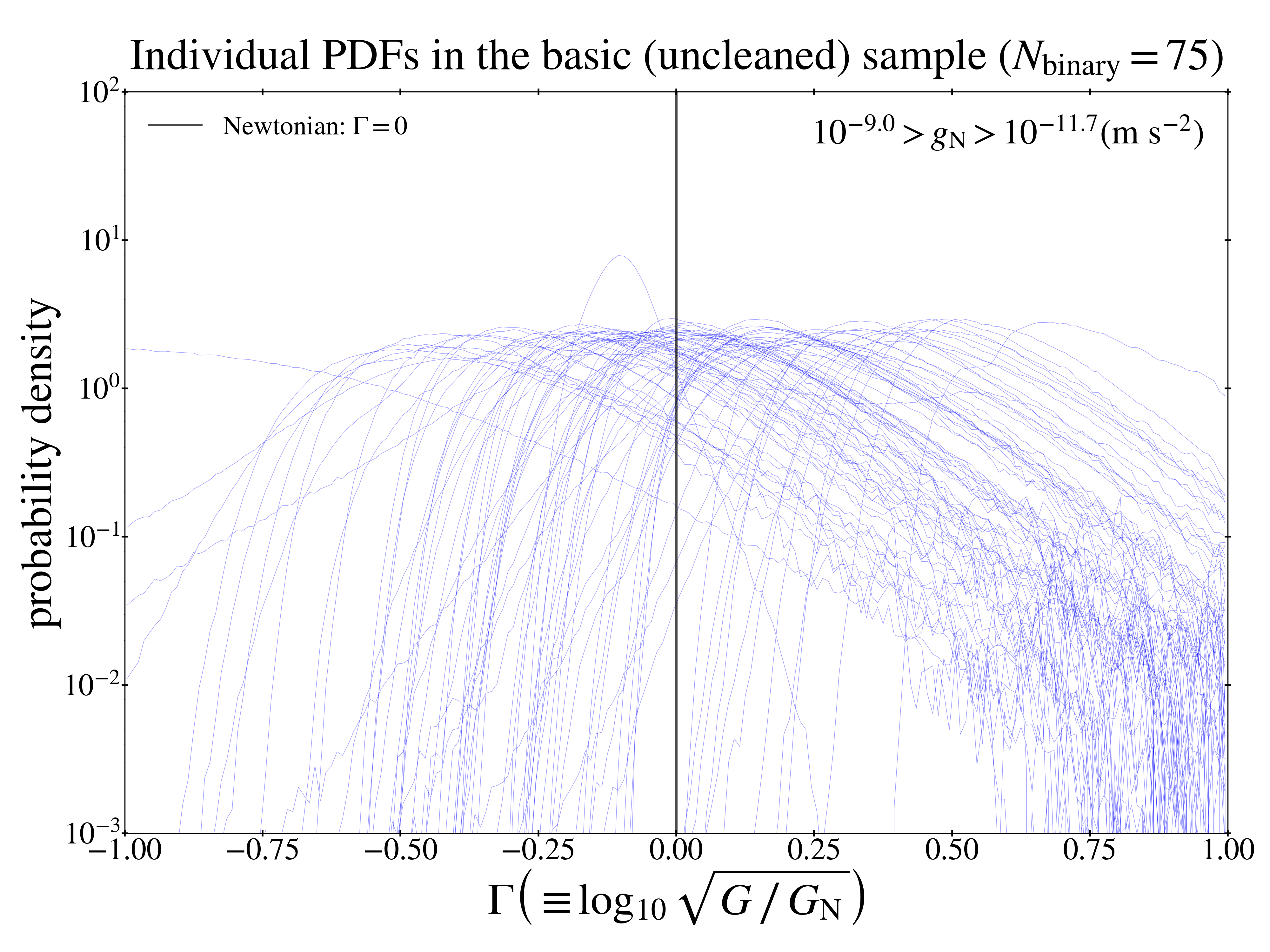}
    \caption{
    This figure shows the individual PDFs of $\Gamma$ for the entire sample of Table~\ref{tab:sample}. Most PDFs overlap one another reasonably well. However, there are several individual cases peaked at large positive values of $\Gamma$ that do not overlap with one narrowly peaked at a negative value. It turns out that none of these mutually excluding systems are included in the clean sample.
    }
    \label{fig:Gamma_basic}
\end{figure}

\begin{figure*}[!htb]
    \centering
    \includegraphics[width=1.\linewidth]{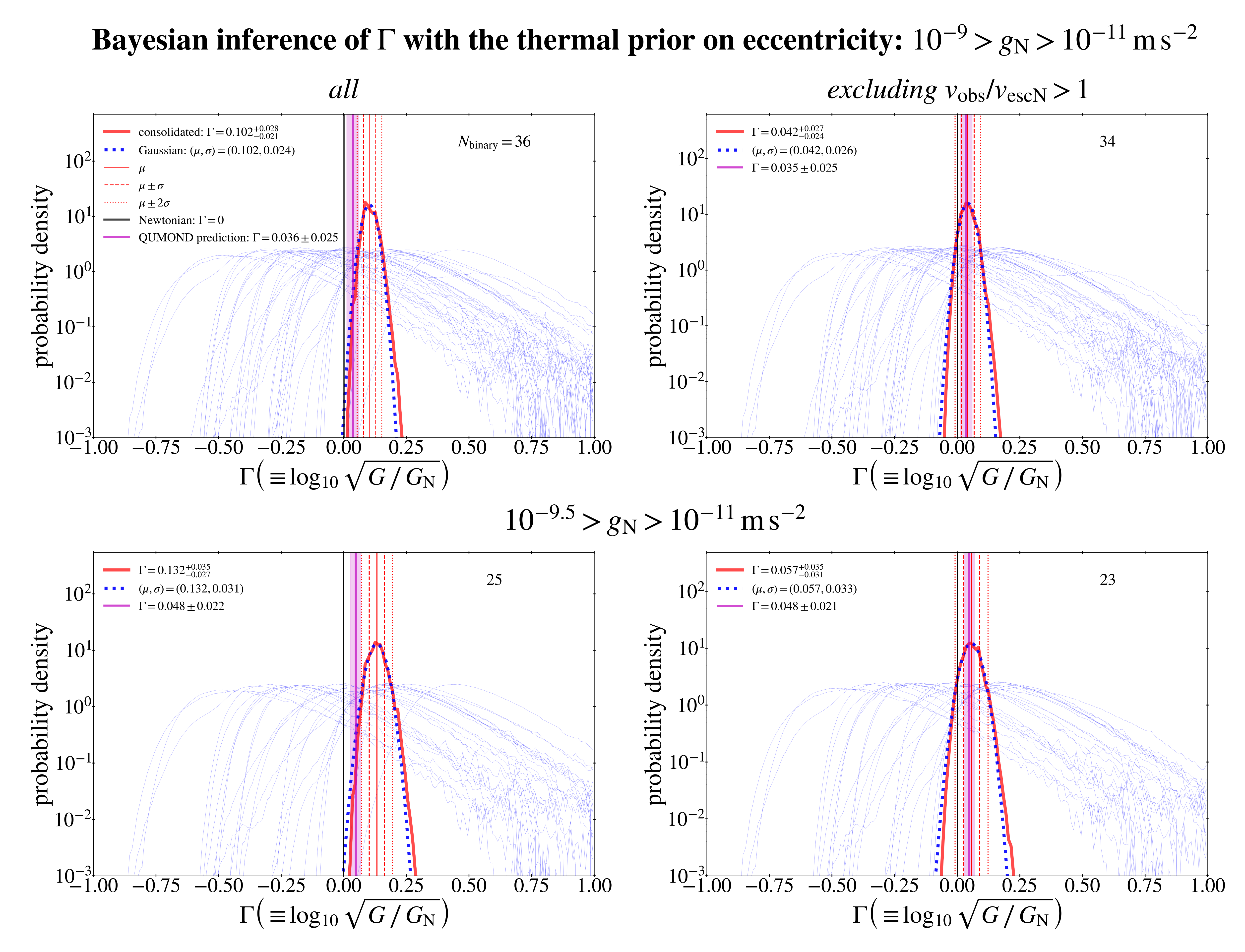}
    \caption{
    This figure shows results on the Bayesian inference of $\Gamma$ for the clean sample. In each panel thin blue curves represent individual PDFs and thick red curve is the consolidated PDF. The left panels show results for all wide binaries including those with $\eta_r({\rm Equation}~(\ref{eq:eta_r}))>1$ while the right panels show results excluding them. The bottom row shows the results excluding a transition regime ($10^{-9}>g_{\rm N}>10^{-9.5}\,{\rm m}\,{\rm s}^{-2}$). The QUMOND prediction is based on Figure~\ref{fig:QUMOND}. }
    \label{fig:Gamma}
\end{figure*}

In Figure~\ref{fig:Gamma_basic}, we first examine the individual PDFs of $\Gamma$ for all 75 wide binaries of the basic-cut sample listed in Table~\ref{tab:sample}. This sample includes both the clean sample and the rest. Since the systems not included in the clean sample cannot be verified to be uncontaminated on the basis of the currently available observational information, this figure may include kinematically contaminated cases. In Section~\ref{sec:basic_sample}, we noted cases of $\eta_r/\eta_s\le0.98$ in fixed gravity models ($G=G_{\rm N}$ or $1.4G_{\rm N}$). In particular, the five cases (i.e., Binary {\#}18, {\#}68, {\#}71, {\#}73, and {\#}74) for $G=1.4G_{\rm N}$ are those with median $\Gamma>0.5$, most of which do not overlap well with the one narrowly peaked at a negative value. This means that the former and the latter are mutually exclusive and cannot obey the same gravity law. This strongly indicates that at least the five cases from the basic-cut sample are likely to be kinematically contaminated. None of such suspicious cases are included in the clean sample. 

\subsection{Main results} \label{sec:main_result}

The top left panel of Figure~\ref{fig:Gamma} shows the PDFs of $\Gamma$ for the clean sample of 36 wide binaries. In contrast to Figure~\ref{fig:Gamma_basic}, all individual PDFs behave well and overlap well with one another without exception. This is expected if all wide binaries obey the same gravity law and all velocities are uncontaminated and reliably measured. In other words, individual PDFs will represent only random scatters around the underlying gravity law. The properties of the individual PDFs are reassuring of the reliability of all the data involved and the process of selecting the clean sample. Then, the underlying gravity (parameterized by $\Gamma$ in the present study) can be recovered by the statistical consolidation described in Section~2.2 of \cite{Chae:2025} based on \cite{Hill:2011}. As the Newtonian simulations in Figures~4 and 8 of \cite{Chae:2025} demonstrate, the underlying gravity can be correctly recovered by combining all individual PDFs from various orbit true anomalies, in particular including rightmost ones which are from near the periastron phase. If rightmost PDFs based on uncontaminated data are excluded just because they appear to be significantly away from $\Gamma=0$, the consolidated value of $\Gamma$ is bound to be biased. Thus, it is crucial to include all valid data for an unbiased inference of gravity.   

The top left panel of Figure~\ref{fig:Gamma} shows the statistical consolidation for all wide binaries of the clean sample. The consolidated PDF for the entire clean sample is well outside the Newtonian value $\Gamma=0$, in stark contrast with the samples with higher internal acceleration ($> 10^{-9}$~m\,s$^{-2}$) presented in \cite{Chae:2025,Chae:2025b}. As can be seen from the curve displayed on the logarithmic scale, the consolidated PDF is approximately Gaussian but not quite exactly. There is a mild left-right asymmetry, and the curve does not decline exactly like Gaussian. The mild asymmetry is expected from simulations (in particular, see the top left panel of Figure~8 of \cite{Chae:2025}, which is most relevant to the present sample). We find $\Gamma=0.102_{-0.021}^{+0.028}$, a $60_{-14}^{+24}$\% boost to Newtonian gravity, where the quoted formal uncertainties represent the halves of the 95.4\% bounds (rather than the 68.3\% bounds) from the central median. Based on the Gaussian-like formal uncertainties or the actual consolidated PDF, Newton is ruled out at $\ga 5\sigma$. In contrast, the realistic QUMOND prediction (see Figure~\ref{fig:QUMOND}) $\Gamma=0.036\pm 0.025$ for the binary sample is in significantly less tension with the consolidated PDF.

In the bottom left panel of Figure~\ref{fig:Gamma}, we consider a subsample with $g_{\rm N}<10^{-9.5}\,{\rm m}\,{\rm s}^{-2}(\approx 2.6a_0)$ excluding systems in a transition range $10^{-9}>g_{\rm N}>10^{-9.5}\,{\rm m}\,{\rm s}^{-2}$, so that the QUMOND prediction $\Gamma=0.048\pm0.022$ is better distinguished from Newton. We find $\Gamma=0.132_{-0.027}^{+0.035}$, which is higher than the case including the transition regime. The increase of $\Gamma$ is qualitatively consistent with the the generic expectation from MOND gravity, but the value is higher than the specific QUMOND prediction as is the case for the full sample. Below we will consider bins of $\log_{10}g_{\rm N}$, so that the trend of $\Gamma$ with $g_{\rm N}$ can be investigated within the current sample.  

In the right panels of Figure~\ref{fig:Gamma}, we consider the subsamples obtained by excluding systems with unusually high 3D velocity $\eta_r>1$ from the corresponding left-panel samples, where $\eta_r$ (Equation~(\ref{eq:eta_r})) is the value with a fixed gravity of $G=1.4G_{\rm N}$ (see Figure~\ref{fig:v_over_vescN_clean}). Here we are concerned with only two clear cases with $\eta_r > 1$ (Binary {\#}3 and {\#}59) as the other cases are ambiguous. We consider the option of excluding them because of their big contribution to the rather large consolidated value of $\Gamma$.  

The right panels of Figure~\ref{fig:Gamma} show that the consolidated value of $\Gamma$ is significantly reduced, as expected from numerical experiments. For example, modeling and consolidating simulated data (assuming Newtonian gravity) shown in Figures 4 and 8 of \cite{Chae:2025} demonstrate that selectively removing individual PDFs covering only higher values of $\Gamma$ (based on simulated data near the periastron) will result in a biased consolidated PDF that is shifted from the assumed gravity. However, the consolidated values of $\Gamma = 0.042_{-0.024}^{+0.027}$ for $g_{\rm N}<10^{-9.0}\,{\rm m}\,{\rm s}^{-2}$ and $\Gamma = 0.057_{-0.031}^{+0.035}$ for $g_{\rm N}<10^{-9.5}\,{\rm m}\,{\rm s}^{-2}$ are still inconsistent with Newton at a $\approx 1.8\sigma$ level. In other words, the sample as a whole appears to have an anomaly or MOND(-like) signal that cannot be naturally removed. If one wants to interpret our results for the full clean sample within a Newtonian framework, one needs to assume that all $v_{\rm obs}>v_{\rm escN}$ cases, however unlikely they are (see below), must be unbound systems, the removal of which should yield agreement with Newtonian expectations, as all remaining binaries are individually consistent with Newtonian expectations. However, this is not the case. Even removing all $v_{\rm obs}>v_{\rm escN}$ binaries leaves us with a sample that collectively is still $\approx 1.8\sigma$ discrepant with the Newtonian hypothesis. Since the included binaries are individually consistent with Newton, it is clear that being inconsistent with Newton as a sample stems from requiring orbital parameter distributions (typically inclination and orbital phase) which are at odds with the isotropy and Keplerian assumptions of the method.

Interestingly, the consolidated PDFs excluding systems with $\eta_r>1$ agree well with the gravity boost range predicted by realistic QUMOND numerical simulations (\citealt{Pf-A:2025}; see Figure~\ref{fig:QUMOND}) although the distinction from Newton is less pronounced. Thus, it is tempting to exclude the systems with $\eta_r>1$ even from a MOND point of view, but we believe that such a choice is not warranted for several reasons. 

First of all, the two systems with $\eta>_r1$ have passed all predefined observational selection criteria for pure wide binaries as the other systems. There is no compelling reason to exclude them on the basis of the currently available observational information. Table~\ref{tab:WB_Newtonunbound} summarizes some properties of interest for them. Both systems have relatively precise values of $v_r$ from Gaia DR3 (Gaia uncertainties smaller than $\approx 0.3$~km\,s$^{-1}$ are rare: see \citealt{Chae:2025}) and they agree excellently with the more precise independent values measured at different epochs. In both cases, the radial separation is consistent with zero within the measurement errors. In both systems, the projected separation is less than 20~kau and the metallicity difference between the two stars is nearly zero or not significant considering the Gaia DR3 [M/H] realistic errors (see Section~\ref{sec:metal}). Other systems with $\eta_r\ga 0.95$ have no clear issues in their observed properties either. One system (ID \#33) may have a moderate issue. The projected separation is relatively large with $s=30.34$~kau and the metallicity difference is the largest among the clean sample with $\Delta[M/H]\equiv[M/H]_{A}-[M/H]_B\approx -0.7$. Because the two stars of this system have effective temperatures $>4600$K, the Gaia DR3 metallicity difference cannot be specifically explained by the known bias. However, the $1\sigma$ range of realistic errors of [M/H] can be as large as $\approx 0.7$ (see Figure~11 of \cite{Gaia:GSPP}). Moreover, $v_r$ of this system was measured twice by the LCO and it is consistent with the relatively precise Gaia DR3 value from about 10 years earlier epoch. Thus, even this system needs not be excluded.

Here it is interesting to note that the presence of $\eta_r>1$ was first noticed in the \cite{Saglia:2025} sample with the occurrence rate of $1/8$ among low-acceleration systems \citep{Chae:2025b}. In our clean sample, we have identified one clear and two ambiguous cases from the newly added 28 wide binaries. Thus, these newly added systems suggest that cases with $\eta_r>1$ (gravitationally-unbound from the Newtonian perspective) are common with an occurrence rate of about $\approx 0.05-0.10$, confirming that the system discovered by \cite{Saglia:2025} is not a fluke.

Secondly, the (mildly) Newtonian-unbound systems are not particularly different from the Newtonian-bound systems with $0.9\la \eta_r<1$. Table~\ref{tab:WB_Newtonunbound} summarizes the key properties of the data and the inferred values of $\Gamma$ for 10 wide binaries with $\eta_r \ga 0.9$ that are most responsible for the amplitude of the gravitational anomaly in the clean sample. The values of $v_p$ (the scalar sky-plane relative velocity) and $|v_r|$ are statistically similar between the two subsamples with $\eta_r>1$ or $0.9\la \eta_r <1$. The inferred PDFs of $\Gamma$ are all similar with similar median values in the range $0.22\le\Gamma\le0.49$ and similar confidence widths. No individual PDF is abnormal compared with the other PDFs (although Binary~{\#}3 is somewhat extreme). This means that the presence of gravitational anomaly is not driven by a particular abnormal system but by the collective property of the sample as a whole.

\startlongtable
\begin{deluxetable*}{lcccccccl}
\tablecaption{\textbf{Some properties of wide binaries with $v_{\rm obs}/v_{\rm escN}\ga 0.9$ from the clean sample}: the first two are the clearly Newtonian-unbound systems with $v_{\rm obs}/v_{\rm escN} > 1$ while the rest are ambiguous .}\label{tab:WB_Newtonunbound} 
\centerwidetable
\tabletypesize{\footnotesize}
\startdata
ID\tablenotemark{a} & $v_p$\tablenotemark{b} & $v_r$\tablenotemark{c} & $v_{\rm obs}$\tablenotemark{d}  & $s$\tablenotemark{e}  & $z^\prime$\tablenotemark{f} & $v_{\rm obs}/v_{\rm escN}$\tablenotemark{g} & $\Gamma$\tablenotemark{h}  & $v_r$ source \\
 &  (km\,s$^{-1}$) & (km\,s$^{-1}$) & (km\,s$^{-1}$) & (kau) & (kau) &   &   & \\
\hline
3 & $0.374\pm0.019$ & $0.656\pm0.103$ & $0.755\pm0.091$ & $19.322\pm0.046$ & $74.0\pm94.4$ & $1.682_{-0.205}^{+0.216}$ & $0.491_{-0.127}^{+0.207}$ &   APOGEE  \\ 
59 & $0.389\pm0.009$ & $-0.861\pm0.043$ & $0.945\pm0.039$ & $5.786\pm0.010$ & $46.1\pm52.0$ & $1.152_{-0.055}^{+0.059}$ & $0.339_{-0.113}^{+0.246}$ &   HARPS  \\ 
\hline
25 & $0.577\pm0.009$ & $-0.345\pm0.184$ & $0.672\pm0.090$ & $6.524\pm0.007$ & $13.3\pm31.4$ & $0.964_{-0.134}^{+0.164}$ & $0.241_{-0.123}^{+0.229}$ &   LCO  \\ 
32 & $0.351\pm0.008$ & $-0.331\pm0.092$ & $0.483\pm0.063$ & $13.101\pm0.021$ & $5.0\pm52.0$ & $0.975_{-0.132}^{+0.167}$ & $0.229_{-0.124}^{+0.241}$ &   LCO  \\ 
33 & $0.216\pm0.010$ & $-0.355\pm0.189$ & $0.415\pm0.140$ & $30.340\pm0.060$ & $41.1\pm37.1$ & $1.284_{-0.437}^{+0.508}$ & $0.198_{-0.155}^{+0.268}$ &   LCO  \\ 
36 & $0.346\pm0.006$ & $-0.565\pm0.162$ & $0.662\pm0.136$ & $6.695\pm0.007$ & $22.7\pm23.6$ & $0.984_{-0.205}^{+0.228}$ & $0.194_{-0.153}^{+0.236}$ &   LCO  \\ 
39 & $0.474\pm0.006$ & $-0.055\pm0.099$ & $0.477\pm0.018$ & $11.055\pm0.014$ & $42.9\pm20.2$ & $0.917_{-0.062}^{+0.103}$ & $0.264_{-0.133}^{+0.245}$ &   LCO  \\ 
42 & $0.205\pm0.005$ & $-0.592\pm0.146$ & $0.627\pm0.135$ & $7.870\pm0.007$ & $-7.3\pm7.3$ & $1.141_{-0.250}^{+0.301}$ & $0.224_{-0.188}^{+0.252}$ &   LCO  \\ 
53 & $0.339\pm0.013$ & $-0.458\pm0.043$ & $0.570\pm0.036$ & $4.839\pm0.008$ & $98.8\pm56.1$ & $0.989_{-0.086}^{+0.116}$ & $0.314_{-0.154}^{+0.254}$ &   HARPS  \\ 
54 & $0.662\pm0.009$ & $0.452\pm0.040$ & $0.802\pm0.024$ & $4.933\pm0.007$ & $-18.5\pm50.6$ & $0.975_{-0.048}^{+0.081}$ & $0.222_{-0.108}^{+0.233}$ &   HARPS  \\ 
\enddata
\tablenotetext{a}{See Table~\ref{tab:sample}.}
\tablenotetext{b}{Magnitude of the relative sky-plane velocity $v_p\left(\equiv\sqrt{v_{x^\prime}^2+v_{y^\prime}^2}\right)$ corrected for the perspective effect.}
\tablenotetext{c}{Relative radial velocity between the pair $v_r\equiv{\rm RV}_A-{\rm RV}_B$ corrected for the GR+CB effect (if available) and the perspective effect} 
\tablenotetext{d}{Magnitude of the relative 3D velocity $v_{\rm obs}\left(\equiv\sqrt{v_{p}^2+v_{r}^2}\right)$ based on the corrected $v_p$ and $v_r$.}
\tablenotetext{e}{Sky-plane 2D separation between the pair taken from Table~\ref{tab:physparameters}.} 
\tablenotetext{f}{Relative distance (radial separation) between the pair, $z^\prime=-(d_B-d_A)$, based on the distances given in Table~\ref{tab:physparameters}.} 
\tablenotetext{g}{The Bayesian inferred value of $\eta(= v_{\rm obs}/v_{\rm escN})$ for the pseudo-Newtonian case with $G=1.4G_{\rm N}$ taken from Table~\ref{tab:physparameters}.} 
\tablenotetext{h}{The Bayesian inferred value of the gravitational anomaly parameter taken from Table~\ref{tab:sample}.} 
\end{deluxetable*}

Thirdly, these wide binaries selected from the solar neighborhood of $d< 150$~pc are extremely unlikely to be random chance associations. With a local number density of stars $n$ estimated from the Gaia DR3 database and the well-known distribution of peculiar velocity components in the solar neighborhood, we estimate that a randomly selected star to have a fly-by with projected separation $s<s_{\rm max}$, radial separation $l<l_{\rm max}$, and scalar relative 3D velocity $v(\equiv\sqrt{v_{x^\prime}^2+v_{y^\prime}^2+v_{z^\prime}^2})< v_{\rm max}$ is
\begin{eqnarray}
    p_{\rm chance} \approx & & (1.4-3.4)\times 10^{-8} \frac{n}{0.15\text{ pc}^{-3}} \nonumber\\ 
      & &\times\left( \frac{s_{\rm max}}{30\text{ kau}} \right)^2 \frac{l_{\rm max}}{100\text{ kau}}  \left( \frac{v_{\rm max}}{1\text{ km s}^{-1}} \right)^3,
      \label{eq:probFB}
\end{eqnarray}
where the 1-dimensional velocity dispersion of peculiar velocities is assumed to be in the range $30-40$~km\,s$^{-1}$. For the clean sample, the nominal choices of Equation~(\ref{eq:probFB}) are conservative, and thus for a sample of 2 million stars within 150 pc, we expect $\la 0.04$ cases, meaning that practically no fly-by cases meeting the thresholds should be observed. Thus, the observed 2 - 4 cases from the clean sample are extremely unlikely to be chance associations unrelated to gravity. Moreover, these numbers are a lower limit to the true number of binaries with $v_{\rm obs}>v_{\rm escN}$ because they were selected from samples of wide binaries that meet extremely strict observational constraints (e.g., many from the basic-cut sample were excluded for now, but some of them could be included in the future, once more high quality multi-epoch radial velocities are available). In other words, not all pairs within 150~pc have sufficient data to check whether they meet the selection criteria of the clean sample.

Fourthly, it is unlikely that any of these pairs is in the process of gravitationally escaping each other from its birth, because the expected time to reach the observed 2D separation or the inferred 3D separation is too small compared with the expected ages of the stars of mass $\approx 1{\rm M}_\odot$. As a specific example, we consider Binary~{\#}3 which has a relative 3D velocity of $v\approx0.8\,{\rm km}\,{\rm s}^{-1}$ and a 3D separation of $r\approx 20\,{\rm kau}$, which gives an upper bound of $t_{\rm max}\approx r/v \approx 1.2\times 10^5\,{\rm yr}$ because $v$ has been decreasing during the course of dispersing. This time is orders of magnitude smaller than any conceivable ages of the stars or the Galactic disk. Also, as \cite{BanikZhao:2018} discussed in their section~8.1, the probability that a passing star or molecular cloud in the solar neighborhood has perturbed a wide binary with $r=20\,{\rm kau}$ is in the order of $10^{-4}$, far smaller than the observed occurrence rate of $\approx 10^{-1}$. When a system is significantly perturbed by a passing star and thus becomes gravitationally unbound, it will so quickly disperse that it is unlikely for such a pair to be caught in action at the observed small separations.

\begin{figure}[!htb]
    \centering
    \includegraphics[width=1.\linewidth]{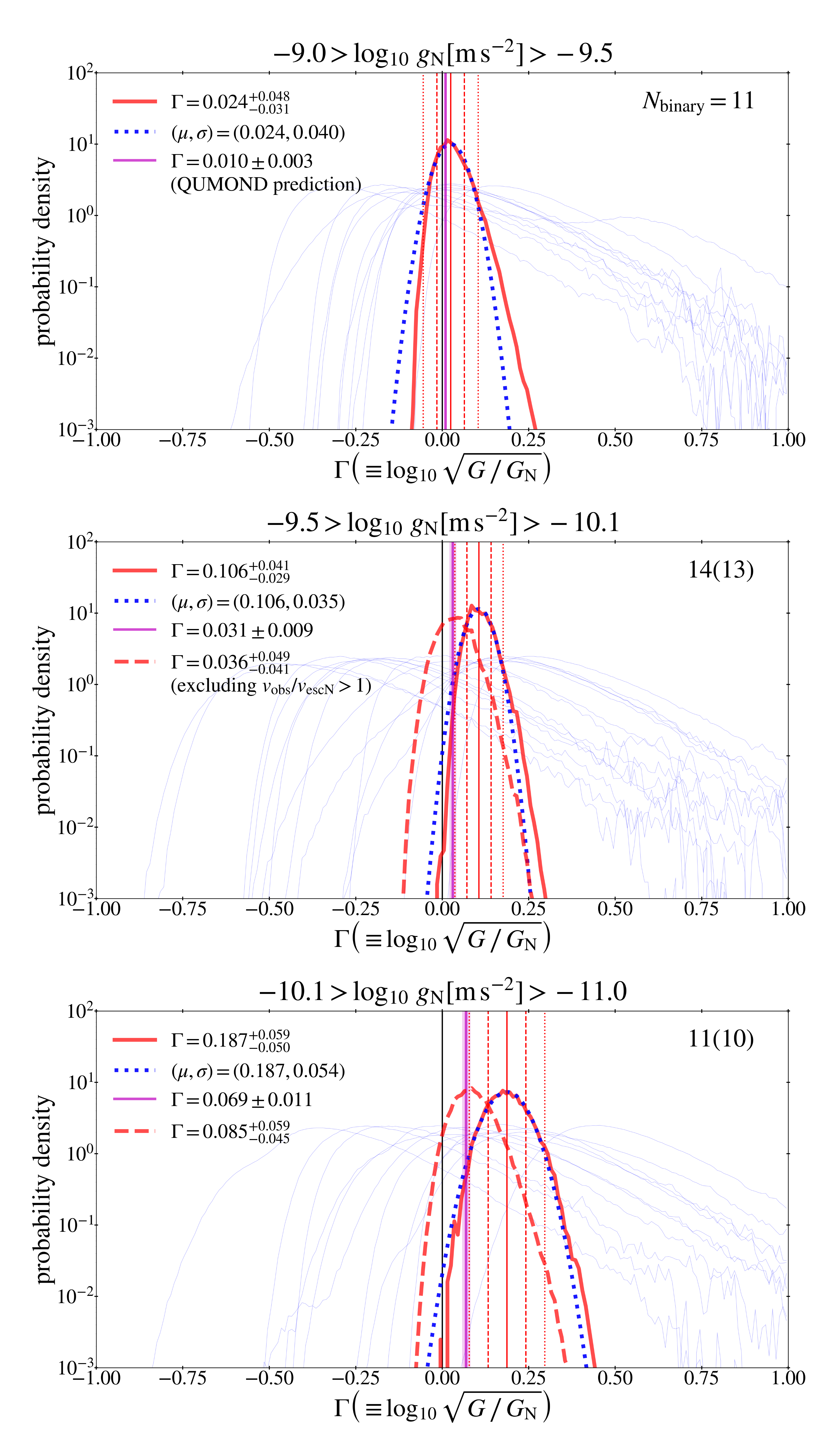}
    \caption{
    Each panel shows the Bayesian inference of $\Gamma$ for a specific range of $\log_{10}g_{\rm N}$ based on the individual PDFs shown in the top left panel of Figure~\ref{fig:Gamma}. In the middle and bottom panels, the statistical consolidation of $\Gamma$ excluding the system with $\eta>1$ is represented by the dashed red curve.
    }
    \label{fig:Gamma_3bins_loggN}
\end{figure}

Lastly, existing MOND gravity models \textit{do} predict\footnote{We thank Cezary Migaszewski for sharing numerical simulations results prior to publication.} the range of $v_{\rm obs}/v_{\rm escN}$ observed in the clean sample save the particular case of Binary~{\#}3 which is discrepant with the MOND-predicted range at about $2\sigma$. Moreover, whatever MOND gravity models predict, they (i.e.\ nonrelativistic gravity models) are far from an established theory. Thus, even the predictions of the current MOND models need to be taken with a grain of salt. Above all, because this work is about a measurement of $\Gamma$, it should not be controlled by any existing gravity models.

\begin{figure*}[!htb]
    \centering
    \includegraphics[width=1.03\linewidth]{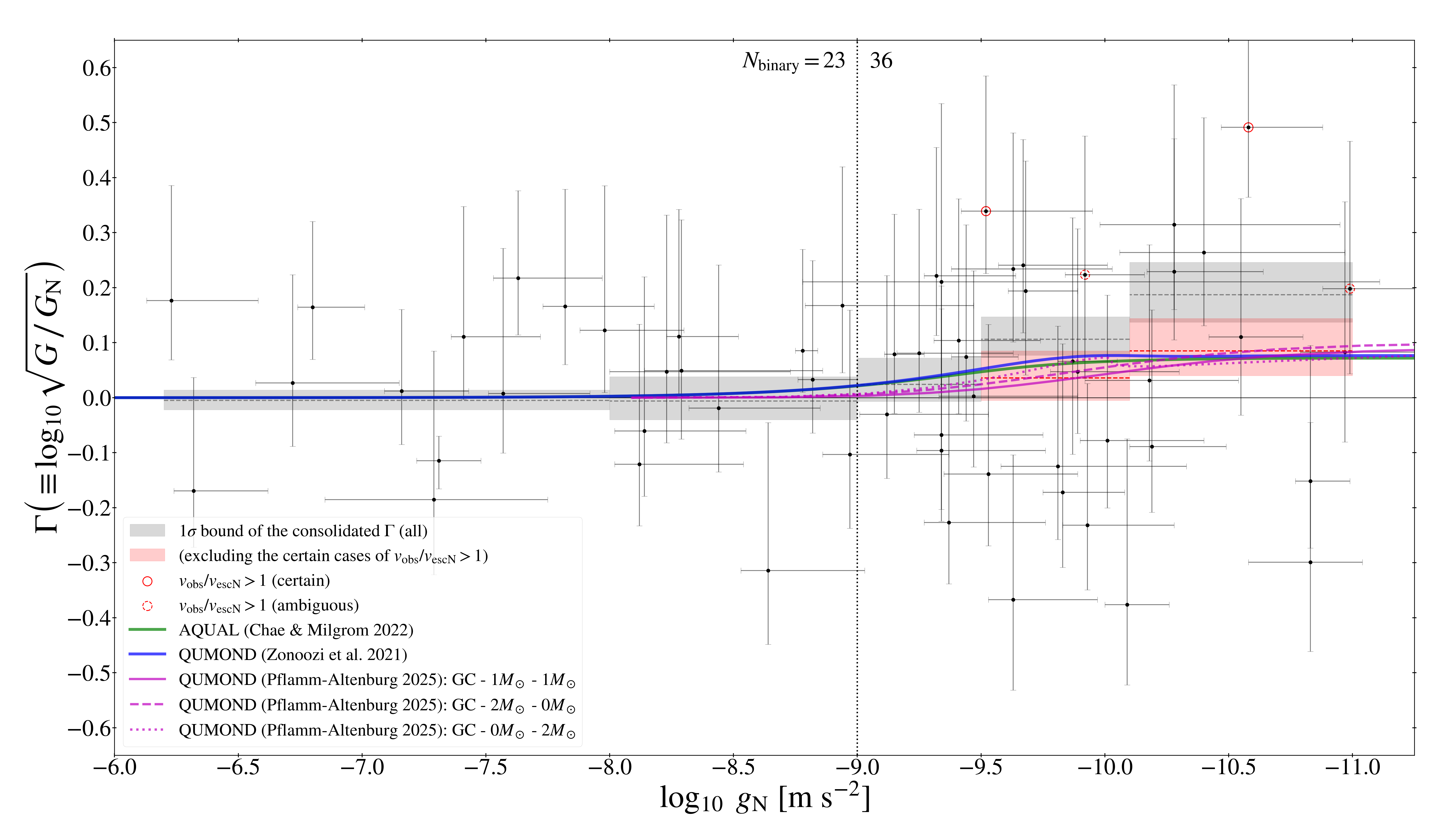}
    \caption{
    This figure shows the inferred values of $\Gamma$ against the internal accelerations $g_{\rm N}$ for the 36 wide binaries with $g_{\rm N}<10^{-9}\,{\rm m}\,{\rm s}^{-2}$ (shown in the top left panel of Figure~\ref{fig:Gamma}) and 23 wide binaries in the range $10^{-6}>g_{\rm N}>10^{-9}\,{\rm m}\,{\rm s}^{-2}$ (shown in \cite{Chae:2025b}). The consolidated values of $\Gamma$ in five bins are indicated by gray bands. The two red bands represent the results excluding the two systems with $\eta_r > 1$. Various MOND numerical curves \citep{ChaeMilgrom:2022,Zonoozi:2021,Pf-A:2025} from the literature are also shown (see Section~4 of \cite{ChaeYoon:2026} for the details).
    }
    \label{fig:Gamma_loggN}
\end{figure*}
 
 Having noted in the above that the systems with $\eta_r>1$ need not be excluded from observational and theoretical points of view, we now examine whether $\Gamma$ shows any trend with the internal acceleration $g_{\rm N}$ within our considered limit $g_{\rm N}<10^{-9}$~m\,s$^{-2}$. Figure~\ref{fig:Gamma_3bins_loggN} shows the consolidated values of $\Gamma$ in three bins of $\log_{10}g_{\rm N}$. Although each bin has a relatively small number ($\ge 11$) of wide binaries, there appears to be a clear trend. The first bin with $-9.0>\log_{10}g_{\rm N}\,[{\rm m}\,{\rm s}^{-2}] > -9.5$ has $\Gamma=0.024_{-0.031}^{+0.048}$, which is consistent with zero, and interestingly such a low value of $\Gamma$ in this transition regime is predicted by the realistic QUMOND simulation shown in Figure~\ref{fig:QUMOND}. The other two bins are clearly different from the first bin and have high positive values of $\Gamma$ with $>3\sigma$ significance respectively. Interestingly, the lowest-acceleration bin has the highest value of $\Gamma$. The high values of the consolidated $\Gamma$ in the latter two bins are largely contributed by the two systems with $\eta_r>1$. The values are so high that they are in $\approx 2\sigma$ tension with the corresponding QUMOND predictions. Without them, the second bin has a significantly reduced value consistent with both zero and the QUMOND prediction within the confidence internal, but the last bin ($g_{\rm N}<10^{-10.1}\,{\rm m}\,{\rm s}^{-2}$) has a value $\Gamma = 0.085_{-0.045}^{+0.059}$ agreeing well only with the QUMOND prediction while in $\approx 2\sigma$ tension with zero.

Figure~\ref{fig:Gamma_loggN} shows a panoramic view of a broad dynamic range $-6>\log_{10}g_{\rm N}\,[{\rm m}\,{\rm s}^{-2}] > -11$ based on the above binned results in the regime $g_{\rm N}<10^{-9}\,{\rm m}\,{\rm s}^{-2}$ along with two binned results in the high-acceleration regime $g_{\rm N}>10^{-9}$~m\,s$^{-2}$ based on the Bayesian outputs from \cite{Chae:2025b}.  The trend of the five bins indicates a MOND-type gravitational anomaly as $g_{\rm N}$ decreases, i.e., no deviation down to $10^{-9}\,{\rm m}\,{\rm s}^{-2}$ (we note that the recovery of Newtonian dynamics at the high acceleration end shown here and \cite{Chae:2025,Chae:2025b} validates the Bayesian methodology) and then deviation increasing gradually in the range $-9>\log_{10}g_{\rm N}\,[{\rm m}\,{\rm s}^{-2}] \ga -10.5$. We note that individual values of $\Gamma$ in the low-acceleration regime provide an insight why the consolidated value of $\Gamma$ must be positive in the regime. All individual values of $\Gamma<0$ are consistent with zero within about $1.2\sigma$. However, there are 7 individual values of $\Gamma>0$ that are $\ga 2\sigma$ away from zero. These values shift the consolidated value by $\approx +0.1$ so that all values agree with the shifted consolidated value within $\approx 2\sigma$.

\begin{figure}[!htb]
    \centering
    \includegraphics[width=1.03\linewidth]{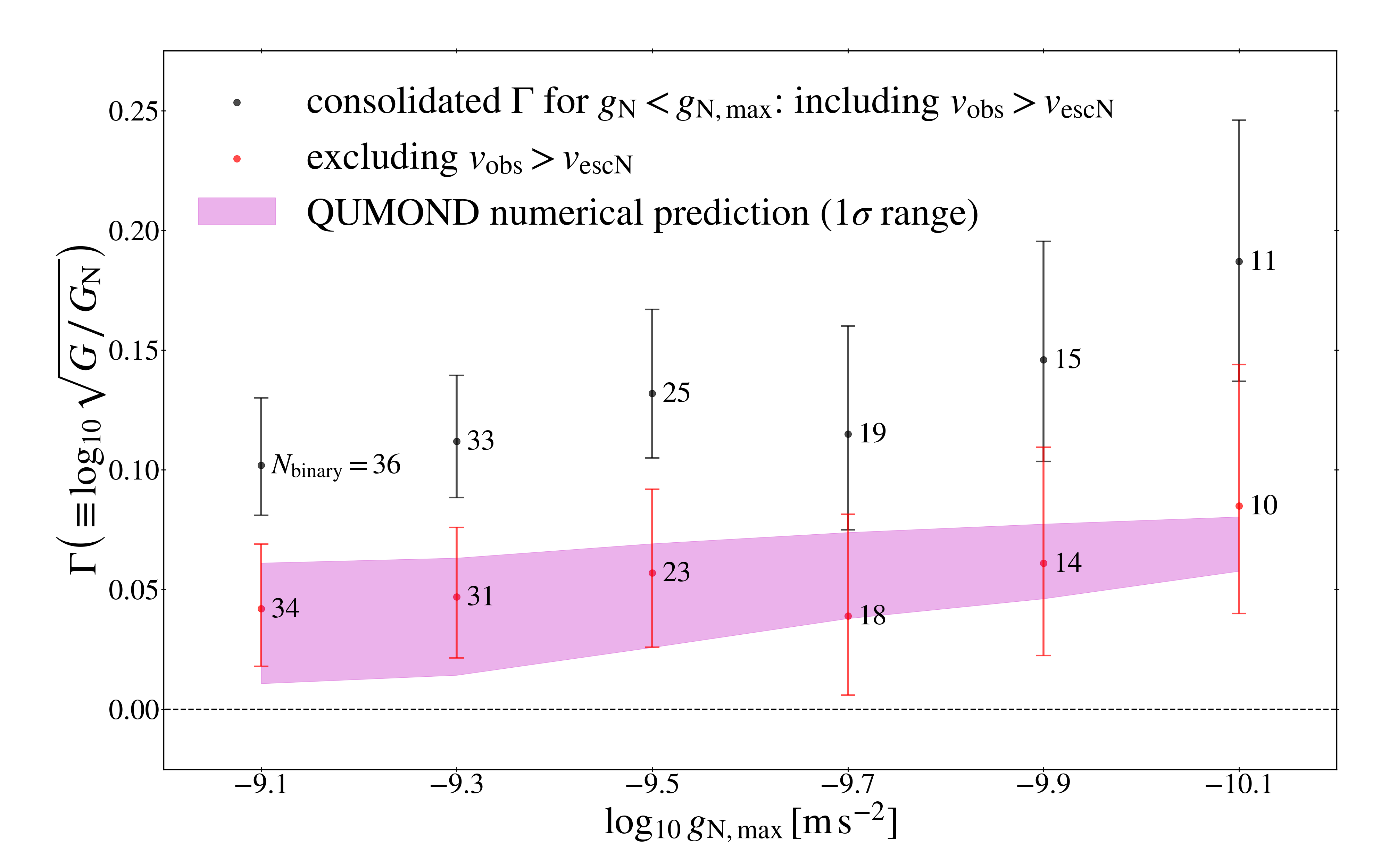}
    \caption{
    This figure shows the trend of consolidated $\Gamma$ in cumulative bins of $g_{\rm N}$ for the clean sample of 36 wide binaries shown in the top left panel of Figure~\ref{fig:Gamma}. The QUMOND prediction is based on Figure~\ref{fig:QUMOND}.
    }
    \label{fig:Gamma_cumul}
\end{figure}

\begin{figure*}[!htb]
    \centering
    \includegraphics[width=1.0\linewidth]{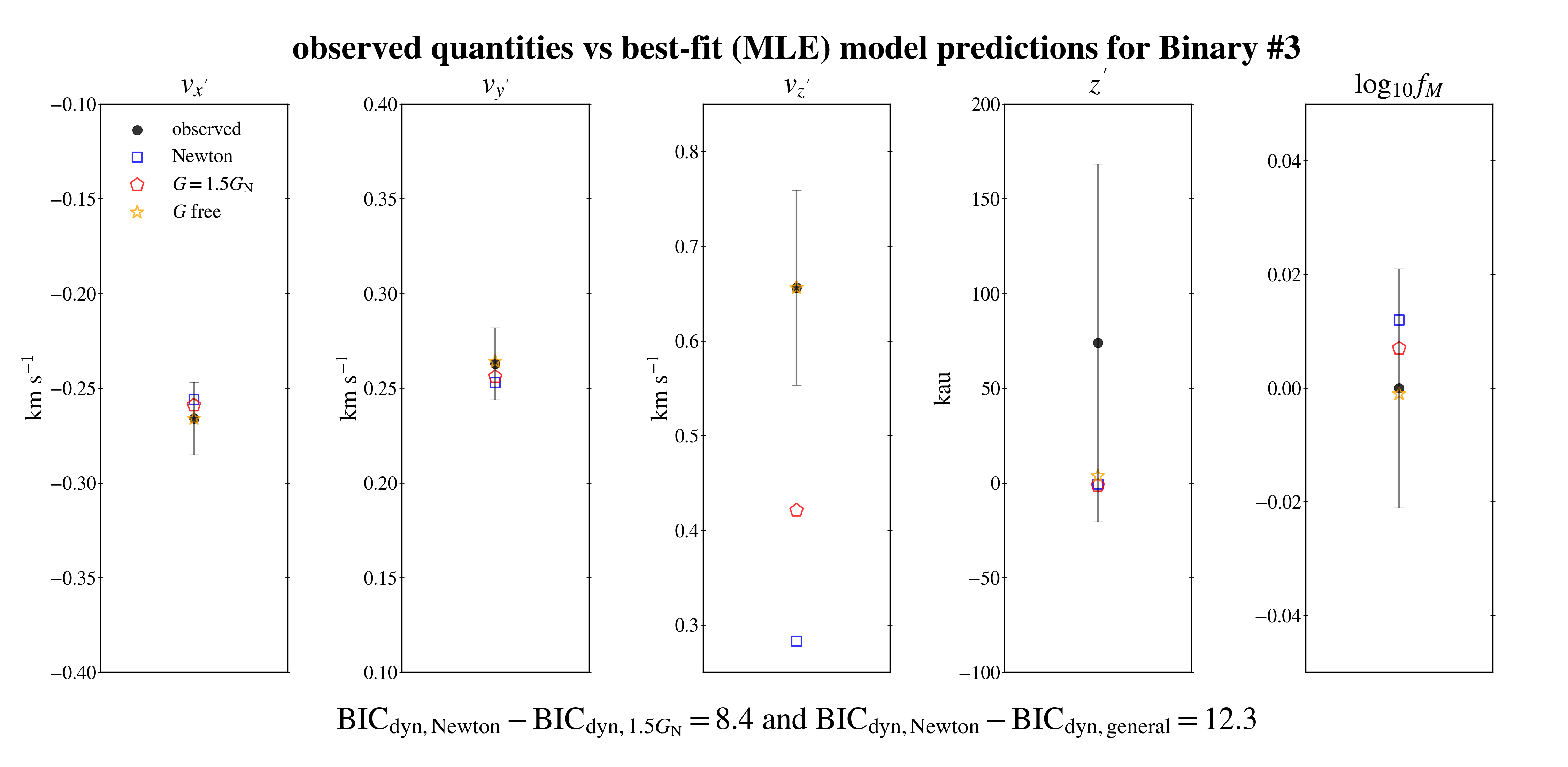}
    \caption{
    For Binary {\#}3, the observed quantities are compared with the best-fit (maximum likelihood estimate: see the Appendix of \cite{Chae:2025b}) predictions of three models: Newtonian, pseudo-Newtonian with $G = 1.5G_{\rm N}$, and generalized gravity with free $\Gamma$. The uncertain radial separation $z^\prime$ is not used in defining the best-fit models. Newtonian prediction of $v_{z^\prime}=0.283\,{\rm km}\,{\rm s}^{-1}$ is $3.6\sigma$ away from the observed value $v_r=0.656\pm0.103\,{\rm km}\,{\rm s}^{-1}$ while the pseudo-Newtonian (with $G=1.5G_{\rm N}$) prediction of $v_{z^\prime}=0.421\,{\rm km}\,{\rm s}^{-1}$ is $2.2\sigma$ away. The calculated values of $\Delta{\rm BIC}>8$ between Newton and the other models are significant.
    }
    \label{fig:binary3}
\end{figure*}

Because our clean sample is not sufficiently large, how the bins of $\log_{10}g_{\rm N}$ are defined can affect the trend of consolidated $\Gamma$ with median $\log_{10}g_{\rm N}$. For example, if the first bin is enlarged to include Binary~{\#}59, it will have a significant positive value of $\Gamma$. Moreover, the inferred confidence interval of each $\log_{10}g_{\rm N}$ is rather broad, so that there is a significant uncertainty in how to distribute systems into bins. Bearing this caveat in mind in interpreting the trend seen in Figure~\ref{fig:Gamma_loggN}, we also consider the trend of $\Gamma$ in cumulative bins of $\log_{10}g_{\rm N}$ (starting from the entire sample to the lowest acceleration bin) rather than split bins. Figure~\ref{fig:Gamma_cumul} shows the trend of consolidated $\Gamma$ in cumulative bins of $g_{\rm N}<g_{\rm N,max}$ as a function of $\log_{10}g_{\rm N,max}$. As $\log_{10}g_{\rm N,max}$ decreases incrementally from left to right along the horizontal axis, the QUMOND-predicted median $\Gamma$ increases as expected, and the predicted range gets narrower due to the allowed narrower range of $\log_{10}g_{\rm N}$. The consolidated $\Gamma$ in the sample overall increases systematically as $\log_{10}g_{\rm N,max}$ decreases with its uncertainty increasing because the number of systems decreases. We note that there occurs a decrease from the 3rd to the 4th bin due to fluctuations in the data. Nevertheless, the systematic rising trend in the cumulative bins is evident and confirms the MOND-type gravitational anomaly. In line with the various results shown in this subsection, the amplitude of the varying deviation better agrees with the QUMOND prediction without the two systems with $\eta_r>1$, although the distinction with Newton is weaker.

Before we close this subsection, we examine the modeling results of Binary~{\#}3 as it is a newly discovered system that is \emph{individually} discrepant with Newton. Another such system (Binary~{\#}59 in our list) was recently discovered by \cite{Saglia:2025} and analyzed in detail by \cite{Chae:2025b}. Following the format of Figure~9 of \cite{Chae:2025b}, we compare the observed velocity components with the predictions of three gravity models: $G=G_{\rm N}$, $G = 1.5G_{\rm N}$, and individually fitted $G$. Here we consider $G = 1.5G_{\rm N}$ as a proxy of MOND gravity since the \cite{Pf-A:2025} QUMOND simulation predicts such a boost factor at low acceleration near $10^{-11}\,{\rm m}\,{\rm s}^{-2}$ (we note that the allowed range of $\log_{10}g_{\rm N}$ for this binary is rather broad including $-11$ within about $1.5\sigma$). The model with $G = 1.5G_{\rm N}$ is significantly better than Newton in matching the observed value of $v_r(=v_{z^\prime})$ with $\Delta{\rm BIC}=8.4$ where BIC refers to the Bayesian information criterion specified in the Appendix of \cite{Chae:2025b}. For the fitted-$G$ gravity with a higher boost factor, $\Delta{\rm BIC}$ is even higher. In Table~\ref{tab:fitparameters_lowestgN}, we give fitted and derived parameter values with $G = 1.5G_{\rm N}$ for all wide binaries with $g_{\rm N}<10^{-10.1}\,{\rm m}\,{\rm s}^{-2}$ from the basic-cut sample (our lowest acceleration bin defined in Figure~\ref{fig:Gamma_3bins_loggN}).

\startlongtable
\begin{deluxetable*}{rccccccccc}
\tablecaption{\textbf{Fitted and derived parameters\tablenotemark{a} (with $G=1.5G_{\rm N}$) of wide binaries in the lowest acceleration bin}\label{tab:fitparameters_lowestgN}} 
\centerwidetable
\tabletypesize{\scriptsize}
\startdata
ID\tablenotemark{b} & $G$   &  $r$   &  $a$ &  $e$ & $i$ & $\phi_0$ & $\Delta\phi$ &  $\log_{10}f_M$ & $v_{\rm mod}$ \\
  &    &  (kau) & (kau)  &    &  (deg) &  (deg)  &  (deg)  &    &  (${\rm km}\,{\rm s}^{-1}$)  \\
 \hline
 2 & $1.5G_{\rm N}$ & $56.59_{-7.40}^{+28.43}$ & $36.54_{-9.15}^{+25.55}$ & $0.860_{-0.277}^{+0.109}$ & $90.1_{-17.5}^{+18.0}$ & $152.2_{-67.6}^{+93.5}$ & $97.6_{-278.9}^{+87.5}$ & $-0.001_{-0.021}^{+0.021}$ & $0.074_{-0.040}^{+0.056}$ \\ 
\hline 
3 & $1.5G_{\rm N}$ & $19.71_{-0.35}^{+1.31}$ & $61.01_{-23.75}^{+58.51}$ & $0.812_{-0.132}^{+0.095}$ & $123.5_{-6.4}^{+9.7}$ & $102.3_{-14.5}^{+14.1}$ & $83.4_{-33.2}^{+28.6}$ & $0.010_{-0.021}^{+0.021}$ & $0.500_{-0.034}^{+0.029}$ \\ 
\hline 
6 & $1.5G_{\rm N}$ & $20.21_{-0.33}^{+1.21}$ & $68.15_{-25.74}^{+62.32}$ & $0.780_{-0.123}^{+0.102}$ & $54.8_{-40.6}^{+110.9}$ & $198.3_{-149.5}^{+114.3}$ & $62.1_{-129.9}^{+208.7}$ & $0.011_{-0.021}^{+0.020}$ & $0.369_{-0.021}^{+0.018}$ \\ 
\hline 
7 & $1.5G_{\rm N}$ & $18.08_{-5.04}^{+25.32}$ & $10.67_{-3.45}^{+18.52}$ & $0.866_{-0.192}^{+0.087}$ & $89.5_{-29.0}^{+30.3}$ & $117.9_{-61.8}^{+176.9}$ & $170.3_{-352.7}^{+13.7}$ & $-0.001_{-0.021}^{+0.021}$ & $0.105_{-0.030}^{+0.053}$ \\ 
\hline 
8 & $1.5G_{\rm N}$ & $14.22_{-0.37}^{+1.07}$ & $22.39_{-7.07}^{+17.09}$ & $0.623_{-0.239}^{+0.189}$ & $112.5_{-3.9}^{+6.8}$ & $81.6_{-31.6}^{+29.1}$ & $270.0_{-37.0}^{+53.9}$ & $0.003_{-0.021}^{+0.021}$ & $0.393_{-0.046}^{+0.040}$ \\ 
\hline 
9 & $1.5G_{\rm N}$ & $25.16_{-2.91}^{+10.84}$ & $20.36_{-4.76}^{+16.10}$ & $0.758_{-0.307}^{+0.124}$ & $97.8_{-55.4}^{+40.2}$ & $181.5_{-128.7}^{+128.0}$ & $-90.7_{-111.7}^{+291.0}$ & $-0.000_{-0.021}^{+0.021}$ & $0.197_{-0.031}^{+0.035}$ \\ 
\hline 
11 & $1.5G_{\rm N}$ & $18.64_{-0.48}^{+2.22}$ & $32.02_{-11.42}^{+28.83}$ & $0.620_{-0.238}^{+0.181}$ & $121.8_{-5.0}^{+6.8}$ & $61.2_{-23.0}^{+30.0}$ & $276.9_{-37.6}^{+50.0}$ & $0.004_{-0.021}^{+0.021}$ & $0.360_{-0.047}^{+0.040}$ \\ 
\hline 
13 & $1.5G_{\rm N}$ & $11.20_{-0.31}^{+0.76}$ & $10.23_{-1.82}^{+4.52}$ & $0.236_{-0.113}^{+0.146}$ & $142.3_{-9.5}^{+12.6}$ & $134.4_{-111.3}^{+36.1}$ & $205.6_{-32.4}^{+105.1}$ & $0.001_{-0.021}^{+0.021}$ & $0.350_{-0.037}^{+0.049}$ \\ 
\hline 
17 & $1.5G_{\rm N}$ & $18.35_{-2.66}^{+9.19}$ & $13.05_{-2.51}^{+11.62}$ & $0.770_{-0.263}^{+0.106}$ & $75.3_{-40.5}^{+64.8}$ & $191.9_{-166.4}^{+145.8}$ & $-114.2_{-49.5}^{+276.4}$ & $-0.000_{-0.021}^{+0.021}$ & $0.343_{-0.014}^{+0.015}$ \\ 
\hline 
19 & $1.5G_{\rm N}$ & $27.27_{-4.10}^{+15.05}$ & $17.91_{-3.64}^{+14.75}$ & $0.648_{-0.221}^{+0.093}$ & $104.9_{-70.7}^{+41.2}$ & $162.0_{-99.7}^{+132.0}$ & $165.7_{-353.1}^{+26.4}$ & $-0.001_{-0.021}^{+0.021}$ & $0.176_{-0.015}^{+0.026}$ \\ 
\hline 
23 & $1.5G_{\rm N}$ & $26.34_{-5.83}^{+21.04}$ & $19.85_{-6.47}^{+26.02}$ & $0.759_{-0.417}^{+0.190}$ & $93.2_{-33.0}^{+14.6}$ & $223.5_{-71.3}^{+95.5}$ & $77.5_{-271.9}^{+111.1}$ & $-0.001_{-0.021}^{+0.021}$ & $0.173_{-0.028}^{+0.047}$ \\ 
\hline 
24 & $1.5G_{\rm N}$ & $27.06_{-2.15}^{+6.69}$ & $16.00_{-1.97}^{+4.88}$ & $0.893_{-0.198}^{+0.062}$ & $99.3_{-55.4}^{+24.4}$ & $62.6_{-42.4}^{+262.0}$ & $176.1_{-5.1}^{+11.0}$ & $-0.001_{-0.021}^{+0.021}$ & $0.142_{-0.020}^{+0.043}$ \\ 
\hline 
27 & $1.5G_{\rm N}$ & $20.47_{-2.61}^{+6.86}$ & $15.68_{-3.88}^{+9.58}$ & $0.482_{-0.171}^{+0.159}$ & $99.1_{-2.1}^{+3.2}$ & $126.9_{-59.3}^{+38.7}$ & $193.1_{-19.1}^{+49.6}$ & $-0.000_{-0.021}^{+0.021}$ & $0.283_{-0.057}^{+0.059}$ \\ 
\hline 
28 & $1.5G_{\rm N}$ & $20.07_{-3.38}^{+14.19}$ & $13.25_{-3.63}^{+11.85}$ & $0.831_{-0.235}^{+0.116}$ & $98.0_{-27.8}^{+15.1}$ & $97.9_{-59.5}^{+218.7}$ & $176.2_{-38.8}^{+10.7}$ & $-0.001_{-0.021}^{+0.021}$ & $0.206_{-0.094}^{+0.116}$ \\ 
\hline 
32 & $1.5G_{\rm N}$ & $14.07_{-0.89}^{+3.09}$ & $16.46_{-4.20}^{+11.05}$ & $0.715_{-0.295}^{+0.186}$ & $118.8_{-10.3}^{+17.9}$ & $247.5_{-31.0}^{+49.1}$ & $132.3_{-61.7}^{+19.5}$ & $0.001_{-0.021}^{+0.021}$ & $0.446_{-0.045}^{+0.052}$ \\ 
\hline 
33 & $1.5G_{\rm N}$ & $33.55_{-2.94}^{+11.01}$ & $32.11_{-8.91}^{+25.99}$ & $0.758_{-0.350}^{+0.171}$ & $77.8_{-19.8}^{+45.2}$ & $182.4_{-98.7}^{+91.9}$ & $81.4_{-294.7}^{+122.6}$ & $0.000_{-0.021}^{+0.021}$ & $0.254_{-0.033}^{+0.061}$ \\ 
\hline 
38 & $1.5G_{\rm N}$ & $13.25_{-2.08}^{+7.59}$ & $8.06_{-1.59}^{+6.10}$ & $0.889_{-0.277}^{+0.088}$ & $85.6_{-23.3}^{+30.4}$ & $200.8_{-155.8}^{+119.6}$ & $177.1_{-7.9}^{+13.2}$ & $-0.001_{-0.021}^{+0.021}$ & $0.220_{-0.016}^{+0.045}$ \\ 
\hline 
39 & $1.5G_{\rm N}$ & $13.49_{-2.08}^{+3.62}$ & $15.89_{-4.71}^{+16.27}$ & $0.508_{-0.094}^{+0.076}$ & $139.3_{-15.0}^{+22.1}$ & $197.6_{-45.0}^{+47.4}$ & $98.7_{-50.4}^{+25.1}$ & $0.002_{-0.021}^{+0.021}$ & $0.478_{-0.007}^{+0.009}$ \\ 
\hline 
53 & $1.5G_{\rm N}$ & $5.62_{-0.72}^{+2.07}$ & $4.11_{-0.73}^{+2.62}$ & $0.554_{-0.093}^{+0.087}$ & $56.5_{-3.8}^{+5.1}$ & $115.2_{-39.0}^{+42.9}$ & $205.7_{-16.8}^{+16.8}$ & $-0.001_{-0.021}^{+0.021}$ & $0.567_{-0.035}^{+0.035}$ \\ 
\hline 
56 & $1.5G_{\rm N}$ & $14.77_{-3.89}^{+5.98}$ & $10.60_{-3.58}^{+7.44}$ & $0.534_{-0.068}^{+0.094}$ & $94.5_{-1.0}^{+1.5}$ & $252.9_{-38.8}^{+37.6}$ & $160.5_{-28.9}^{+20.3}$ & $-0.001_{-0.021}^{+0.021}$ & $0.312_{-0.041}^{+0.044}$ \\ 
\hline 
58 & $1.5G_{\rm N}$ & $31.27_{-7.13}^{+7.44}$ & $17.12_{-4.19}^{+4.68}$ & $0.880_{-0.037}^{+0.041}$ & $96.8_{-1.7}^{+2.4}$ & $252.5_{-9.0}^{+8.4}$ & $176.2_{-4.7}^{+3.0}$ & $-0.001_{-0.021}^{+0.021}$ & $0.117_{-0.026}^{+0.032}$ \\ 
\hline 
63 & $1.5G_{\rm N}$ & $11.67_{-2.23}^{+6.47}$ & $8.05_{-2.27}^{+6.55}$ & $0.707_{-0.270}^{+0.129}$ & $70.1_{-22.4}^{+60.4}$ & $252.1_{-151.6}^{+61.2}$ & $186.4_{-18.3}^{+22.6}$ & $-0.001_{-0.021}^{+0.021}$ & $0.248_{-0.036}^{+0.094}$ \\ 
\hline
\enddata

\tablenotetext{a}{All parameters are given in the same format as Table~\ref{tab:fitparameters}.} 
\tablenotetext{b}{All entries satisfy $g_{\rm N}<10^{-10.1}\,{\rm m}\,{\rm s}^{-2}$.}

\end{deluxetable*}

\subsection{Additional and auxiliary results} \label{sec:additional_result}

\begin{figure*}[!htb]
    \centering
    \includegraphics[width=1.\linewidth]{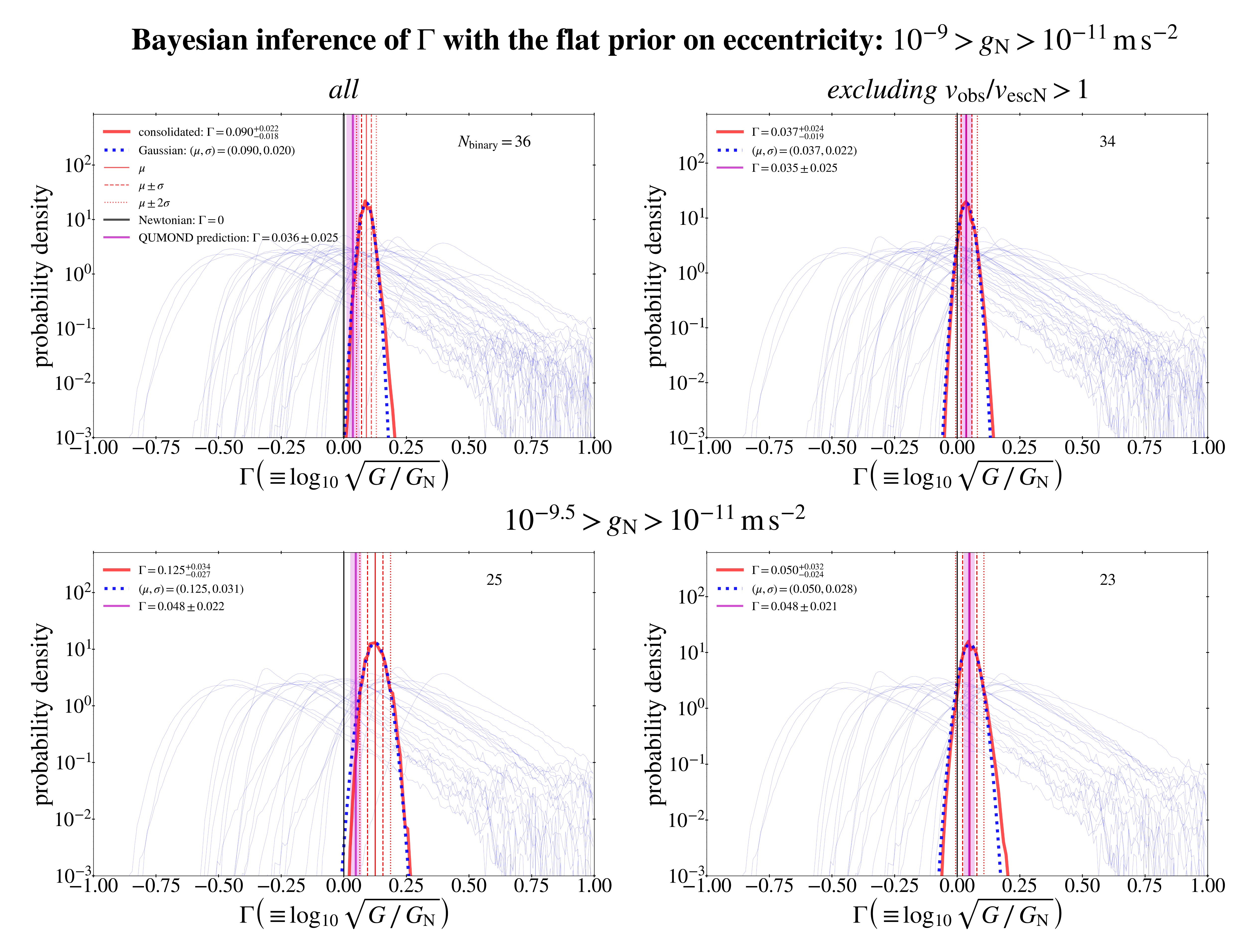}
    \caption{
    Same as Figure~\ref{fig:Gamma} but with the flat prior on eccentricity.
    }
    \label{fig:Gamma_flat}
\end{figure*}
In this subsection we present additional results obtained by varying the prior on eccentricity from the nominal choice of the thermal prior. We also check the posterior distributions of the orbit and orientation parameters with respect to the prior distributions required by the randomness of the sample.

Although our nominal choice is the thermal distribution, the most relevant distribution of eccentricities for wide binaries with $s>1$~kau would be a superthermal distribution of $\alpha\approx 1.3$ in $f_{\rm pr}(e)=(1+\alpha)e^\alpha$, according to the \cite{Hwang:2022} Bayesian study of a large number of wide binaries with precise projected displacements and velocities. Thus, we consider a superthermal prior with $\alpha=1.3$. It turns out that the results with the superhermal prior are nearly indistinguishable from the results with the thermal prior. For the whole clean sample, we have $\Gamma=0.104_{-0.021}^{+0.030}$. This is not surprising because the moderate difference between the thermal distribution and the specific superthermal distribution can make only a tiny effect on the parameter inference when the precise 3D velocities along with the nearly exact two sky-projected displacements have some constraining power on the orbital parameters. We also consider the uninformative flat prior on eccentricity to investigate the role of the eccentricity prior in the gravity inference and at the same time to gauge the power of our sample in constraining the eccentricity distribution.

Figure~\ref{fig:Gamma_flat} shows the results on $\Gamma$ with the flat prior on eccentricity. Compared with the nominal results, the inferred value of $\Gamma$ is slightly reduced to $\Gamma=0.090_{-0.018}^{+0.021}$ but the significance of the deviation from Newton remains the same (a $5.0\sigma$ deviation) because individual PDFs get somewhat narrower, and consequently the consolidated PDF is also narrower. Thus, perhaps contrary to expectation, the flat prior on eccentricity does not make Newtonian gravity less strongly discrepant with our wide binary sample.

\begin{figure*}[!htb]
    \centering
    \includegraphics[width=1.\linewidth]{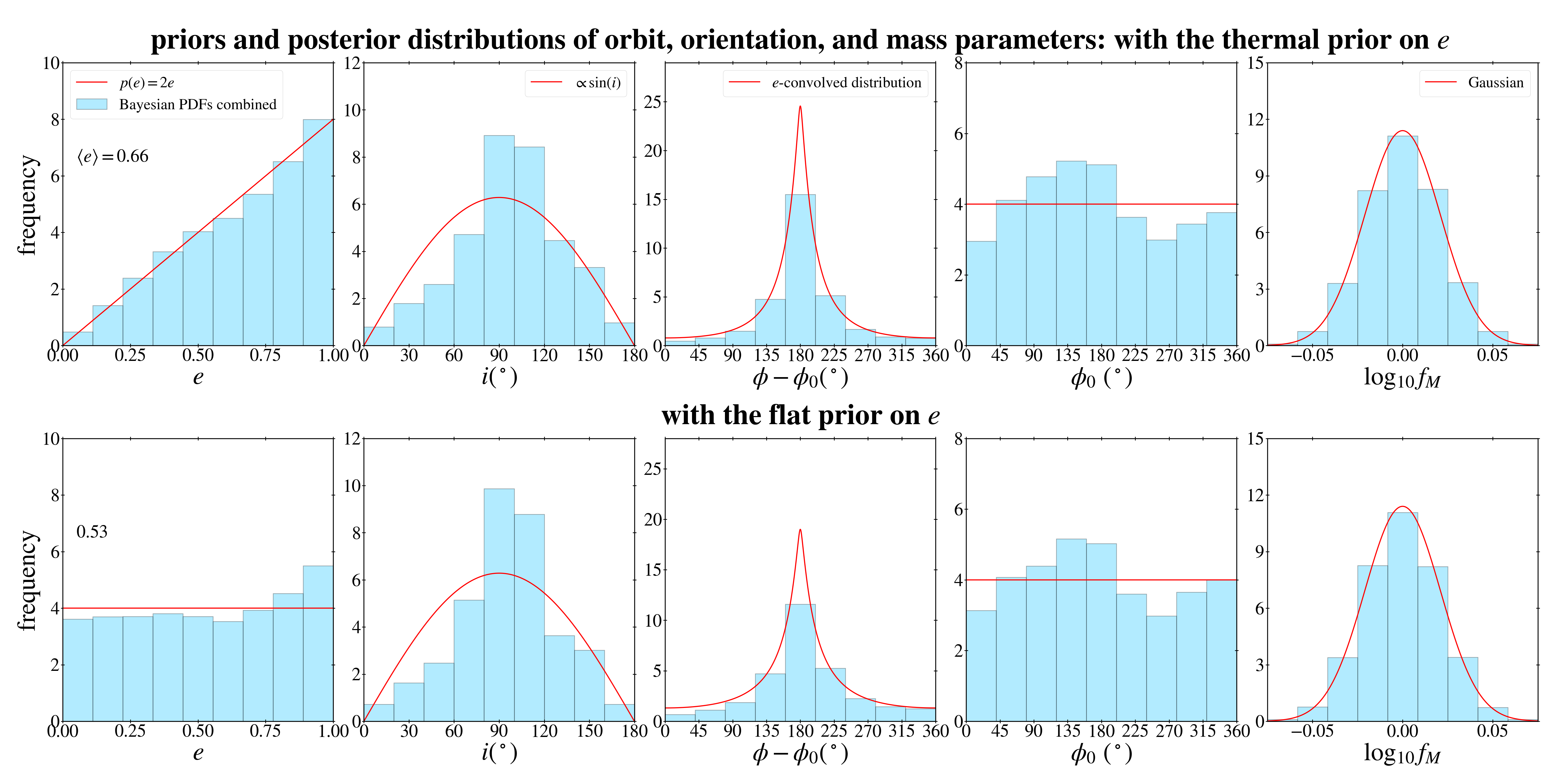}
    \caption{
    Posterior distributions of the orbit, inclination, and mass parameters are compared with the priors represented by red lines/curves. The histogram of a parameter represents the composite of the individual posterior PDFs of the parameter for all 36 wide binaries in the clean sample. The prior distribution of $\Delta\phi(=\phi-\phi_0)$ is the function (with a rescaled amplitude) given by Equation~(\ref{eq:PrDelphi}) convolved with $e$ shown in the leftmost column. 
    }
    \label{fig:distribution_5prmt}
\end{figure*}

Figure~\ref{fig:distribution_5prmt} shows the posterior distributions of the orbit, inclination, and mass parameters for the two Bayesian results with the thermal or flat prior on eccentricity. Due to the degeneracy between the gravity parameter $\Gamma$ and the other parameters in fitting the data $(\mathbf{r},\mathbf{v})$, when $\Gamma$ is free while the other parameters are constrained by their priors, the posterior distributions of the latter are expected to collectively follow the prior distributions as a population. If the posterior distribution of a parameter is significantly different from the prior individually or collectively, that would strongly indicate that the kinematic data prefer a value or distribution different from the prior.

Figure~\ref{fig:distribution_5prmt} shows that the parameters overall follow the prior distributions. In particular, the posterior distribution of the orbit true anomaly parameter $\Delta\phi(=\phi-\phi_0)$ is well consistent with the $e$-convolved distribution fulfilling the required self-consistency. This means that the inferred PDFs of $\Gamma$ are based on orbital inclination and occupancy distributions consistent with isotropy and random presentations.

For the thermal prior, the posterior distribution of $e$ matches well the prior distribution. However, for the flat prior, the posterior distribution of $e$ is tilted toward the upper limit indicating that the data prefer a non-flat distribution with $\alpha>0$. However, the relatively moderate tilt may indicate that the kinematic data alone (when $\Gamma$ is free) are not sufficient to constrain well the orbit parameters, especially due to the large uncertainty of the radial separation $z^\prime$. 

Finally, we note that if gravity is fixed (e.g., Newtonian or a boosted gravity), the orbit and orientation parameters can be significantly constrained by the data $(\mathbf{r},\mathbf{v})$, and the posterior distributions of the parameters, in particular the orbit true anomaly parameter $\Delta\phi$, can be used to test the assumed gravity model, as was investigated in the pilot study by \cite{Chae:2025b}. In such a test, priors should not be imposed on the orbit and orientation parameters because gravity is fixed (so the degeneracy between gravity and orbit parameters is not a concern) and we seek to test the distributions required by the assumed gravity without the influence of any priors. As already noted with a small sample of 8 or 9 wide binaries with $g_{\rm N}<10^{-9}$~m\,s$^{-2}$ by \cite{Saglia:2025} and \cite{Chae:2025b}, Newtonian gravity requires a biased distribution of $\Delta\phi$ towards the periastron to compensate for the boosted gravity obtained for the correct prior distribution (see Figure~\ref{fig:distribution_5prmt}), violating Kepler's second law in a statistical sense for the population. With a 4 times larger low-acceleration sample than the \cite{Saglia:2025} sample, the statistical significance is now clearly stronger. Thorough statistical tests of Newtonian gravity as well as control boosted gravity (toy) models will be presented in a separate work as those work involves statistical methods such as the Kolmogorov-Smirnov test, the Anderson-Darling test, and Bayesian information criterion (see \citealt{Chae:2025b}). We stress that this work is mainly devoted to the construction of the wide binary sample and the ``measurement'' of $\Gamma$, which is the most reliable and straightforward representation of the data (without requiring additional statistical methods) that can be readily carried out with the available code \citep{Chae:2025b}. 

\subsection{Exploring possible systematic errors: Can the gravitational anomaly be removed?} \label{sec:systematic}

So far we have presented the main and additional results on the gravity inference along with relevant discussions based on the currently available observational information. Here we further consider extreme possibilities to check whether the low-acceleration gravitational anomaly from the clean sample can be removed. The gravity inference may be affected by variation in modeling inputs or sample selection criteria.

\begin{figure}[!htb]
    \centering
    \includegraphics[width=1.\linewidth]{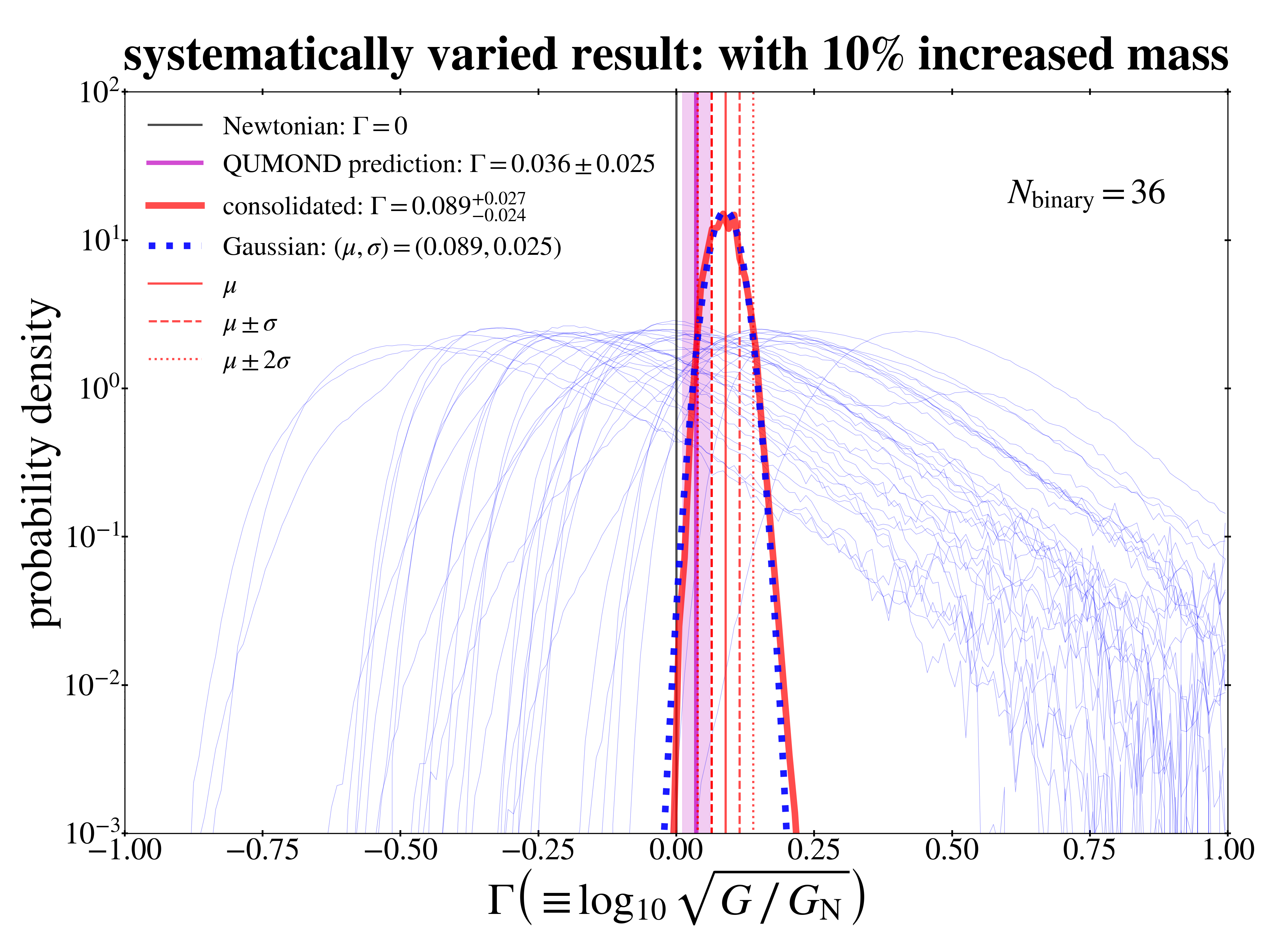}
    \caption{
    Same as the top left panel of Figure~\ref{fig:Gamma} but with the masses of all the stars increased by 10\%.
    }
    \label{fig:Gamma_massplus10pt}
\end{figure}

As for modeling inputs, we have already considered a sufficiently wide range of possibilities for the orbit eccentricity. It was also shown that our adopted stellar masses agree well with the Gaia FLAME masses (Figure~\ref{fig:mass}) whenever the latter are available. However, given that gravity inference is directly and significantly affected by the input mass, here we consider the possibility that the masses of all our stars are systematically larger by 10\% than the values we have adopted. This is in part motivated from the study by \cite{Saglia:2025} who quote a maximum of 10\% systematic error for their masses based on the HARPS spectra. Our sample is a composite of several subsamples and masses were estimated through independent methods for different subsamples. So it is unlikely that all stellar masses are systematically biased in the same direction by 10\%. Thus, the uniform increase of 10\% for all stars is an extreme possibility. Figure~\ref{fig:Gamma_massplus10pt} shows the result on $\Gamma$ with the increased mass. The inferred value is reduced to $\Gamma=0.089_{-0.024}^{+0.027}$, which is however still inconsistent with Newtonian at a $3.7 \sigma$ level. This shows that any reasonable adjustment of the currently estimated stellar masses cannot remove the gravitational anomaly.

As for sample selection criteria, we have employed the philosophy of sacrificing statistical significance to maximize the purity of the sample, by taking only wide binaries that pass all the stringent selection criteria \emph{and} have additional observational indicators for kinematically uncontaminated pure binaries. This is why only 17\% of the raw sample with relatively precise $v_r$ (Figure~\ref{fig:vr_err}) survived in the final clean sample. The purity of the clean sample may be reaffirmed by the fact that none of the posterior PDFs of $\Gamma$ are abnormal, and in particular, our recovered distributions of orbital phases and inclinations are consistent with isotropy and angular momentum conservation expectations, as shown in Figure~\ref{fig:distribution_5prmt}. However, it is interesting to note that some component stars of our pure binaries are classified as unresolved binaries according to the Gaia DR3 multiple-star classifier \citep[MSC; ][]{Gaia_Apsis:StellarParameters}, making binaries triples or quadruples. In line with \cite{Saglia:2025}, we did not use the MSC because it is not reliable on an individual basis.

\begin{figure}[!htb]
    \centering
    \includegraphics[width=1.\linewidth]{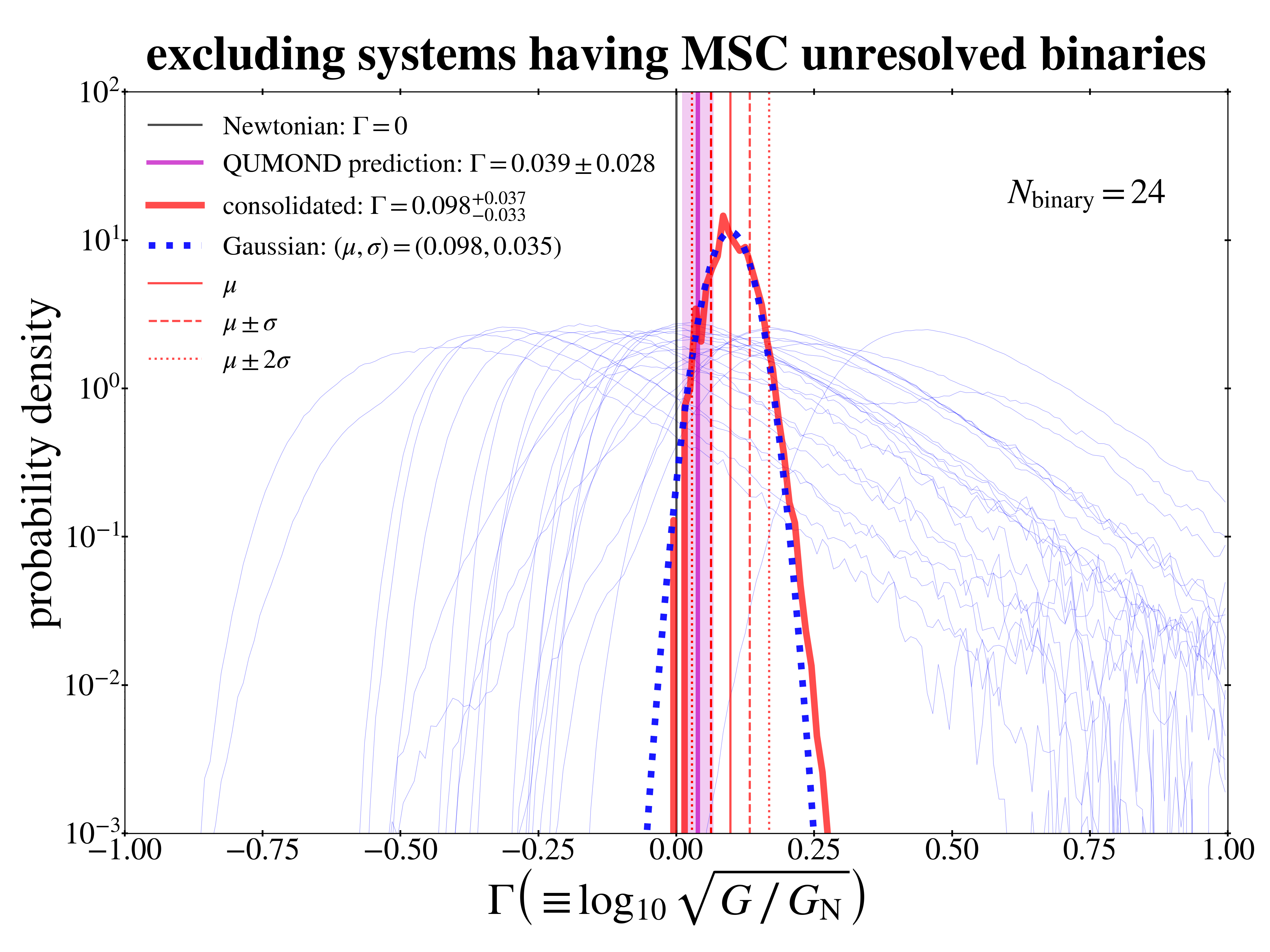}
    \caption{
    Same as the top left panel of Figure~\ref{fig:Gamma} but for a subsample excluding systems marked as unresolved binaries (potentially triples or quadruples) in the Gaia DR3 multiple-star classifier.
    }
    \label{fig:Gamma_excludingMSC}
\end{figure}

It turns out that the \cite{Saglia:2025} sample and our clean sample include systems listed as triples or quadruples in the MSC-based \texttt{Object Type} given in the Gaia DR3 online archive. This classification is known to be significant only in a qualitative and statistical sense. Considering our general philosophy of this work, we consider removing systems marked as triples/quadruples by the MSC. Figure~\ref{fig:Gamma_excludingMSC} shows the result on gravity after excluding the 12 systems (i.e.\ one third of our clean sample) that include MSC-based unresolved binary stars. The inferred value of $\Gamma=0.098_{-0.033}^{+0.037}$ for the remaining 24 binaries is nearly indistinguishable from that for the entire sample other than the slightly increased statistical errors consistent with the reduced sample size. This indicates that the excluded systems are not special but random. If they were triples/quadruples, the inferred $\Gamma$ for them would be abnormally large because of the systematically underestimated masses and the kinematic contamination due to multiplicity. In that hypothetical case, our inference would shift into accordance with Newtonian expectations once such cases were removed, which is not what happens. This exercise demonstrates that the gravitational anomaly found cannot be removed by invoking hidden unresolved close binaries as per the Gaia DR3 MSC in our clean sample.

Regarding the sample selection, it is of interest to consider only wide binaries whose $v_r$ are stable over more than several years from direct multi-epoch observations. Speckle observations show that our targets (before applying the additional cuts to select the clean sample) have a low chance of about 5\% to have a faint Speckle resolved companion. Since only 10 wide binaries (including 6 cases with $\log_{10}g_{\rm N}<10^{-9.5}\,{\rm m}\,{\rm s}^{-2}$) from the clean sample were directly verified to be free of resolved faint companion from our Speckle observations, it is in principle possible that the remaining 26 wide binaries may include at most (noting that our clean sample is more strictly defined than the Speckle observation samples) one or two cases that may have a faint companion more than tens of kau apart from a component star. However, removing \emph{any} one or two wide binaries from our clean sample cannot remove the anomaly. Even in the extreme case that both {\#}3 and {\#}59 are selectively removed, about $2\sigma$ anomaly remains.  Thus, the remaining concern is any unresolved companion within tens of kau from a component star.

\begin{figure*}[!htb]
    \centering
    \includegraphics[width=0.8\linewidth]{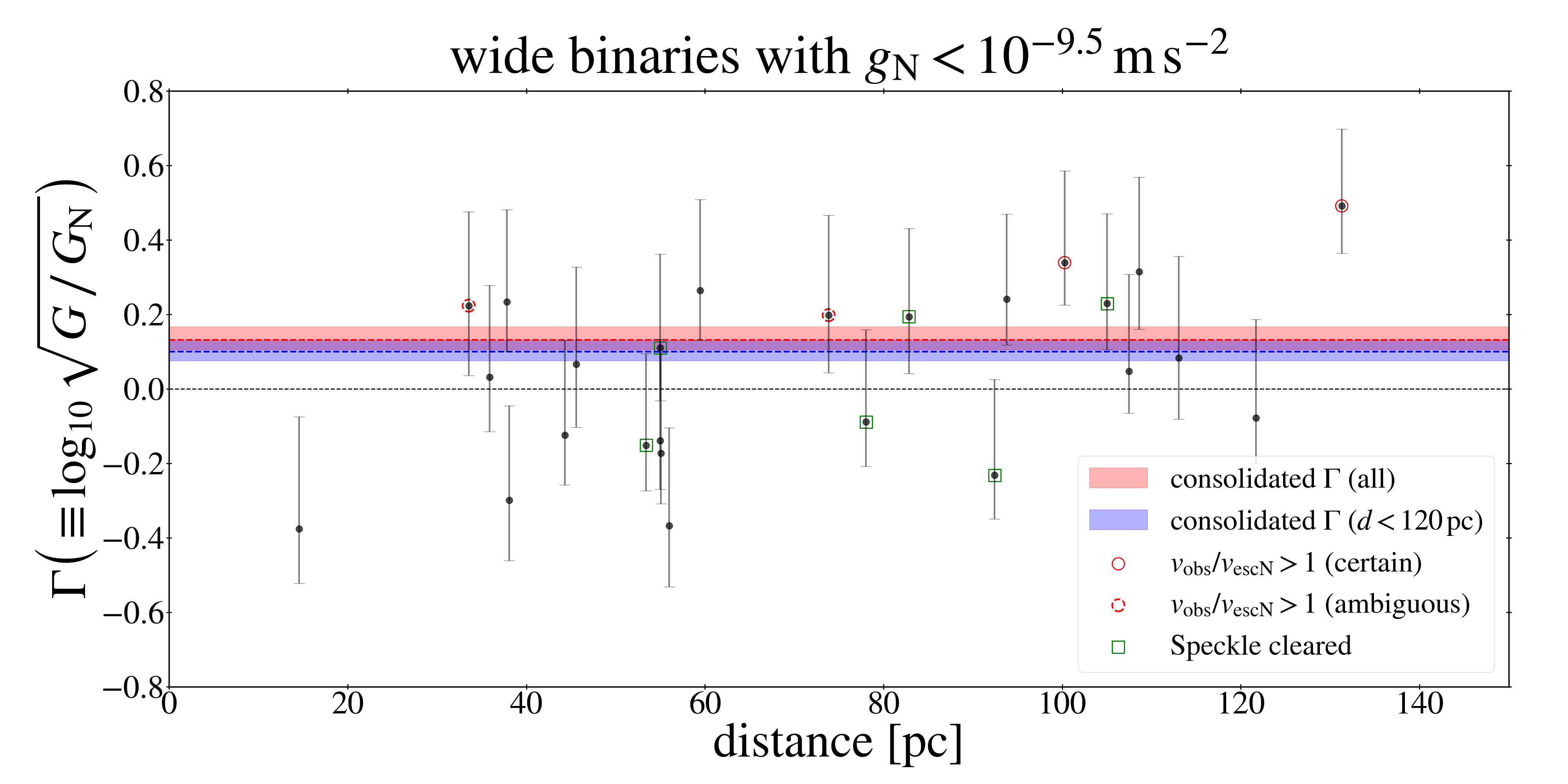}
    \caption{
    Individual values of $\Gamma$ for wide binaries with $g_{\rm N}<10^{-9.5}\,{\rm m}\,{\rm s}^{-2}$ against their distances from the Sun. Six Speckle-cleared systems are indicated by green squares, which alone indicate a mild anomaly.
    }
    \label{fig:Gamma_dist}
\end{figure*}

Companions of relatively close separation less than several tens au may be missed by Speckle observations with a probability increasing with distance because the same angular resolution limit means an increasing limit of physical separation. Our clean sample already excludes a relatively larger distance of $>150$~pc. However, if kinematic contamination in the intermediate separation (that cannot be flagged by Gaia's \texttt{ruwe} values: see below) from $\approx 10$~au to several tens au is present, it is expected to cause a trend in the inferred $\Gamma$ with distance in our clean sample as wide binaries are distributed over a broad distance range of $10 <d <150$~pc. Figure~\ref{fig:Gamma_dist} exhibits the individual values of $\Gamma$ with respect to the distances of the binaries with $g_{\rm N}<10^{-9.5}\,{\rm m}\,{\rm s}^{-2}$ in the clean sample. There is no obvious trend of $\Gamma$ with $d$ and the four (including the two ambiguous ones) systems with $v_{\rm obs}/v_{\rm escN}>1$ are distributed over the entire distance range. Binary~{\#}3 happens to have the largest distance of $\approx 131\,{\rm pc}$, but the consolidated value of $\Gamma$ is reduced only by a small amount if we limit the distance to 120\,pc, as shown in the figure. This indicates that significant kinematic contamination in the intermediate range is unlikely.

\begin{figure}[!htb]
    \centering
    \includegraphics[width=1.\linewidth]{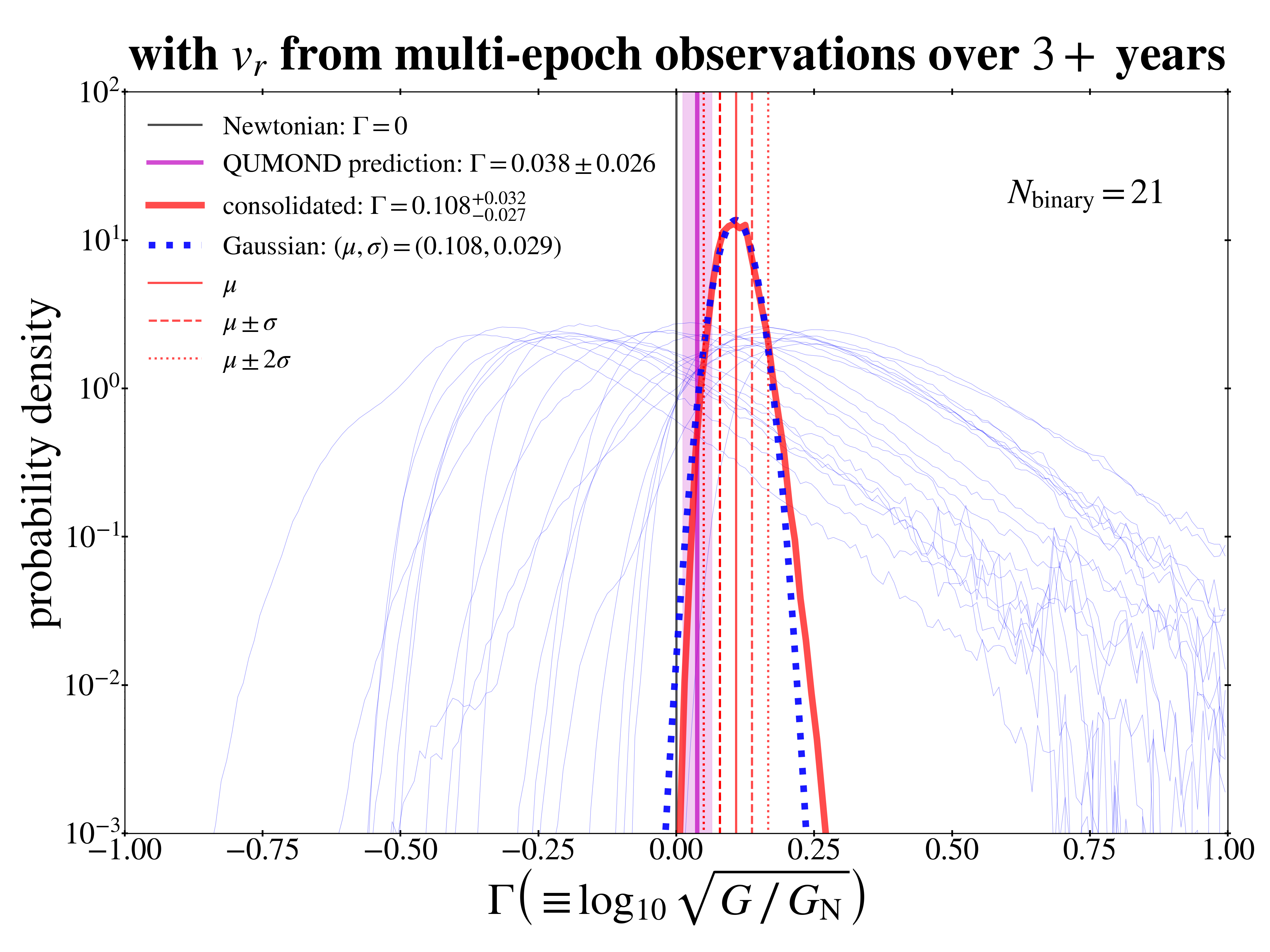}
    \caption{
    Same as the top left panel of Figure~\ref{fig:Gamma} but for a subsample with $v_r$ stable from two observations separated by 3 - 11 years (except for 4 systems from those shown in the right panel of Figure~\ref{fig:vr_compare}). 
    }
    \label{fig:Gamma_multiepoch}
\end{figure}

Closer companions ($\la 10$~au) will have kinematic effects on either tangential velocities (preferentially when the orbit of the companion is close to face-on) or radial velocities (preferentially when the orbit of the companion is close to edge-on). The kinematic effects on tangential velocities can be flagged by Gaia's \texttt{ruwe} \citep{Belokurov:2020} while those on radial velocities can be flagged through a monitoring observation covering a significant fraction of the orbital period. Because we have already required $\texttt{ruwe}<1.25$, requiring an observed stability of $v_r$ over more than several years can complete the observational requirement for pure binaries. Figure~\ref{fig:Gamma_multiepoch} shows the inference of $\Gamma$ for 21 wide binaries with time baselines of 3 - 11 years taken (excluding shorter baseline cases) from those shown in the right panel of Figure~\ref{fig:vr_compare}. Again, the inferred value of $\Gamma=0.108_{-0.027}^{+0.032}$ is strongly discrepant with Newton.

\begin{figure}[!htb]
    \centering
    \includegraphics[width=1.\linewidth]{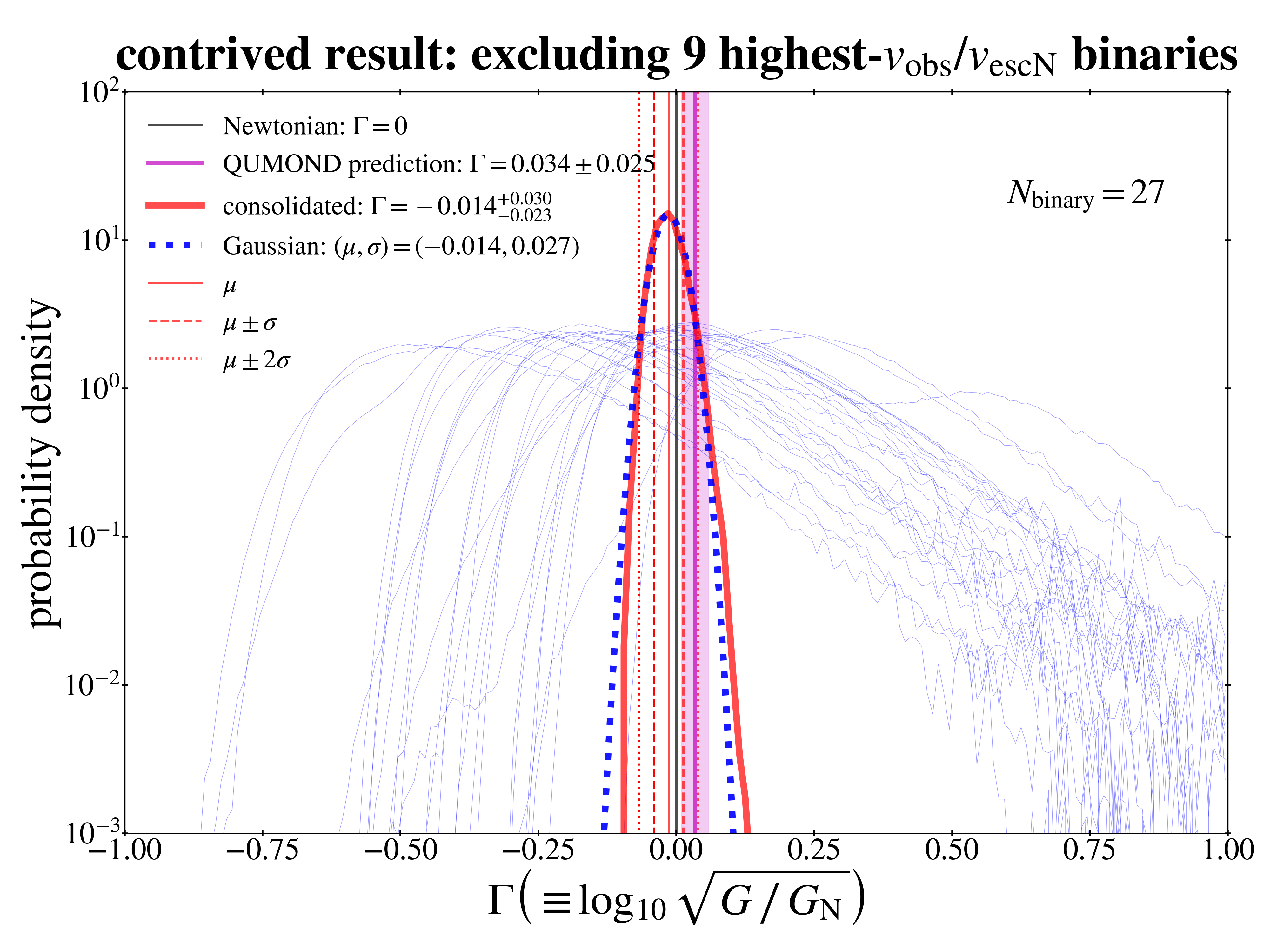}
    \caption{
    Same as the top left panel of Figure~\ref{fig:Gamma} but for a subsample selectively excluding a quarter of the sample. The excluded 9 wide binaries are those having the highest values of $v_{\rm obs}/v_{\rm escN}$ (see Tables~\ref{tab:physparameters} and~\ref{tab:WB_Newtonunbound}).
    }
    \label{fig:Gamma_contrived}
\end{figure}

Therefore, it seems impossible to remove the gravitational anomaly seen in the clean sample within any conceivable variation of the currently available observational information. Yet, one might be still skeptical of the gravitational anomaly and want to find a way to remove it. As a cautionary tale of any fabricated result through an intentional or unintentional distortion of our sample (or any other samples present and future), we carry out an exercise of deliberately removing choice systems in order to produce a Newtonian result. For this we consider selectively removing wide binaries that are responsible for the gravitational anomaly. Figure~\ref{fig:Gamma_contrived} shows a contrived result obtained by removing 9 wide binaries (a quarter of the clean sample) that have highest-$v_{\rm obs}/v_{\rm escN}$ values. This is a false result that appears to agree with Newton remarkably well. The removal of the 9 systems with the smallest values of $v_{\rm obs}/v_{\rm escN}$ would correspondingly shift our results to artificially large values of $\Gamma$, while removing 9 random binaries on average leaves our inference unchanged, beyond the expected increase of the confidence intervals due to the reduced sample.

The Gaia limiting magnitude of 20.5 out to the limit of our sample of 150 pc implies
that stellar companions at separations beyond the Gaia resolution of $\approx 0.5$ arcsec \citep{Gaia:2018,Lindegren:2021} can be excluded as contaminating our sample. At closer separations, the combination of only 5\% close companions identified by our Speckle campaign and the clearing of a subsample of our final clean sample through this technique, exclude a further range of potential stellar companions. In order to bring our results into consistency with Newtonian expectations we would require, as detailed above, about 9 of our binaries to be contaminated, something which would require a probability of about $(0.05)^{9}$ if unresolved stellar companions were to blame. At even closer separation, where close binary periods become comparable to the Gaia DR3 temporal range of 34 months, the use of \texttt{ruwe} restrictions for all (e.g. see \cite{Castro-Ginard:2024} where the presence of unresolved stellar companions of all types has been shown to be incompatible with stars in DR3 for values of $\texttt{ruwe}> 1.15$, a limit only one of our binaries in the clean sample crosses) and consistent multi-epoch high quality radial velocity measurements, in practice eliminate stellar companion contamination as a concern.

\begin{figure}[!htb]
    \centering
    \includegraphics[width=1.\linewidth]{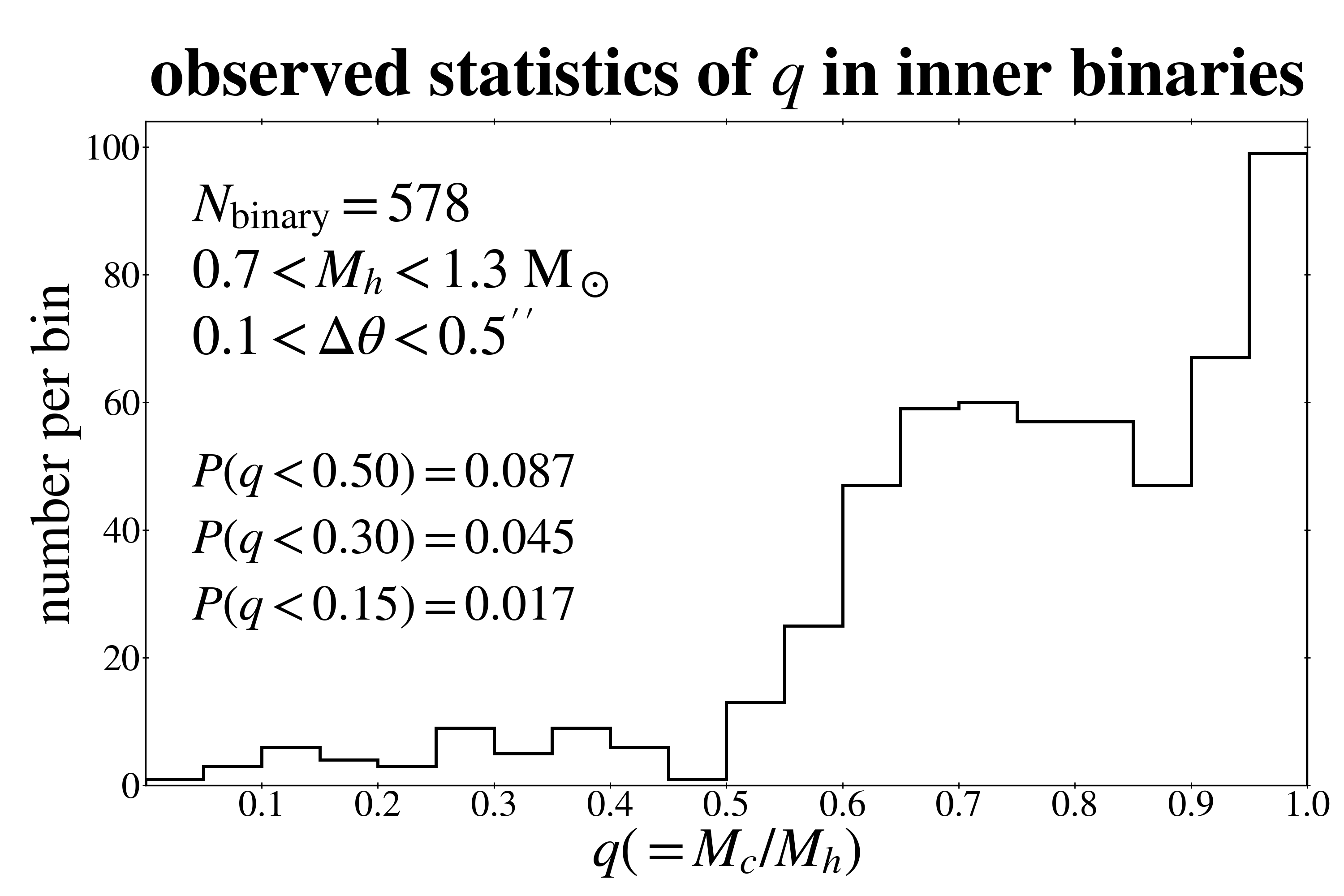}
    \caption{
    Statistics of mass ratio in close inner binaries satisfying the specified mass and angular separation range based on the multiple-star catalog by A.\,Tokovinin.
    }
    \label{fig:q_distribution}
\end{figure}

Nevertheless, let us consider numerical simulations to further assess whether hidden binaries with separation less than several tens au can have substantial impact on our conclusion of the gravitational anomaly in our clean sample. We specifically consider Binary~{\#}3 and {\#}59 for this purpose, as they are \emph{individually} discrepant with Newton and contribute most to the $5\sigma$ anomaly. For this, we use empirical statistics of inner binaries from a public large compilation of multiple-star systems established by A.\,Tokovinin.\footnote{https://www.ctio.noirlab.edu/$\sim$atokovin/stars/index.htm} Given that the four stars of the two systems have masses in the range $0.9 \la M \la 1.3{\rm M}_\odot$ and we are concerned with separations $\ga 10\,{\rm au}$ (or about $0.1{^\prime}{^\prime}$ for their distances) up to angular separation of $0.5{^\prime}{^\prime}$, we first obtain an empirical distribution of mass ratio $q(\equiv M_c/M_h)$ (where $M_h$ and $M_c$ denote the masses of the host and companion stars) for $0.7\la M_h \la 1.3\,{\rm M}_\odot$ (which covers typical masses in our sample) and angular separation $0.1<\Delta\theta<0.5{^\prime}{^\prime}$. Figure~\ref{fig:q_distribution} shows the distribution of $q$ in 578 binaries. Notably, $q<0.5$ is quite rare and probability of $q<0.3$ is only 4.5\%. Probability of $q<0.15$ is just 1.7\%.

Although low values of $q$ are quite rare, we will consider only two cases of low values, $q=0.15$ and $0.3$, as higher values may have too large kinematic effects to be compatible with observed small 3D velocities $<1\,{\rm km}\,{\rm s}^{-1}$ in the binaries under consideration. For the simulation we consider a host star of $M_h=1.0{\rm M}_\odot$ and explore possibilities that a companion star is hidden in an angular-separation range of $0.1<\Delta\theta<0.5{^\prime}{^\prime}$. Noting that the corresponding physical separation is in the range $10\la r \la 65$\,au, we consider eccentricities $e\la 0.55$ based on the empirical statistics summarized in Figure~24 of \cite{Chae:2023}. Assuming that $M_c$ is too faint to make a contribution to the observed light of the system, the velocity induced on $M_h$ due to $M_c$ is given by
\begin{equation}
\begin{array}{ccl}
    v_{\rm{inn}} & = &\sqrt{G\frac{M_h+M_c}{r}}\frac{M_c}{M_h + M_c} f(e) \\ 
    &= &\sqrt{\frac{GM_h}{r}}\frac{q}{\sqrt{1+q}} f(e), \\
\end{array}
\label{eq:vinn}
\end{equation}
where $f(e)$ is the factor due to non-circular orbits, and we take $0.7\le f(e)\le 1.0$ that can cover from the apastron phase to near the periastron at $e\approx 0.5$. 

\begin{figure*}[!htb]
    \centering
    \includegraphics[width=1.\linewidth]{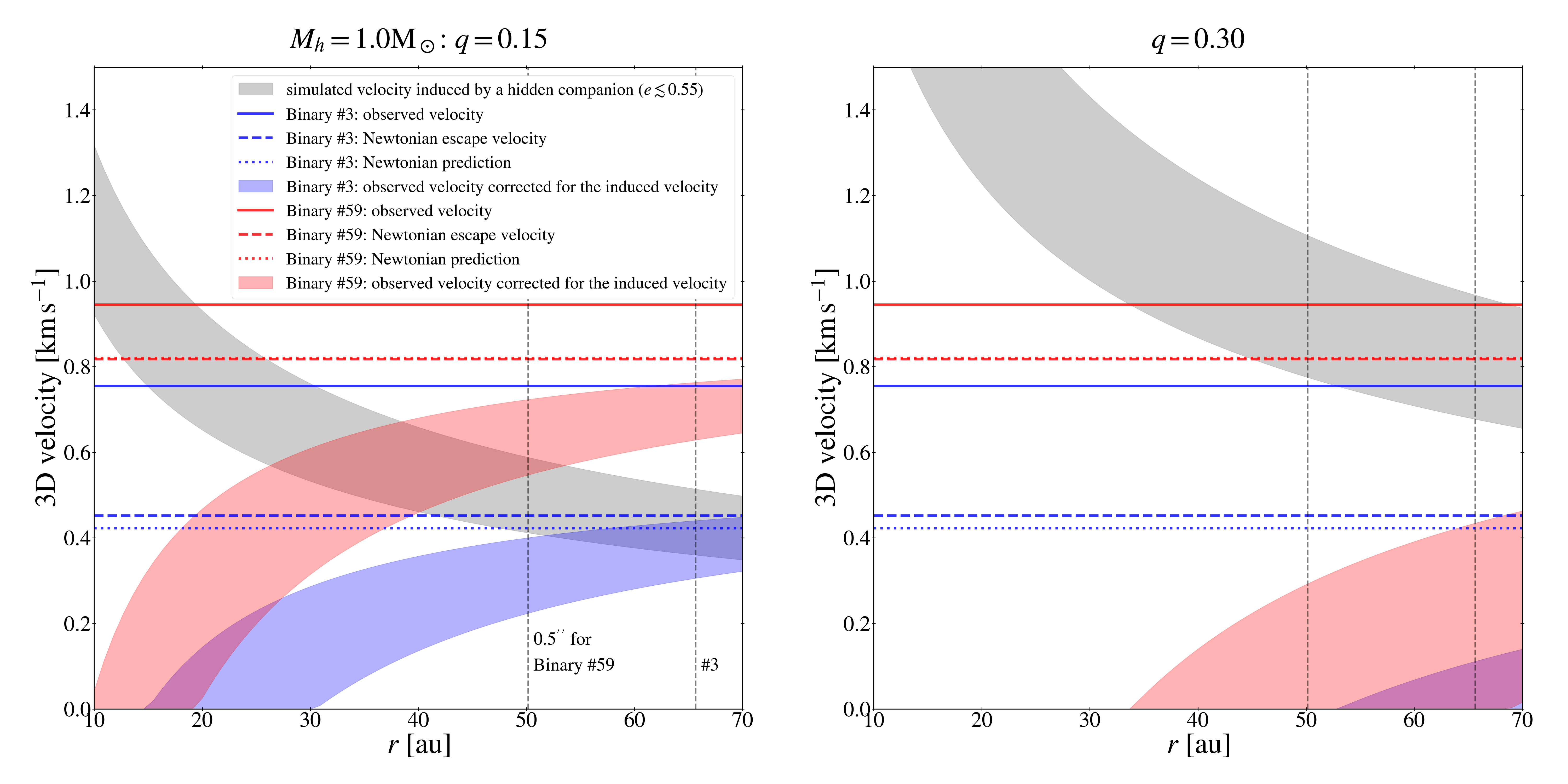}
    \caption{
    This figure shows the predicted kinematic contamination (gray band) due to a hypothetical inner binary. We use the observed 3D velocities in Binary~{\#}3 and {\#}59 as inputs of the simulations. The predicted 3D velocity ranges corrected for the contamination are indicated by blue and red bands for Binary~{\#}3 and {\#}59, respectively. See the text for the details.
    }
    \label{fig:closebinary}
\end{figure*}

When the observed (scalar) 3D velocity is $v_{\rm obs}$, the presence of the hidden companion would have contributed to its value on average as $v_{\rm obs}^2=v_{\rm out}^2 + v_{\rm inn}^2$, where $v_{\rm out}$ is the velocity of $M_h$ relative to the outer pure binary companion when the inner binary is absent. Thus, we have $v_{\rm out}=\sqrt{v_{\rm obs}^2 - v_{\rm inn}^2}$. Figure~\ref{fig:closebinary} shows the simulated distributions of $v_{\rm inn}$ and $v_{\rm out}$ for the two binaries. With $q=0.15$, there is a window of $50\la r \la 60$\,au in which the simulated $v_{\rm out}$ for Binary~{\#}3 can be compatible with its Newtonian prediction. However, the simulated $v_{\rm out}$ for Binary~{\#}59 is always lower than the Newtonian prediction for $\Delta\theta<0.5{^\prime}{^\prime}$ meaning that it would require a smaller value of $q$ to match Newton. With $q=0.3$, the predicted ranges of $v_{\rm out}$ are too low to match the Newtonian prediction for both binaries because $v_{\rm inn}$ is too large. Thus, while there is some chance that Binary~{\#}3 can agree with Newton by invoking an inner binary with $q\la 0.15$, such a chance seems extremely low for Binary~{\#}59. However, if an inner binary is present in one (or both) of the binaries, Figure~\ref{fig:closebinary} indicates that the more likely chance is that $v_{\rm obs}$ is slightly reduced to lessen the current large anomaly (which is even in moderate tension with the MOND prediction). Above all, values of $q\la 0.15$ are rare in reality, and thus invoking two such cases in a sample of 36 systems seems contrived.

The only remaining potential contaminants to reconcile our observations with Newtonian gravity would be brown dwarfs at separations above about 3 au, where the inner orbital periods would be larger than 34 months. Here we enter the `brown dwarf dessert', where independent observations have consistently identified an absence of brown dwarf companions to stars of the type we are using, with masses a little below one solar mass (e.g., \citealt{Kraus:2011}). Direct searches using eclipse and lensing surveys have shown this `dessert' to extend out to a few tens of au. Dynamical models of brown dwarf formation have shown instabilities in fragmenting disks to preclude the formation of stable brown dwarfs about stars of the type used here out to about 200 pc (e.g., \citealt{Vorobyov:2013}). Beyond such distances the kinematic signal of such a hypothetical contaminant becomes comparable to the noise level in our study, and the possibility ceases to be relevant. Hence, invoking a hypothetical distribution of undetected brown dwarf companions of sufficient frequency to explain the gravitational anomaly we detect within a Newtonian framework would not only be a contrived and {\it ad hoc} proposal lacking any independent evidential support, but would also run counter to all observational and theoretical knowledge on the point.

\section{Meanings and Implications of the Results} \label{sec:meaning}

In this work, we have assembled a sample of pure binaries and then focused on testing Newtonian gravity at low internal acceleration $\la 10^{-9}$~m\,s$^{-2}$ by measuring $\Gamma$ assuming elliptical orbits. If observed wide binaries represent random phases of Newtonian elliptical orbits, the consolidated PDF of $\Gamma$ must be consistent with zero as demonstrated for wide binaries with relatively small separation $\la 1$~kau or relatively strong internal acceleration $\ga 10^{-8}$~m\,s$^{-2}$ \citep{Chae:2025,Chae:2025b}. If not, it simply means that Newtonian gravity is broken at low acceleration (whatever the implication for gravity may be) because the result derived assuming the Newtonian framework does not agree with the Newtonian prediction.

The inferred value of $\Gamma$ is a pure measurement/quantification of the degree by which the data deviate from Newton under the specific assumption of elliptical orbits and angular momentum conservation. Thus, while it provides a direct test of Newtonian gravity, it is not straightforward to test specific nonstandard gravity such as MOND models \citep{Bekenstein:1984,Milgrom:2010} using $\Gamma$. The orbits of wide binaries in a nonstandard model can only be obtained by numerically solving the nonlinear field equation \citep{Pf-A:2025}. Mock data $(\mathbf{r},\mathbf{v})$ can be obtained from the mock orbits, and the mock data can be modeled in the same manner as the real data. Then, the inferred values of $\Gamma$ for the mock data can be compared with those for the real data obtained here to test the assumed gravity model. 

Simply obtaining a boost factor for the gravitational parameter through $\gamma = G/G_{\rm N} = 10^{2\Gamma}$ and testing a MOND model with it should be regarded as only a first-order approach. In concordance with the results mostly based on the 2D sky-plane velocities $v_p$ from recent works led independently by two of us \citep{Chae:2023,Chae:2024a,Chae:2024b,Chae:2025,Chae:2025b,Hernandez:2023,Hernandez:2024a,HernandezKroupa:2025,Yoon:2025,ChaeYoon:2026}, the first-order test shows that the MOND predicted gravitational boost is supported by our clean sample.

In the literature, there have been efforts (e.g., see \cite{Hernandez:2024review,ChaeYoon:2026} for reviews) to discriminate between Newton and MOND (or MOND-like gravity) in the low-acceleration regime using general samples of wide binaries that include triples and higher-order multiples (and even gravitationally unbound fly-bys). In principle, such an approach can work and should be concordant with the direct measurement of $\Gamma$. For that approach to work, two conditions must be met. First, the prediction of MOND models must be accurately correct. As \cite{Pf-A:2025} showed, analytical or numerical calculations under simplifying assumptions (e.g., one-particle equivalent description) can be misleading for wide binary orbital motions under the strong ($\approx 1.8a_0$) external field of the Milky Way. None of the existing studies used correct numerical solutions to date. Second, the degeneracy among gravity, the fraction of triples and higher-order multiples ($f_{\rm multi}$) in a sample of apparent binaries, and the fraction of fly-bys should be well under control. The fraction $f_{\rm multi}$ can be calibrated using a control subsample of small-separation binaries whose individual stars are required to satisfy the same observational criteria (e.g., $S/N$ of velocities, \texttt{ruwe}, etc) as those of the main subsample of wide binaries of interest. Fly-bys can be removed by requiring a threshold on the scalar relative velocity between the pair. Unfortunately, the studies in question failed to do either calibrate $f_{\rm multi}$ or remove fly-bys. Our clean sample of wide binaries (and future extended samples) have no fly-bys and satisfy $f_{\rm multi}=0$, so it may be used to discriminate between Newton and MOND in the future.

As identified by recent independent studies \citep[e.g.][]{Pittordis:2025}, the most serious systematic up to our present study was the potential presence of undetected multiple systems. Here we have very substantially reduced that possibility through a series of independent approaches such as the use of multi-epoch radial velocity observations, the consistency of Hipparcos and Gaia proper motion observations, the presence of consistent metallicities for the two components of a binary, and the use of Speckle imaging to rule out extra components. Note that \cite{Manchanda:2023} recently explored a range of follow-up observational techniques towards solving this issue, and identify Speckle interferometry as the leading technique towards guaranteeing pure binary samples.

 {In this work we are concerned with measuring $G$ at low accelerations based solely on wide binaries at sub-parsec scales. The gravitational anomaly we infer is a pure measurement with respect to Newtonian gravity, independent of any MOND or modified gravity models. In this context, it is interesting to note recent results challenging classical MOND models (AQUAL by \cite{Bekenstein:1984} and QUMOND by \cite{Milgrom:2010}) using solar-system tests, such as Cassini radio tracking data analyses of the quadrupole moment of the gravitational potential \citep{Park:2026} and numerical simulations of trans-Neptunian objects (TNOs) and comets \citep{Vokrouhlicky:2024} over giga year scales (but see also \cite{Brown:2023}). The above studies find no deviations from Newtonian gravity, results inconsistent with the MOND variants tested there. The quadrupole moment test probes second-order predictions of specific MOND models at relatively high accelerations and significantly smaller length scales. Thus, if their results are correct, they would provide useful constraints on gravity complementary to wide binary anomalies (which probe significantly lower accelerations and larger length scales), to narrow theoretical possibilities. From an observational point of view, two tests may constrain different aspects of gravity at distinct acceleration and length scales. We note that numerical simulations of TNOs and comets are quite complex and purely statistical while 3D modeling of wide binaries allows individually and directly ``seeing'' and quantifying anomalies. The continual exploration of complementary gravity tests at various acceleration and length scales is clearly important towards developing a more complete picture of gravity.}

\section{Summary, Conclusion, and outlook} \label{sec:conclusion}

We have collected a purest possible statistical sample of 36 solar neighborhood wide binaries having low internal accelerations ($\la 10^{-9}$~m\,s$^{-2}$) based on an unprecedented combination of various observational data including our crucial new observations of Speckle interferometric imaging and spectroscopic measurements of radial velocities. We have then carried out the most through and complete study of the internal dynamics of wide binaries to date, through the Bayesian 3D modeling methodology.

We started from assembling a new collection of 306 Gaia DR3 wide binaries with relatively precise radial velocities. Speckle interferometric imaging of 390 wide binaries of similar qualities shows that about 5\% can have resolvable faint companions. From the initial collection, we carefully selected a high-quality sample of 75 isolated wide binaries, based on limits such as distance $<150$~pc from the Sun, Gaia's $\texttt{ruwe}<1.25$, a narrow locus in the CM diagram, and relative radial velocity ($v_r$) error $<100$~m\,s$^{-1}$.

We then included in the final statistical clean sample only those wide binaries for which multi-epoch measurements of $v_r$ showed no detectable variability, in most cases over more than several years, and/or Speckle observations have not detected any resolved faint companion. Wide binaries in the sample have highest-quality 3D relative velocities: the scalar sky-plane relative velocity $v_p$ (between the pair) has an error smaller than 20~m\,s$^{-1}$ with a median of 7.5~m\,s$^{-1}$, and the relative radial velocity $v_r$ has an error smaller than 100~m\,s$^{-1}$ with a median of 47~m\,s$^{-1}$. The clean sample is further verified by a consistency check between Hipparcos and Gaia relative proper motions over the $\approx 25$-year period covered by the two observations, and the metallicity consistency between the component stars of the binaries.

The Bayesian inference method of gravity \citep{Chae:2025,Chae:2025b} using 3D relative velocities is applied to recover the effective value of the gravitational constant $G(= \gamma G_{N})$ in the low acceleration regime, under the assumption of elliptical orbits and angular momentum conservation. Despite the relatively small sample of 36 wide binaries, the 3D velocity modeling algorithm allows strong constraints on gravity. Our findings can be summarized as follows:
\begin{itemize}
    \item We find $\Gamma \equiv \log_{10}\sqrt{\gamma}=0.102_{-0.021}^{+0.028}$ for the whole clean sample (satisfying $g_{\rm N}<10^{-9.1}\,{\rm m}\,{\rm s}^{-2}$) and $\Gamma = 0.132_{-0.027}^{+0.035}$ for $g_{\rm N}<10^{-9.5}\,{\rm m}\,{\rm s}^{-2}$, a 5$\sigma$ falsification of the hypothesis that Newtonian gravity can be extrapolated to the low acceleration regime. This result cannot be significantly changed by a reasonable systematic variation of stellar masses or any other reasonable variation in sample or modeling inputs such as orbital eccentricity prior.
    \item We find that $\Gamma$ gradually increases from $g_{\rm N}=10^{-9}\,{\rm m}\,{\rm s}^{-2}$ to $10^{-11}\,{\rm m}\,{\rm s}^{-2}$ both in split bins and cumulative bins. This trend is qualitatively consistent with the QUMOND numerical prediction by \cite{Pf-A:2025}, but the amplitude of the trend of $\Gamma$ with $g_{\rm N}$ is higher than the QUMOND-predicted curve (Figure~\ref{fig:QUMOND}). 
    \item We find that our clean sample of 36 wide binaries contains two clear cases of $v_{\rm obs}/v_{\rm escN}> 1$ (Newtonian-unbound), one of which is a newly discovered system ({\#}3: the pair of TYC~259-236-1 and TYC~259-906-1) from this work, while the other ({\#}59: the pair of HD~189739 and HD~189760) was previously discovered by \cite{Saglia:2025}. Our sample also contains two ambiguous cases. Thus, the occurrence rate of Newtonian-unbound systems is 5 - 10\%. Because $v_{\rm obs}<1$~km\,s$^{-1}$ (i.e., they have essentially the same 3D velocities within the Milky Way with a small velocity gradient consistent with a relative motion under mutual gravity) in these systems, the probability that any of these systems is a chance association (fly-by) in 3D space is negligibly small. It will be interesting to see whether this occurrence rate persists in larger future samples.
    \item Without the two systems with $v_{\rm obs}/v_{\rm escN}> 1$, we find $\Gamma = 0.057_{-0.031}^{+0.035}$ for $g_{\rm N}<10^{-9.5}\,{\rm m}\,{\rm s}^{-2}$ and $\Gamma = 0.085_{-0.045}^{+0.059}$ for $g_{\rm N}<10^{-10.1}\,{\rm m}\,{\rm s}^{-2}$, which interestingly agree with the realistic QUMOND prediction by \cite{Pf-A:2025}.
\end{itemize}

Our $2\sigma$ range of the boost factor $1.32<\gamma<2.07$ can be consistent with some nonstandard theories of gravity including MOND gravity models \citep{Bekenstein:1984,Milgrom:2010}. However, the current precision is not sufficient to distinguish between such models, and a proper test of such models requires additional analyses based on mock data from numerical solutions of the nonlinear gravitational field equations (see Section~\ref{sec:meaning}). Ongoing high quality radial velocity and Speckle interferometry observational campaigns will result in larger samples to further refine the results presented here.

\begin{acknowledgments}
We thank Cezary Migaszewski for sharing the distributions of $v_{\rm obs}/v_{\rm escN}$ obtained from numerical solutions of MOND gravity, and Jan Pflamm-Altenburg for providing us with the distributions of $\gamma$ from the QUMOND solutions of wide binary orbits. We thank R. A. M. Cort\'{e}s for the assistance with the Hipparcos archive data. We thank Arthur Kosowsky, Cezary Migaszewski, and Jan Pflamm-Altenburg for discussions. We thank two anonymous referees for the constructive comments that led to significantly improved results and presentations. This work was supported by the National Research Foundation of Korea (NRF) grant funded by the Korea government (MSIT) (RS-2026-25492976). X.H. acknowledges financial assistance from SECIHTI SNII and UNAM DGAPA PAPIIT grant IN-102624. V.O. acknowledges financial assistance from UNAM DGAPA PAPIIT grant IN-114123. D.L. acknowledges support from Basic Science Research Program through the National Research Foundation of Korea funded by the Ministry of Education (RS-2025-25419201 and RS-2022-NR070872). Y.-W.L. acknowledges support from the NRF of Korea to the Center for Galaxy Evolution Research (RS-2022-NR070872, RS-2022-NR070525). This work has made use of data from the European Space Agency (ESA) mission Gaia ({https://www.cosmos.esa.int/gaia}), processed by the Gaia Data Processing and Analysis Consortium (DPAC, {https://www.cosmos.esa.int/web/gaia/dpac/consortium}). Funding for the DPAC has been provided by national institutions, in particular the institutions participating in the Gaia Multilateral Agreement. The LCO measurements of radial velocities of stars were carried out with telescopes at McDonald Observatory in the USA, South African Astronomical Observatory (SAAO) in South Africa, and WISE Observatory in Israel. The Speckle observations reported were acquired at the Observatorio Astron\'omico Nacional in the Sierra San Pedro M\'artir (OAN-SPM), Baja California, M\'exico. We thank the daytime and night support staff at the OAN-SPM for facilitating and helping obtain our observations. The MAROON-X observations were supported by K-GMT Science Program (PID: GN-2024B-Q-122) of Korea Astronomy and Space Science Institute (KASI).

\end{acknowledgments}

\bibliographystyle{aasjournalv7.bst}
\bibliography{ms.bib}{}

%\vspace{5mm}

\newpage

\appendix

\section{Description of LCO observations and data reduction of radial velocities} 
\label{sec:LCO}

A sample of 60 wide binaries were selected from the \cite{ElBadry:2021} catalog with the following requirements: (1) Both stars of a binary are brighter than $G=11$~mag and are in the absolute magnitude range $3\la M_G < 8$, (2) the sky-plane separation $s>3.5$~kau, and (3) the relative Gaia DR3 RV satisfies 
\begin{equation}
    |v_r|\equiv|{\rm RV}_A - {\rm RV}_B|<\sqrt{9(\sigma_A^2 +\sigma_B^2) + (\Delta V)^2} 
    \label{eq:RVthreshold}
\end{equation}
where $\Delta V = 0.9419\sqrt{M_{\rm tot}/s}\times 1.3\times 1.2$~km\,s$^{-1}$ for the binary total mass $M_{\rm tot}$ given in $M_\odot$ and the sky-plane separation $s$ given in kau. This threshold is slightly relaxed from that given by Equations (2) and (3) of \cite{Chae:2024a}. We note that we impose only a requirement on the relative RV because the \cite{ElBadry:2021} catalog was obtained with a requirement on the relative PM.

Spectroscopic observations were made from December 2024 to March 2025 using the fiber-fed high-resolution Las Cumbres Observatory (LCO; \citealt{2013PASP..125.1031B}) Network of Robotic Echelle Spectrographs (NRES; \citealt{2018SPIE10702E..6CS}) attached to telescopes at McDonald Observatory, South African Astronomical Observatory (SAAO), South Africa, and WISE Observatory, Israel. The spectrograph provides a spectral resolution of 53,000 and covers a wavelength range of 3800 $–$ 8600~$\rm{\AA}$. The exposure time of the observations was adjusted between 600 and 1800 seconds, depending on the brightness of the target. To prevent spectral line blending due to star rotation and maintain precision, the maximum exposure time was limited to 1800 seconds. The signal-to-noise ratio (SNR) in LCO/NRES is typically estimated per resolution element within the spectral order that includes the Mg b lines (5167 $–$ 5184~$\rm{\AA}$). The SNRs of the observed targets ranged from approximately 15 to 100. An SNR of at least 25 is recommended to obtain accurate radial velocity (RV) measurements; however, achieving this level is challenging for targets with V $>$ 10.

The data extracted from the LCO archive were bias and flat-field corrected images processed with the BANZAI pipeline \citep{2018SPIE10707E..0KM}. BANZAI-NRES is designed to handle all data from the NRES of the LCO network.  Stellar RVs in LCO/NRES are estimated by comparing the BLAZ-extracted spectrum with a corresponding ZERO file, derived from PHOENIX stellar models \citep{2013A&A...553A...6H}. Each target star is assigned a specific ZERO file to ensure consistency. The pipeline first determines an initial redshift by cross-correlating the BLAZ and ZERO spectra. It then interpolates the ZERO spectrum to this redshift and divides each spectral order into wavelength segments. The residual redshift for each segment is then estimated along with formal errors. Finally, multiple estimates of the mean redshift are computed using different averaging methods.

The BANZAI-NRES pipeline provides extracted and wavelength-corrected spectra. If the target is a star, the pipeline also delivers RV measurements and stellar classification parameters, such as effective temperature and surface gravity. The RV precision of the BANZAI-NRES pipeline has been demonstrated to be as good as 10~m s$^{-1}$ for bright (V $\approx$ 6) standard stars. For this study, we used the cross-correlation function (CCF) implemented in the pipeline, with approximate errors of 71~m\,s$^{-1}$ for the entire sample.
A summary of the observations can be found in Table~\ref{tab:LCO}.

\startlongtable
\begin{deluxetable}{lllrllcccl}
\tablecaption{\textbf{Summary of Wide Binaries Observed with LCO/NRES}\label{tab:LCO}}
\centerwidetable
\tabletypesize{\scriptsize}
\startdata
 ID & Name & Gaia DR3 identifier & Sep\tablenotemark{a}  & Mass\tablenotemark{b} & Spec Type\tablenotemark{c} & RV(DR3)\tablenotemark{d} & RV(LCO)\tablenotemark{e} & $v_r$(LCO)\tablenotemark{f} & site\tablenotemark{g} \\ 
   &     &      &  [kau] & [$M_{\odot}$] &   & [km\,s$^{-1}$]  &[km\,s$^{-1}$] & [km\,s$^{-1}$]  &      \\
 \hline 
1A & BD+65 14 & 528186492427483392 & 5.246 &  1.18  &        F6V  & $  7.967 \pm   0.261 $ & $     8.116 \pm      0.064 $ & $   -0.652 \pm   0.087 $ &     (1) \\ 
1B & BD+65 12 & 528187248343141376 &      &     1  &        G1V  & $   8.442 \pm   0.342 $ & $     8.768 \pm      0.059 $ &       &     (1)  \\ 
\hline 
2A & HD 236377 & 428811394564909568 & 5.063 &   1.2  &         F8  & $ -66.402 \pm   0.234 $ & $   -66.074 \pm      0.037 $ & $   -0.647 \pm   0.149 $ &     (1)  \\  
2B & BD+59 32B & 428811428924644352 &      & \textit{0.85}  &         G0  & $  -65.447 \pm   0.245 $ & $   -65.427 \pm      0.144 $ &       &     (1)  \\ 
\hline 
3A & BD-09 71 & 2426718786382002304 & 9.547 &  0.97  &             & $ -14.245 \pm   0.570 $ & $   -13.311 \pm      0.084 $ & $   -0.772 \pm   0.142 $ &     (1)  \\ 
3B & TYC 5262-512-1 & 2426718614583009280 &      &  0.99  &             & $ -13.280 \pm   0.487 $ & $   -12.539 \pm      0.115 $ &       &     (2) \\
\enddata
\tablenotetext{a}{2D separation: Sky-plane physical separation from \cite{ElBadry:2021}.}
\tablenotetext{b}{Mass: Stellar masses, along with their 1$\sigma$ uncertainties, were retrieved from the Gaia DR3 StarHorse catalog using ADQL queries. \textit{Mass:} The stellar masses estimate corresponds to the median (50th percentile) value from the Gaia EDR3 StarHorse catalog, following the photo-astrometric methodology described by \citet{EDR3:2022}.}
\tablenotetext{c}{Spectral Type: Spectral classifications adopted from SIMBAD.}
\tablenotetext{d}{RV DR3: Radial velocity from Gaia DR3.}
\tablenotetext{e}{RV LCO: Radial velocity measured with NRES at LCO.}
\tablenotetext{f}{$v_r(\equiv{\rm RV}_A-{\rm RV}_B)$ LCO: Radial velocity difference between the binary components, as measured with LCO/NRES.}
\tablenotetext{g}{Site: Observational site, listed in chronological order — (1) McDonald Observatory; (2) South African Astronomical Observatory (SAAO); (3) WISE Observatory, Israel.}
\end{deluxetable}

\subsection{Systems that were observed more than once with LCO}\label{sec:LCOdouble}

Eighteen of the 60 binary systems were observed more than once. Their respective properties are listed below. \\

6A (HD 8745) was observed twice at McDonald Observatory. RVs of $-7.955 \pm 0.040$~km\,s$^{-1}$ and $-7.846 \pm 0.037$~km\,s$^{-1}$ were measured on 2024 December 14 (BJD 2460659.703) and 2025 February 9 (BJD 2460716.58759), respectively. The mean RV is $-7.901 \pm 0.039$~km\,s$^{-1}$.
6B (BD+36 251) was observed twice at McDonald Observatory. RVs of $-7.545 \pm 0.031$~km\,s$^{-1}$ and $-7.411 \pm 0.056$~km\,s$^{-1}$ were measured on 2024 December 14 (BJD 2460659.764) and 2025 February 9 (BJD 2460716.6159), respectively. The mean RV is $-7.478 \pm 0.048$~km\,s$^{-1}$.\\

8A (HD 11584) was observed twice at the South African Astronomical Observatory (SAAO). RVs of $22.761 \pm 0.030$~km\,s$^{-1}$ and $22.784 \pm 0.044$~km\,s$^{-1}$ were measured on 2024 December 14 (BJD 2460659.437) and 2025 February 11 (BJD 2460718.29242), respectively. The mean RV is $22.773 \pm 0.038$~km\,s$^{-1}$.
8B (CD–50 524) was observed twice at the WISE Observatory. RVs of $22.398 \pm 0.056$~km\,s$^{-1}$ and $22.374 \pm 0.060$~km\,s$^{-1}$ were measured on 2024 December 18 (BJD 2460663.566) and 2025 February 17 (BJD 2460724.5223), respectively. The mean RV is $22.386 \pm 0.058$~km\,s$^{-1}$.\\

15A (HD 29356) was observed twice at the WISE Observatory. RVs of $38.938 \pm 0.050$~km\,s$^{-1}$ and $39.039 \pm 0.030$~km\,s$^{-1}$ were measured on 2025 January 8 (BJD 2460684.54458) and 2025 February 9 (BJD 2460716.5881), respectively. The mean RV is $38.989 \pm 0.042$~km\,s$^{-1}$.
15B (HD 29355) was observed twice at the WISE Observatory. RVs of $39.036 \pm 0.033$~km\,s$^{-1}$ and $39.093 \pm 0.043$~km\,s$^{-1}$ were measured on 2025 January 7 (BJD 2460683.71539) and 2025 February 7 (BJD 2460714.62182), respectively. The mean RV is $39.065 \pm 0.038$~km\,s$^{-1}$.\\

16A ($\zeta$ Dor) was observed twice at SAAO. RVs of $-0.966 \pm 0.103$~km\,s$^{-1}$ and $-0.985 \pm 0.100$~km\,s$^{-1}$ were measured on 2024 December 14 (BJD 2460659.584) and 2025 February 28 (BJD 2460735.37616), respectively. The mean RV is $-0.976 \pm 0.102$~km\,s$^{-1}$.
16B (CD$-$57 1079) was observed twice at the WISE Observatory. RVs of $-0.185 \pm 0.074$~km\,s$^{-1}$ and $-0.605 \pm 0.061$~km\,s$^{-1}$ were measured on 2024 December 18 (BJD 2460663.73559) and 2025 March 4 (BJD 2460739.55929), respectively. The mean RV is $-0.395 \pm 0.068$~km\,s$^{-1}$.\\

24A (HD 53566) was observed twice at SAAO. RVs of $-14.822 \pm 0.045$~km\,s$^{-1}$ and $-14.896 \pm 0.047$~km\,s$^{-1}$ were measured on 2025 January 10 (BJD 2460686.47896) and 2025 February 12 (BJD 2460719.37888), respectively. The mean RV is $-14.859 \pm 0.046$~km\,s$^{-1}$.
24B (TYC 170-2913-1) was observed twice at SAAO. RVs of $-14.510 \pm 0.024$~km\,s$^{-1}$ and $-14.519 \pm 0.049$~km\,s$^{-1}$ were measured on 2025 January 7 (BJD 2460683.4967) and 2025 February 14 (BJD 2460721.37678), respectively. The mean RV is $-14.515 \pm 0.039$~km\,s$^{-1}$.\\

27A (TYC 1364-1623-1) was observed twice, at SAAO and McDonald Observatory, respectively. RVs of $-27.039 \pm 0.026$~km\,s$^{-1}$ and $-26.946 \pm 0.047$~km\,s$^{-1}$ were measured on 2024 December 19 (BJD 2460664.47497) and 2025 February 16 (BJD 2460723.81325), respectively. The mean RV is $-26.993 \pm 0.038$~km\,s$^{-1}$.
27B (TYC 1364-1760-1) was observed twice at SAAO. RVs of $-26.961 \pm 0.039$~km\,s$^{-1}$ and $-26.916 \pm 0.049$~km\,s$^{-1}$ were measured on 2024 December 19 (BJD 2460664.47497) and 2025 February 14 (BJD 2460721.35215), respectively. The mean RV is $-26.939 \pm 0.044$~km\,s$^{-1}$.\\

29A (HD 72584) was observed twice, at McDonald Observatory and WISE Observatory, respectively. RVs of $12.749 \pm 0.043$~km\,s$^{-1}$ and $12.599 \pm 0.055$~km\,s$^{-1}$ were measured on 2024 December 14 (BJD 2460659.941) and 2025 March 3 (BJD 2460738.70737), respectively. The mean RV is $12.674 \pm 0.049$~km\,s$^{-1}$.
29B (TYC 4874-1253-1) was observed twice at WISE Observatory. RVs of $12.854 \pm 0.077$~km\,s$^{-1}$ and $12.885 \pm 0.146$~km\,s$^{-1}$ were measured on 2024 December 15 (BJD 2460660.700) and 2025 March 2 (BJD 2460737.69814), respectively. The mean RV is $12.870 \pm 0.117$~km\,s$^{-1}$.\\

30A (HD 78796) was observed twice at WISE Observatory and SAAO, respectively. RVs of $-3.313 \pm 0.047$~km\,s$^{-1}$ and $-3.316 \pm 0.037$~km\,s$^{-1}$ were measured on 2024 December 14 (BJD 2460659.772) and 2025 February 11 (BJD 2460718.34031), respectively. The mean RV is $-3.315 \pm 0.042$~km\,s$^{-1}$.
30B (CD-23 8096) was observed twice at McDonald Observatory and WISE Observatory, respectively. RVs of $-3.195 \pm 0.057$~km\,s$^{-1}$ and $-3.195 \pm 0.079$~km\,s$^{-1}$ were measured on 2024 December 18 (BJD 2460663.93401) and 2025 February 11 (BJD 2460718.71122), respectively. The mean RV is $-3.195 \pm 0.069$~km\,s$^{-1}$.\\

31A (HD 81268) was observed twice at McDonald Observatory and WISE Observatory, respectively. RVs of $14.892 \pm 0.026$~km\,s$^{-1}$ and $14.815 \pm 0.021$~km\,s$^{-1}$ were measured on 2024 December 16 (BJD 2460661.891) and 2025 February 13 (BJD 2460720.64292), respectively. The mean RV is $14.854 \pm 0.024$~km\,s$^{-1}$.
31B (BD-08 2665) was observed twice at McDonald Observatory and SAAO, respectively. RVs of $15.214 \pm 0.036$~km\,s$^{-1}$ and $15.147 \pm 0.031$~km\,s$^{-1}$ were measured on 2024 December 16 (BJD 2460661.908) and 2025 February 15 (BJD 2460722.53986), respectively. The mean RV is $15.181 \pm 0.034$~km\,s$^{-1}$.\\

32A (HD 85137) was observed twice at SAAO. RVs of $3.960 \pm 0.048$~km\,s$^{-1}$ and $3.874 \pm 0.048$~km\,s$^{-1}$ were measured on 2024 December 13 (BJD 2460658.486) and 2025 February 9 (BJD 2460716.33343), respectively. The mean RV is $3.917 \pm 0.048$~km\,s$^{-1}$.
32B (CD-57 2838) was observed twice at WISE Observatory. RVs of $4.353 \pm 0.081$~km\,s$^{-1}$ and $4.155 \pm 0.060$~km\,s$^{-1}$ were measured on 2024 December 13 (BJD 2460658.715) and 2025 February 9 (BJD 2460716.70867), respectively. The mean RV is $4.254 \pm 0.071$~km\,s$^{-1}$.\\

33A (HD 88418) was observed twice at SAAO. RVs of $-42.705 \pm 0.096$~km\,s$^{-1}$ and $-42.861 \pm 0.072$~km\,s$^{-1}$ were measured on 2025 February 15 (BJD 2460722.37206) and 2025 March 4 (BJD 2460739.43796), respectively. The mean RV is $-42.783 \pm 0.085$~km\,s$^{-1}$.
33B (TYC 251-458-1) was observed twice at McDonald Observatory and WISE Observatory, respectively. RVs of $-42.430 \pm 0.081$~km\,s$^{-1}$ and $-42.458 \pm 0.059$~km\,s$^{-1}$ were measured on 2025 February 15 (BJD 2460722.84552) and 2025 March 3 (BJD 2460738.78113), respectively. The mean RV is $-42.444 \pm 0.071$~km\,s$^{-1}$.\\

35A (HD 92677) was observed twice at SAAO. RVs of $1.398 \pm 0.030$~km\,s$^{-1}$ and $1.418 \pm 0.056$~km\,s$^{-1}$ were measured on 2024 December 13 (BJD 2460658.572) and 2025 February 10 (BJD 2460717.46585), respectively. The mean RV is $1.408 \pm 0.045$~km\,s$^{-1}$.
35B (HD 92652) was observed twice at WISE Observatory. RVs of $1.030 \pm 0.047$~km\,s$^{-1}$ and $1.019 \pm 0.066$~km\,s$^{-1}$ were measured on 2024 December 13 (BJD 2460658.842) and 2025 February 9 (BJD 2460716.72902), respectively. The mean RV is $1.025 \pm 0.057$~km\,s$^{-1}$.\\

38A (HD 101574) was observed twice at McDonald Observatory and SAAO, respectively. RVs of $1.665 \pm 0.045$~km\,s$^{-1}$ and $-0.013 \pm 0.073$~km\,s$^{-1}$ were measured on 2024 December 15 (BJD 2460734.79018) and 2025 February 10 (BJD 2460717.43468), respectively. The mean RV is $0.826 \pm 0.061$~km\,s$^{-1}$.
38B (BD$-$01 2557) was observed twice at WISE Observatory. RVs of $0.573 \pm 0.047$~km\,s$^{-1}$ and $0.559 \pm 0.090$~km\,s$^{-1}$ were measured on 2024 December 15 (BJD 2460660.840) and 2025 February 10 (BJD 2460717.67514), respectively. The mean RV is $0.566 \pm 0.072$~km\,s$^{-1}$.\\

41A (HD 105350) was observed twice at WISE Observatory. RVs of $29.135 \pm 0.074$~km\,s$^{-1}$ and $29.000 \pm 0.045$~km\,s$^{-1}$ were measured on 2024 December 14 (BJD 2460659.794) and 2025 March 3 (BJD 2460738.86556), respectively. The mean RV is $29.068 \pm 0.061$~km\,s$^{-1}$.
41B (TYC 7763-590-1) was observed twice at SAAO. RVs of $29.453 \pm 0.107$~km\,s$^{-1}$ and $29.182 \pm 0.069$~km\,s$^{-1}$ were measured on 2024 December 24 (BJD 2460669.51094) and 2025 March 4 (BJD 2460739.56682), respectively. The mean RV is $29.318 \pm 0.090$~km\,s$^{-1}$.\\

42A (HD 107434) was observed twice at SAAO. RVs of $-9.966 \pm 0.084$~km\,s$^{-1}$ and $-10.169 \pm 0.076$~km\,s$^{-1}$ were measured on 2024 December 13 (BJD 2460658.555) and 2025 February 28 (BJD 2460735.54704), respectively. The mean RV is $-10.067 \pm 0.080$~km\,s$^{-1}$.
42B (CD$-$37 7822) was observed twice at WISE Observatory. RVs of $-9.649 \pm 0.068$~km\,s$^{-1}$ and $-9.348 \pm 0.077$~km\,s$^{-1}$ were measured on 2024 December 12 (BJD 2460657.841) and 2025 February 28 (BJD 2460735.65839), respectively. The mean RV is $-9.499 \pm 0.073$~km\,s$^{-1}$.\\

44A (HD 123033) was observed twice at McDonald Observatory and WISE Observatory. RVs of $-19.677 \pm 0.069$~km\,s$^{-1}$ and $-20.018 \pm 0.064$~km\,s$^{-1}$ were measured on 2025 February 17 (BJD 2460724.83569) and 2025 March 3 (BJD 2460738.8245), respectively. The mean RV is $-19.847 \pm 0.066$~km\,s$^{-1}$.
44B (BD+26 2522) was observed twice at McDonald Observatory and WISE Observatory. RVs of $-19.314 \pm 0.042$~km\,s$^{-1}$ and $-19.117 \pm 0.054$~km\,s$^{-1}$ were measured on 2025 February 8 (BJD 2460715.83734) and 2025 March 5 (BJD 2460740.82101), respectively. The mean RV is $-19.215 \pm 0.048$~km\,s$^{-1}$.\\

47A (BD+08 2889) was observed twice at SAAO and WISE Observatory. RVs of $-17.980 \pm 0.025$~km\,s$^{-1}$ and $-17.948 \pm 0.033$~km\,s$^{-1}$ were measured on 2025 February 11 (BJD 2460718.58422) and 2025 February 28 (BJD 2460735.77632), respectively. The mean RV is $-17.964 \pm 0.029$~km\,s$^{-1}$.
47B (BD+08 2887) was observed twice at WISE Observatory. RVs of $-17.954 \pm 0.085$~km\,s$^{-1}$ and $-17.921 \pm 0.055$~km\,s$^{-1}$ were measured on 2025 February 11 (BJD 2460718.83524) and 2025 February 26 (BJD 2460733.78058), respectively. The mean RV is $-17.937 \pm 0.071$~km\,s$^{-1}$.\\

48A (HD 129171) was observed twice at McDonald Observatory. RVs of $-11.419 \pm 0.022$~km\,s$^{-1}$ and $-11.393 \pm 0.030$~km\,s$^{-1}$ were measured on 2024 December 23 (BJD 2460668.97809) and 2025 February 5 (BJD 2460712.89457), respectively. The mean RV is $-11.406 \pm 0.026$~km\,s$^{-1}$.
48B (HD 129209) was observed twice at McDonald Observatory. RVs of $-11.129 \pm 0.032$~km\,s$^{-1}$ and $-11.118 \pm 0.032$~km\,s$^{-1}$ were measured on 2024 December 14 (BJD 2460660.018) and 2025 February 5 (BJD 2460712.99954), respectively. The mean RV is $-11.123 \pm 0.032$~km\,s$^{-1}$.\\

\section{Description of MAROON-X observations and data reduction of radial velocities} \label{sec:MAROONX}

A sample of 6 wide binaries were selected from the \cite{ElBadry:2021} catalog mainly with the requirement of Equation~(\ref{eq:RVthreshold}) for measurements of RVs with MAROON-X. MAROON-X is a high-resolution fiber-fed optical echelle spectrograph on the Gemini North telescope, designed for detecting Earth-size planets \citep{MAROONX}. It delivers a resolving power of R $\sim$ 80,000 over 5000 $-$ 9200~~$\rm{\AA}$. Our observations were obtained in 2024B semester under the K-GMT Science Program (Program ID: GN-2024B-Q-122). Exposure times ranged from 90 to 1200 seconds, set to achieve a peak SNR of roughly 100. Data were reduced with the MAROON-X team's custom pipeline. We then extracted continuous 1D spectra and applied barycentric corrections. Relative radial velocities between the two components of each wide binary were measured via the CCF using the IRAF task \texttt{fxcor}. A summary of the observations is provided in Table~\ref{tab:MAROONX}.

\startlongtable
\begin{deluxetable}{lllccccc}
\tablecaption{\textbf{MAROON-X measurements of relative radial velocities in 6 wide binaries} \label{tab:MAROONX}}
\centerwidetable
\tabletypesize{\footnotesize}
\startdata
 Name & Gaia DR3 identifier & Sep\tablenotemark{a} & Mass\tablenotemark{b} & {\tt ruwe}  & Spec Type & RV(DR3) & $v_r$(MAROON-X)\tablenotemark{c} \\
        &            & [kau] & [$M_\odot$]  &    &        & [km\,s$^{-1}$]   & [km\,s$^{-1}$]   \\
   \hline
 TYC 4264-485-1 & 2201661297490051968 & 8.370 & 1.02 &  0.963 &  & $-12.734 \pm 0.251$ & $-0.0182 \pm 0.006$ \\
 TYC 4264-252-1 & 2201661091331626752 &       & 0.999 & 0.860 &  & $-12.387 \pm 0.268$ &               \\
\hline
 TYC 2700-274-1 & 1871558941576158464 & 7.750 & 0.791 & 0.877 &  & $-22.921 \pm 0.291$ & $-0.0239 \pm 0.009$ \\
 TYC 2700-210-1 & 1871559697490418816 &       & 0.771 & 0.992 &  & $-23.347 \pm 0.326$ &               \\ 
\hline
 TYC 2743-1157-1 & 1902676117063910016 & 6.271 & 0.899 & 1.282 &  & $3.745 \pm 0.365$ & $-0.9863 \pm 0.005$ \\
 LSPM J2231+3454 & 1902679033343158528 &       & 0.879 & 1.271 &  & $3.330 \pm 0.390$ &               \\ 
\hline
 TYC 3640-273-1 & 1942384773344557184 & 9.209 & 0.84 & 0.956 &  & $-8.452 \pm 0.202$ & $0.2663 \pm 0.084$ \\
 PM J23201+4819 & 1942384872124424832 &       & 0.565 & 1.116 &  & $-9.100 \pm 0.238$ &              \\
\hline
 TYC 1101-86-1 & 1762461893163118464 & 8.828 & 0.941 & 0.931 &  & $-75.799 \pm 0.252$ & $0.2199 \pm 0.004$ \\
 TYC 1101-87-1 & 1762461309047562368 &       & 0.939 & 1.156 &  & $-76.334 \pm 0.238$ &               \\
\hline
 BD+00 4500 & 4230699363889120128 & 5.247 & 1.091 & 1.124 & G0 & $-38.226 \pm 0.165$ & $-0.2781 \pm 0.022$ \\
 BD+00 4497 & 4230699329529382400 &       & 0.878 & 0.949 & K8 & $-38.244 \pm 0.514$ &               \\
\enddata
\tablenotetext{a}{2D separation: Sky-plane physical separation from \cite{ElBadry:2021}.}
\tablenotetext{b}{Stellar mass based on the mass-magnitude relation derived by \cite{Chae:2023}.}
\tablenotetext{c}{Relative radial velocity $v_r(\equiv {\rm RV}_A - {\rm RV}_B)$.}
\end{deluxetable}

\section{Selection of wide binaries from the Scarpa et al.\ sample} 
\label{sec:Scarpa}

Here we describe our selection of wide binary candidates from the \cite{Scarpa:2017} sample of 58 pairs that have precise RVs measured in January and December 2013 with the fiber-fed FIES Echelle spectrograph at the 2.5m Nordic Optical Telescope, the Roque de Los Muchachos observatory in the Canary islands. Although this sample is far from a well-defined sample of wide binaries and contains many unbound pairs, we consider it because it contains true wide binaries with very precise RVs. 

Since we eventually need only uncontaminated pure binaries, we remove obviously unbound, contaminated, or problematic cases by individually examining the pairs. Specifically, we exclude any pair in which (1) Gaia's \texttt{ruwe} has unacceptably large value(s) (e.g., $>1.5$), (2) RV(s) show(s) unacceptably large variation(s) in time or with respect to Gaia DR3, (3) the scalar relative RV ($|v_r|$) is too large (e.g., $|v_r|>2$~km\,s$^{-1}$) indicating an obviously gravitationally-unbound system, or (4) the sky-plane 2D physical separation is too large $>0.22$~pc or $>45$~kau, because any systems beyond the limit are more likely to be gravitationally unbound and require much more precise velocities. We are left with only 24 systems, which are listed in Table~\ref{tab:Scarpa}. All 24 systems are included in the raw sample of 306 systems shown in Figure~\ref{fig:vr_err}. 

Although the majority of the \cite{Scarpa:2017} sample has already been excluded, it turns out that the remaining sample is relatively more contaminated than our other samples such as LCO and MAROON-X described in Appendix~\ref{sec:LCO} and \ref{sec:MAROONX}, which are selected from previous statistical samples used by one of us (e.g., \citealt{Chae:2024a}). This may be because the above exclusion criterion for $|v_r|$ is not tight enough, so that boosted velocities due to unseen perturbers may be present. Indeed, as shown in Section~\ref{sec:Hip_Gaia}, the Hipparcos-Gaia relative PM comparison identified three cases of unacceptably large variation between the two observations, and they all are from the \cite{Scarpa:2017} sample listed in Table~\ref{tab:Scarpa}. 

The relatively high fraction of contaminated cases can be seen by the scalar 3D velocity $v_{\rm obs}$ listed in the penultimate column of Table~\ref{tab:Scarpa}. There are 9 systems with $v_{\rm obs}>1$~km\,s$^{-1}$ including 3 systems with $v_{\rm obs}> 1.5$~km\,s$^{-1}$. Three of them are the above contaminated cases identified by the Hipparcos-Gaia test: HIP54692/HIP54681 ($v_{\rm obs}=1.690$), HIP65602/HIP65574 (1.053), and HIP101916/HIP101932 (1.632). In contrast, the clean statistical sample does not contain any case with $v_{\rm obs}>1$~km\,s$^{-1}$ (see Table~\ref{tab:WB_Newtonunbound} for systems with relatively large values of $v_{\rm obs}$ among the clean sample). 

The systems in Table~\ref{tab:Scarpa} can be further tested with the relative velocity threshold of Equation~(\ref{eq:RVthreshold}) for $v_r$ and $v_p$. The velocity $v_p$ is also considered because the relative PMs were not used in selecting the systems. Three systems fail this test, as indicated in the last column of Table~\ref{tab:Scarpa}. These systems are excluded from gravity tests in this work. These systems exactly match the three systems with $v_{\rm obs}> 1.5$~km\,s$^{-1}$, and two of them are the contaminated cases identified by the Hipparcos-Gaia test. Thus, 4 out of the 9 systems with $v_{\rm obs}>1$~km\,s$^{-1}$ are excluded by the velocity threshold and the Hipparcos-Gaia test. It is unclear whether the remaining 5 cases with $1<v_{\rm obs}<1.5$~km\,s$^{-1}$ from the \cite{Scarpa:2017} sample are also contaminated. It will be interesting to see what future data will reveal about these systems.

Despite the relatively high fraction of contaminated cases, the \cite{Scarpa:2017} sample is useful in this study because it provides independent measurements at different epochs for true pure binaries. Indeed, for 7 systems from the clean sample the \cite{Scarpa:2017} values confirm the stability of $v_r$ in conjunction with other measurements over more than several years (see Table~\ref{tab:WB_overlap}). Also, as demonstrated above, the  sample inadvertently provided useful negative examples of contaminated cases through the Hipparcos-Gaia test and the velocity threshold. The \cite{Scarpa:2017} sample seems to provide a useful testbed for kinematically contaminated cases. 

\startlongtable
\begin{deluxetable}{lllccccccc}
\tablecaption{\textbf{Wide binaries selected from \cite{Scarpa:2017}} \label{tab:Scarpa}}
\centerwidetable
\tabletypesize{\scriptsize}
\startdata
Name & Gaia DR3 identifier & Sep\tablenotemark{a}  & Mass\tablenotemark{b} & RV(DR3)\tablenotemark{c} & RV(Scarpa)\tablenotemark{d} & $v_r$\tablenotemark{e} & $v_p$\tablenotemark{f} & $v_{\rm obs}$\tablenotemark{g} & Test\tablenotemark{h} \\ 
     &     &   [kau] & [$M_{\odot}$] & [km\,s$^{-1}$]  &[km\,s$^{-1}$] & [km\,s$^{-1}$]  &  [km\,s$^{-1}$]  &  [km\,s$^{-1}$] &  \\
\hline
HIP11137 & 76300510625993344 & 2.090 & 0.998 & $27.138\pm0.175$ & $27.129\pm0.014$ & $-0.157\pm0.021$ & $0.535\pm0.007$ & $0.558\pm0.009$ & Pass \\ 
HIP11134 & 76300476266255488 &       & 0.925 & $27.010\pm0.139$ & $27.286\pm0.016$ &       &       &    & \\ 
\hline 
HIP15304 & 10584899657116672 & 7.340 & 1.162 & $31.392\pm0.133$ & $31.860\pm0.019$ & $-0.856\pm0.027$ & $0.547\pm0.009$ & $1.016\pm0.023$ & Pass \\ 
HIP15310 & 10608573516849536 &       & 1.092 & $32.068\pm0.143$ & $32.716\pm0.019$ &       &       &    & \\ 
\hline 
HIP15527 & 5060104351007433472 & 9.082 & 0.960 & $39.849\pm0.118$ & $40.287\pm0.019$ & $-0.377\pm0.024$ & $0.225\pm0.004$ & $0.439\pm0.020$ & Pass \\ 
HIP15526 & 5060105897197110144 &       & 0.880 & $40.270\pm0.120$ & $40.664\pm0.014$ &       &       &    & \\ 
\hline 
HIP19859 & 3285218186904332288 & 1.417 & 1.070 & $-7.277\pm0.137$ & $-6.858\pm0.012$ & $0.572\pm0.020$ & $0.966\pm0.005$ & $1.122\pm0.011$ & Pass \\ 
HIP19855 & 3285218255623808640 &       & 0.960 & $-8.010\pm0.130$ & $-7.430\pm0.016$ &       &       &    & \\ 
\hline 
HIP21537 & 3230677565443833088 & 4.972 & 1.160 & $38.397\pm0.135$ & $38.947\pm0.016$ & $-0.135\pm0.022$ & $0.265\pm0.013$ & $0.298\pm0.015$ & Pass \\ 
HIP21534 & 3230677874682668672 &       & 1.150 & $38.724\pm0.133$ & $39.082\pm0.015$ &       &       &    & \\ 
\hline 
HIP22611 & 4873223829966552192 & 5.959 & 1.503 & $45.755\pm0.120$ & $46.203\pm0.015$ & $-0.241\pm0.022$ & $0.661\pm0.006$ & $0.704\pm0.009$ & Pass \\ 
HIP22604 & 4873226853623529856 &       & 0.997 & $45.859\pm0.165$ & $46.444\pm0.016$ &       &       &    & \\ 
\hline 
HIP25278 & 3400292798990117888 & 10.311 & 1.135 & $37.701\pm0.138$ & $38.350\pm0.047$ & $-0.253\pm0.052$ & $0.106\pm0.006$ & $0.274\pm0.048$ & Pass \\ 
HIP25220 & 3394298532176344960 &       & 0.738 & $38.032\pm0.124$ & $38.603\pm0.023$ &       &       &    & \\ 
\hline 
HIP33705 & 5607190344506642432 & 12.331 & 1.150 & $16.441\pm0.125$ & $16.700\pm0.027$ & $-0.441\pm0.032$ & $0.100\pm0.005$ & $0.452\pm0.031$ & Pass \\ 
HIP33691 & 5607189485513198208 &       & 0.880 & $16.726\pm0.141$ & $17.141\pm0.017$ &       &       &    & \\ 
\hline 
HIP34426 & 3359808231100381312 & 8.189 & 1.106 & $-11.840\pm0.141$ & $-11.544\pm0.013$ & $0.791\pm0.017$ & $1.612\pm0.007$ & $1.796\pm0.010$ & Fail \\ 
HIP34407 & 3359820016490648576 &       & 1.086 & $-12.835\pm0.153$ & $-12.335\pm0.011$ &       &       &    & \\ 
\hline 
HIP39457 & 5595858262287843840 & 4.421 & 1.200 & $26.293\pm0.139$ & $26.807\pm0.015$ & $-0.741\pm0.022$ & $0.791\pm0.008$ & $1.084\pm0.016$ & Pass \\ 
HIP39452 & 5595858159201264768 &       & 1.016 & N/A & $27.548\pm0.016$ &       &       &    & \\ 
\hline 
HIP44858 & 692119656035933568 & 2.524 & 1.014 & $29.976\pm0.139$ & $30.473\pm0.010$ & $-0.292\pm0.015$ & $0.565\pm0.007$ & $0.636\pm0.009$ & Pass \\ 
HIP44864 & 692120029700390912 &       & 1.013 & $30.380\pm0.129$ & $30.765\pm0.011$ &       &       &    & \\ 
\hline 
HIP45836 & 1019361632454363904 & 6.732 & 1.206 & $-7.895\pm0.128$ & $-7.761\pm0.012$ & $-1.337\pm0.016$ & $0.673\pm0.004$ & $1.497\pm0.014$ & Pass \\ 
HIP45859 & 1019174509319377536 &       & 0.919 & $-6.780\pm0.127$ & $-6.424\pm0.010$ &       &       &    & \\ 
\hline 
HIP52787 & 3550081879381593728 & 7.870 & 0.820 & $23.759\pm0.148$ & $24.212\pm0.019$ & $-0.410\pm0.029$ & $0.228\pm0.005$ & $0.469\pm0.026$ & Pass \\ 
HIP52776 & 3550084490721711872 &       & 0.700 & $24.451\pm0.180$ & $24.622\pm0.022$ &       &       &    & \\ 
\hline 
HIP54692 & 777967084390189696 & 6.185 & 1.194 & $11.222\pm0.132$ & $11.696\pm0.011$ & $1.353\pm0.015$ & $1.012\pm0.008$ & $1.690\pm0.013$ & Fail \\ 
HIP54681 & 777967702865481344 &       & 0.996 & $9.949\pm0.152$ & $10.343\pm0.010$ &       &       &    & \\ 
\hline 
HIP58067 & 3975129194660883328 & 2.893 & 0.956 & $5.965\pm0.198$ & $6.461\pm0.015$ & $0.141\pm0.021$ & $0.202\pm0.006$ & $0.246\pm0.013$ & Pass \\ 
HIP58073 & 3975223065466473216 &       & 0.927 & $5.801\pm0.140$ & $6.320\pm0.014$ &       &       &    & \\ 
\hline 
HIP64057 & 3945118265299248128 & 1.458 & 0.948 & $-1.565\pm0.130$ & $-1.247\pm0.014$ & $0.017\pm0.019$ & $0.361\pm0.009$ & $0.362\pm0.009$ & Pass \\ 
HIP64059 & 3945118643256370688 &       & 0.896 & $-1.673\pm0.147$ & $-1.264\pm0.013$ &       &       &    & \\ 
\hline 
HIP65602 & 6193279279612173952 & 9.539 & 0.816 & $-13.492\pm0.144$ & $-12.884\pm0.018$ & $-1.014\pm0.025$ & $0.282\pm0.004$ & $1.053\pm0.025$ & Pass \\ 
HIP65574 & 6193280031230266752 &       & 0.811 & $-12.211\pm0.128$ & $-11.870\pm0.018$ &       &       &    & \\ 
\hline 
HIP71726 & 1282815063829295360 & 16.772 & 1.030 & $-11.925\pm0.121$ & $-11.468\pm0.015$ & $-0.170\pm0.021$ & $0.142\pm0.008$ & $0.221\pm0.017$ & Pass \\ 
HIP71737 & 1282817022334383232 &       & 1.000 & $-11.598\pm0.124$ & $-11.298\pm0.015$ &       &       &    & \\ 
\hline 
HIP74442 & 1274568245587206016 & 2.293 & 1.152 & $-60.315\pm0.141$ & $-60.207\pm0.016$ & $-0.962\pm0.023$ & $0.330\pm0.007$ & $1.017\pm0.022$ & Pass \\ 
HIP74439 & 1274562370068531712 &       & 0.975 & $-59.383\pm0.174$ & $-59.245\pm0.016$ &       &       &    & \\ 
\hline 
HIP85620 & 1440518669436791296 & 8.824 & 1.094 & $-34.155\pm0.127$ & $-33.736\pm0.027$ & $-0.442\pm0.034$ & $0.139\pm0.006$ & $0.463\pm0.033$ & Pass \\ 
HIP85575 & 1440425863783337856 &       & 0.970 & $-33.661\pm0.132$ & $-33.294\pm0.021$ &       &       &    & \\ 
\hline 
HIP99729 & 4249652990144051840 & 2.736 & 1.060 & $-0.464\pm0.129$ & $-0.011\pm0.018$ & $0.061\pm0.026$ & $0.834\pm0.010$ & $0.836\pm0.010$ & Pass \\ 
HIP99727 & 4249652783985617920 &       & 1.050 & $-0.388\pm0.138$ & $-0.072\pm0.019$ &       &       &    & \\ 
\hline 
HIP101082 & 2298101352139398144 & 13.722 & 1.949 & $-14.467\pm0.118$ & $-14.069\pm0.014$ & $-0.322\pm0.020$ & $0.127\pm0.010$ & $0.346\pm0.019$ & Pass \\ 
HIP101166 & 2298101901895214720 &       & 1.042 & $-14.117\pm0.149$ & $-13.747\pm0.014$ &       &       &    & \\ 
\hline 
HIP101916 & 1754191435419155456 & 6.401 & 1.566 & $-54.097\pm0.124$ & $-53.457\pm0.015$ & $-0.339\pm0.022$ & $1.596\pm0.010$ & $1.632\pm0.011$ & Fail \\ 
HIP101932 & 1754191229260708736 &       & 0.841 & $-53.510\pm0.136$ & $-53.118\pm0.016$ &       &       &    & \\ 
\hline 
HIP118254 & 2882262637207289216 & 4.703 & 1.041 & $29.940\pm0.128$ & $30.368\pm0.016$ & $0.747\pm0.021$ & $0.280\pm0.004$ & $0.798\pm0.020$ & Pass \\ 
HIP118251 & 2882262529831237120 &       & 0.973 & $29.271\pm0.128$ & $29.621\pm0.014$ &       &       &    & \\ 
\enddata
\tablenotetext{a}{2D separation: Sky-plane physical separation from \cite{ElBadry:2021}.}
\tablenotetext{b}{Stellar mass based on the mass-magnitude relation derived by \cite{Chae:2023}.}
\tablenotetext{c}{Radial velocity from Gaia DR3.}
\tablenotetext{d}{Radial velocity from \cite{Scarpa:2017}.}
\tablenotetext{e}{Relative radial velocity $v_r(\equiv {\rm RV}_A - {\rm RV}_B)$ from \cite{Scarpa:2017}.}
\tablenotetext{f}{Scalar sky-plane 2D velocity $v_{p} (\equiv\sqrt{v_{x^\prime}^2 + v_{y^\prime}^2}$) from Gaia DR3.}
\tablenotetext{g}{Scalar 3D velocity $v_{\rm obs} (\equiv\sqrt{v_p^2 + v_r^2}$).}
\tablenotetext{h}{Test with $v_r$ and $v_p$ based on the threshold given by Equation~(\ref{eq:RVthreshold}).}
\end{deluxetable}

\section{Selection of wide binaries with radial velocities from SDSS4 APOGEE} 
\label{sec:APOGEE}

The SDSS4 APOGEE survey measured RVs for a large number of stars using high-resolution, multi-object, near-infrared spectroscopy. APOGEE RVs are available on the SDSS DR17 website https://www.sdss4.org/dr17/irspec/use-radial-velocities/. The precision reported for an individual RV is often better than 100m\,s$^{-1}$ and in some cases better than 50m\,s$^{-1}$. Thus, APOGEE RVs potentially provide precise relative RVs for a number of wide binaries. The file {\tt allStar-dr17-synspec-rev1.fits} downloadable from the website provides RVs for 733901 stars, 717925 of which have Gaia DR3 identifications. 

We search for wide binaries from the 717925 stars with Gaia DR3 identifications. Wide binary candidates are selected from the \cite{ElBadry:2021} catalog. Once a candidate binary with APOGEE RVs for both stars is found, we check whether the radial velocity difference between the pair can be consistent with a gravitationally bound system. We use the threshold given by Equation~(\ref{eq:RVthreshold}). %$|v_r|\equiv|{\rm RV}_A - {\rm RV}_B|<\sqrt{9(\sigma_A^2 +\sigma_B^2) + (\Delta V)^2}$ where $\Delta V = 0.9419\sqrt{M_{\rm tot}/s}\times 1.3\times 1.2$~km\,s$^{-1}$ for the binary total mass $M_{\rm tot}$ given in $M_\odot$ and the sky-projected separation $s$ given in kau. This threshold is slightly relaxed from that given by Equations~(2) and (3) of \cite{Chae:2024a}. 
We have found 208 wide binaries within 300~pc from the Sun, which are listed in Table~\ref{tab:APOGEE}. For 15 of them, RVs for both components were measured multiple times at multiple epochs. For these systems, Table~\ref{tab:APOGEE} gives only the error-weighted mean and its error, while Table~\ref{tab:APOGEEmulti} gives all the measured values.

Each measurement of RV for a star is based on \texttt{NVISITS} ($\ge 1$) ``visit(s)'' of the star. The given value of RV refers to a signal-to-noise ratio (SNR)-weighted average if \texttt{NVISITS} $>1$ stored in the APOGEE parameter \texttt{VHELIO}\textunderscore\texttt{AVG}. The given nominal error (Err) refers to the SNR-weighted uncertainty stored in the APOGEE parameter \texttt{VERR}. When multiple observations (each of which may have multiple visits) were made at multiple epochs, the given value of RV and its error refer to the error-weighted mean of RVs and its uncertainty. For each value of RV, two additional errors are given in the table: `Disp' and `Scatt'. Disp simply refers to the standard deviation of RVs from multiple observations at multiple epochs. It is set to zero when only a single-epoch observation (regardless of \texttt{NVISITS}) was made. Scatt refers to the scatter in multiple visits in an observation stored in the APOGEE parameter \texttt{VSCATTER}. When multiple observations were made, the given value of Scatt refers to the maximum value. Scatt is zero when there was only a single observation with a single visit. Disp is provided as a qualitative check of the variability of RV when multiple observations are available while Scatt is used to select reliable RVs. Because Scatt is not available for all RVs, the nominal error `Err' is used in estimating the nominal uncertainty of the relative RV $v_r$ between the two stars. Note that Err is either smaller or larger than Scatt as can be seen in Table~\ref{tab:APOGEE}. The majority of wide binaries have $\sigma_{v_r}<0.1$~km\,s$^{-1}$ and 195 wide binaries with $\sigma_{v_r}<0.35$~km\,s$^{-1}$ are included in the raw sample of the main part shown in Figure~\ref{fig:CM}.

\startlongtable
\begin{deluxetable}{lccccccccc}
\tablecaption{\textbf{Summary of wide binaries with radial velocities from the SDSS DR17 APOGEE database}\label{tab:APOGEE}}
\centerwidetable
\tabletypesize{\scriptsize}
\startdata
identifier & $s$\tablenotemark{a} & Mass\tablenotemark{b} & RV\tablenotemark{c} &  Err\tablenotemark{d}  &  Disp\tablenotemark{e} & Scatt\tablenotemark{f} & $v_r$\tablenotemark{g}& $\sigma_{v_r}$ & multi-epoch\tablenotemark{h} \\
   & [kau] & [$M_{\odot}$] & [km\,s$^{-1}$] & [km\,s$^{-1}$] & [km\,s$^{-1}$] & [km\,s$^{-1}$] & [km\,s$^{-1}$] & [km\,s$^{-1}$] &     \\
  \hline
Gaia DR3 5961072491289480960  &  4.449  &  0.537  &  -1.252  &  0.054  &  0.000  &  0.000  & -0.067 & 0.081 &   \\ 
Gaia DR3 5961072456929737216  &    &  0.524  &  -1.185  &  0.060  &  0.000  &  0.014 &  &  &  \\ 
\hline 
Gaia DR3 466294295706341760  &  6.201  &  1.025  &  -1.956  &  0.034  &  0.000  &  0.000  & -0.660 & 0.065 &   \\ 
Gaia DR3 466291787445425408  &    &  0.392  &  -1.296  &  0.055  &  0.000  &  0.240 &  &  &  \\ 
\hline 
Gaia DR3 750184379766219648  &  10.229  &  0.790  &  19.668  &  0.026  &  0.000  &  0.053  & -0.188 & 0.125 &   \\ 
Gaia DR3 750184311046743168  &    &  0.688  &  19.856  &  0.122  &  0.000  &  0.089 &  &  &  \\
\enddata
\tablenotetext{a}{Sky-plane physical separation from \cite{ElBadry:2021}.}
\tablenotetext{b}{Stellar mass based on the mass-magnitude relation derived by \cite{Chae:2023}.}
\tablenotetext{c}{Radial velocity stored in the APOGEE parameter {\tt VHELIO}\textunderscore{\tt AVG} (the barycentric velocity) which represents a signal-to-noise weighted average if multiple visits took place (i.e., the APOGEE parameter {\tt NVISITS} $>1$): When multiple values from multiple observations (note here that each observation may have multiple visits) are reported, the given value refers to the error-weighted mean $\mu$ calculated by $\mu=\sum({\rm RV}_i/\sigma_i^2)/\sum(1/\sigma_i^2)$.}
\tablenotetext{d}{Reported error: When multiple values from multiple observations are reported, the given value refers to the error of $\mu$ calculated by $\sigma_\mu = 1/\sqrt{\sum(1/\sigma_i^2)}$.}
\tablenotetext{e}{This value refers to the standard deviation of multiple values from multiple observations. It is set to zero when a single observation was made.  }
\tablenotetext{f}{The APOGEE parameter {\tt VSCATTER} which represents the scatter from multiple visits of an observation. It is set to zero when an observation consists of a single visit. When multiple values of {\tt VSCATTER} from observations are available, the maximum value is given.}
\tablenotetext{g}{Relative radial velocity $v_r(\equiv {\rm RV}_A - {\rm RV}_B)$ where ${\rm RV}_A$ and ${\rm RV}_B$ refer respectively to the first and second values.}
\tablenotetext{h}{Whether multiple observations were made at multiple epochs.}
\end{deluxetable}

\startlongtable
\begin{deluxetable}{lccclccc}
\tablecaption{\textbf{Summary of wide binaries with multiple radial velocities of both components from multi-epoch APOGEE observations}\label{tab:APOGEEmulti}} 
\centerwidetable
\tabletypesize{\scriptsize}
\startdata
Star A  & RV\tablenotemark{a} &  Err\tablenotemark{b}  & Scatt\tablenotemark{c} & Star B  & RV\tablenotemark{a} &  Err\tablenotemark{b}  & Scatt\tablenotemark{c} \\
   & [km\,s$^{-1}$] &   &   &   & [km\,s$^{-1}$] &   &     \\
\hline
Gaia DR3 2642922251741817216 & 23.611124 & 0.04403066 & 0.04110121 & 2642922286101533952 & 23.745401 & 0.04636222 & 0.06235123 \\ 
  & 23.493046 & 0.03530691 & 0.04366402 &  & 23.661783 & 0.03730566 & 0.02735676 \\ 
\hline
Gaia DR3 2501323948860951296 & -26.34998 & 0.03360082 & 0.00961134 & 2501325525113734912 & -26.389238 & 0.04771343 & 0.11454166 \\ 
  & -26.28739 & 0.04346129 & 0.06616001 &  & -26.260921 & 0.03940159 & 0.04780705 \\ 
\hline
Gaia DR3 217107577453306112 & 22.248499 & 0.02819021 & 0.07399079 & 217107607515576704 & 21.878296 & 0.03318193 & 0.05599995 \\ 
  & 22.275375 & 0.02912341 & 0.144448 &  & 21.788877 & 0.02936134 & 0. \\
\enddata
\tablenotetext{a}{Reported radial velocity.}
\tablenotetext{a}{Reported error.}
\tablenotetext{c}{Reported value of \texttt{VSCATTER}.}

\end{deluxetable}

\section{Description of Speckle OAN-SPM observations and data reduction}
\label{sec:Speckle}

\subsection{Speckle observations at 2.1m telescope of the Observatorio Astron\'omico Nacional at Sierra de San Pedro M\'artir (SPM), M\'exico. }

Since our main objective is a search for anomalies in either component of a given wide binary system, we aim to identify any component that exhibits a structure different from that of a normal star, such as elongation, a comma-like pattern, or having another component. 
Speckle interferometry is a technique used in astronomy to overcome the limitations of atmospheric turbulence and obtain high-resolution images of astronomical objects. Using this technique with a 2.1m telescope, one can resolve binary stars with small separations of up to 0.05$\arcsec$.
Speckle imaging with the 2.1 m telescope at OAN-SPM routinely covers targets from very bright stars down to $R \approx 11$ mag, delivering contrast sensitivities of about $\Delta m \approx 3$ mag at 0.09$\arcsec$ and up to  $\Delta m \approx 8$ mag at 1.0$\arcsec$. For stars fainter than $R \approx 11$ mag these contrast limits become difficult to achieve, whereas for brighter targets observed under excellent conditions, even higher contrasts can be obtained. 

Observations were carried out with the 2.1 m telescope at the Observatorio Astron\'omico Nacional, San Pedro M\'artir, Mexico, using the Berkut speckle interferometer. The instrument is equipped with an Andor iXon Ultra 888 EMCCD, delivering a plate scale of 0.037$\arcsec$ pixel$^{-1}$  and a maximum field of view of 38$\arcsec$. Due to the small isoplanatic angle, systems with projected separations greater than 8$\arcsec$ were observed as separate pointings.
Five observation runs were conducted between February 2024 and June 2025. The nights were characterized by predominantly variable sky conditions, with seeing values typically ranging from 0.5$\arcsec$ to 2$\arcsec$. Observations were performed through standard Johnson R and I filters. Individual speckle exposures had integration times between 10 and 30 ms, depending on target brightness. For each target, a sequence of 2000 frames was acquired to ensure sufficient statistical averaging. In total, we obtained 1019 data cubes for 391 wide binaries systems.

\begin{table}

    \caption{}
    \begin{tabular}{cccc} 
    \hline
    Run & Cubes & Cubes  & Cubes  \\
    & $256 \times 256$ & $512 \times 512$ & $1024 \times 1024$ \\
    \hline
    Feb-24	&	35	&		&	8	\\
    Sep-24	&	131	&	156	&		\\
    Nov-24	&	70	&	139	&	1	\\
    Mar-25	&	149	&		&		\\
    Jun-25	&	267	&	63	&		\\

    \hline
   Total	&	652	&	358	&	9	\\		

    \hline
    \end{tabular}
    \label{tab:ObsSPM}
\end{table}    

\subsection{Data Processing}

We employed the same data acquisition strategy as in \cite{Orlov2021} and \cite{Luna2020}. The necessary software and expertise are fully in place. Acquiring and processing speckle images requires thousands of short exposures (1–50 ms), with the exact exposure time determined by wavelength, telescope aperture, and atmospheric seeing. Typically, about 2000 short-exposure frames are needed to reconstruct a high-resolution image of an 11th-magnitude star in the I filter. The limiting observable magnitude is strongly constrained by atmospheric turbulence; thus, targets fainter than 11 mag can only be observed under good conditions, with seeing better than 1.5$\arcsec$. A large number of frames is essential to achieve sufficient S/N, particularly at the smallest angular separations.

\begin{figure}[h!]
    \centering
    \begin{subfigure}{0.48\textwidth}
        \includegraphics[width=\textwidth]{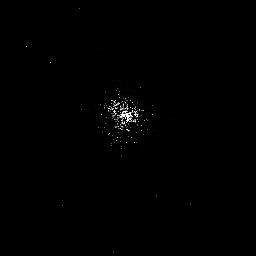}
    \end{subfigure}
    \begin{subfigure}{0.48\textwidth}
        \includegraphics[width=\textwidth]{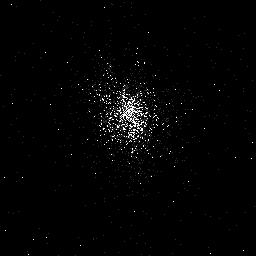}
    \end{subfigure}       
\caption{Data processing. Left to right: Best frame, worst speckle images in a series of 2000 specklegrams}
\label{fig:BestWorst}
\end{figure}

The first step of data processing is the dark field correction of detected images  $i'_n(\vec{x})$:

\begin{equation}
i_n(\vec{x}) =   i'_n(\vec{x}) - Dark(\vec{x}),
\end{equation}

\noindent
where $i_n(\vec{x})  $ is the intensity at point $\vec{x}$ in the corrected image, $ Dark(\vec{x}) $ is the average dark image captured with the closed shutter. Then for each specklegram we calculate the following criteria:
\begin{equation}
Sh1_n = \left[ \sum i_n (\vec{x})  \right]^2 / \sum i_n^2 (\vec{x}),
\end{equation}
where the summations extend over the complete n-th specklegram \citep{McCarthy1986}. Speckle images are then ranked by their $Sh1$, where the ``best'' frames are those with the lowest $Sh1$. Two images in Figure \ref{fig:BestWorst} show the best (a) and worst (b) speckle images in a series of 2000 specklegrams.

\subsection{High-Resolution}
The intensity distribution $i_n(\vec{x})$ of the $n$'th  short exposure image (specklegram) can be described by:

\begin{equation}
    i_n(\vec{x})=o(\vec{x})\otimes PSF_n(\vec{x}),
    \label{eq:conv}
\end{equation}
where  $\vec{x}$ is a 2D spatial coordinate, $o(\vec{x})$ is the object intensity distribution, $PSF_n(\vec{x})$ is the Point Spread Function for the $n$'th specklegram and $\otimes$ denotes convolution.

Then, applying the Fourier transform to Equation~(\ref{eq:conv})

\begin{equation}
I_n (\vec{\omega})=O(\vec{\omega}) \times OTF_n (\vec{\omega}),  
 \label{eq:Furr}
\end{equation}
where $I_n (\vec{\omega})$ and $O(\vec{\omega})$ are the Fourier transform of the image and object intensity distributions respectively, and $OTF_n (\vec{\omega})$ is the Optical Transfer Function (OTF), which is the equivalent of the $PSF$ in Fourier space. The $OTF_n (\vec{\omega})$ is the same for all objects within the isoplanatic region. Equation~(\ref{eq:Furr}) is the base for all high-resolution techniques. 
We applied various image processing techniques to reconstruct the power spectrum and obtain a high-resolution image of the star. This involves analyzing the speckle patterns in each short-exposure frame and combining them. 

\subsection{Power Spectrum}

The next step is to calculate the averaged power spectrum (PS) for each star: 
\begin{equation}
PS(\vec{\omega}) =  \left\langle \left|  I_n(\vec{\omega})   \right|^2 \right\rangle,  
\end{equation}
where  $\vec{\omega}$ is a spatial frequency and $\left\langle...\right\rangle$ denotes averaging over all images.

\begin{figure}[h!]
\begin{center}
    \begin{subfigure}{0.25\textwidth}
        \includegraphics[width=\textwidth]{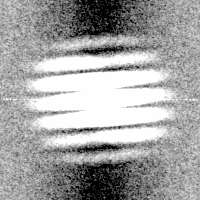}
    \end{subfigure}
    \begin{subfigure}{0.25\textwidth}
        \includegraphics[width=\textwidth]{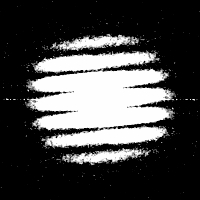}
    \end{subfigure}
    \begin{subfigure}{0.4\textwidth}
        \includegraphics[width=\textwidth]{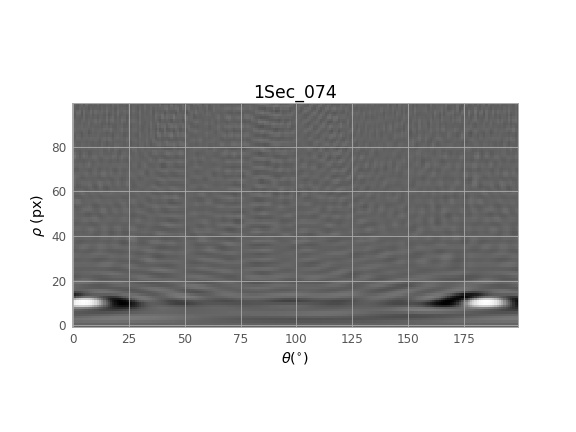}
    \end{subfigure}
\caption{Power Spectrum of a close binary star before photon bias correction (a) and after correction (b). The High resolution ACF in polar coordinates (c). Separation 0.3$\arcsec$. Panels (a), (b) and (c) are from left to right.}
\label{fig:FP}
\end{center}
\end{figure}

In the case of low light images, the averaged power spectrum can be expressed as \citep{Kerp1992}:
\begin{equation}
PS(\vec{\omega}) =  P(\vec{\omega}) \cdot \left| G(\vec{\omega}) \right|^2 + q\left| G(\vec{\omega}) \right|^2,
\end{equation}
where $ P(\vec{\omega}) $ is the unshifted estimation of the power spectrum, $q$ is some constant, and $\left| G(\vec{\omega}) \right|^2$ is the power spectrum of the photon event shape function also known as photon bias. The photon bias  $\left| G(\vec{\omega}) \right|^2$ can be determined as the normalized power spectrum of the night sky.
The power spectrum $\left| G(\vec{\omega}) \right|^2$ is constant in the vertical direction for this camera. Thus, it can be determined directly from $PS(\vec{\omega})$ (Figure~\ref{fig:FP}(a)) by analyzing its part beyond the cut-off frequency of the telescope. 
The unshifted power spectrum of specklegrams $P({\bf f})$ is shown in Figure~\ref{fig:FP}(b). Therefore, it can be presented as: 
\begin{equation}
P(\vec{\omega}) = \left| O(\vec{\omega}) \right|^2\left\langle  \left| S_n(\vec{\omega}) \right|^2 \right\rangle,
\label{Eq:eq4}
\end{equation}
where $\left| O(\vec{\omega}) \right|^2$ is the power spectrum of the object, and $\left\langle  \left| S_n(\vec{\omega}) \right|^2 \right\rangle$ is the speckle interferometric transfer function. The speckle interferometric transfer function can be obtained by observing a reference star, or one can construct universal synthetic speckle interferometric transfer function \citep{Tokovinin_2010}.

We are concerned only with the presence or absence of a close stellar component, and the information can be obtained directly from $P(\vec{\omega})$ without requiring the speckle interferometric transfer function. In particular, we calculated the high resolution autocorrelation function in polar coordinates $ACF_p$:
\begin{equation}
    ACF_p(\rho,\theta) =const\int_{0}^{\infty}\int_{0}^{2\pi} \cos (2\pi r\rho \cos(\theta-\phi))P({r,\phi })W(r,\phi)r d{r}d{\phi},
    \label{Eq:ACFp1}
\end{equation}
where $W(r,\phi)$ is the window that excludes part of $P(r,\phi)$ beyond the cut-off frequency of the telescope $f_T$ and for frequencies lower than the atmospheric cutoff $f_A$. Also taking into account central symmetry of  $P(r,\phi)$, Equation ~(\ref{Eq:ACFp1}) can be rewritten as: 
\begin{equation}
    ACF_p(\rho,\theta) =const \int_{f_A}^{f_T}\int_{0}^{\pi} \cos (2\pi r\rho \cos(\theta-\phi))P({r,\phi })r d{r}d{\phi}.
    \label{Eq:ACFp2}
\end{equation}

A star has a component if the high-resolution $ACF_p$ has a pronounced maximum. $ACF_p$ allows one to find the component even when the power spectrum is distorted by vibrations and strong aberrations of the telescope \citep{Orlov2021}. An example of $ACF_p$ is shown in Figure~\ref{fig:FP}(c).

\subsection{Speckle Holography}

\begin{figure}[h!]
    \centering
    \begin{subfigure}{0.24\textwidth}
        \includegraphics[width=\textwidth]{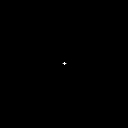}
    \end{subfigure}
    \begin{subfigure}{0.24\textwidth}
        \includegraphics[width=\textwidth]{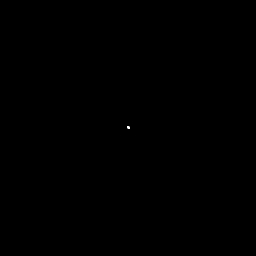}
    \end{subfigure}
    \begin{subfigure}{0.24\textwidth}
        \includegraphics[width=\textwidth]{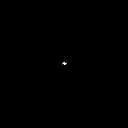}
    \end{subfigure}
    \begin{subfigure}{0.24\textwidth}
        \includegraphics[width=\textwidth]{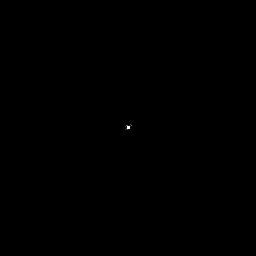}
    \end{subfigure}
\caption{The typical wide binary system, composed of Gaia DR3 1172915990414659328 (G = 9.2) and Gaia DR3 1172920487244742912 (G = 9.4). (a) High-resolution autocorrelation of Gaia DR3 1172915990414659328. (b) High-resolution speckle holography image of Gaia DR3 1172915990414659328. (c) High-resolution autocorrelation of Gaia DR3 1172920487244742912. (d) High-resolution speckle holography image of Gaia DR3 1172920487244742912. Panels (a), (b), (c) and (d) are from left to right. }
\label{fig:SingleACF}
\end{figure}

\begin{figure}[h!]
    \centering
    \includegraphics[width=0.3\columnwidth]{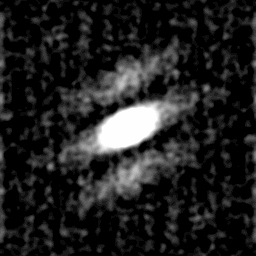}
    \includegraphics[width=0.3\columnwidth]{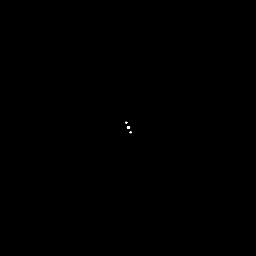}   
    \includegraphics[width=0.3\columnwidth]{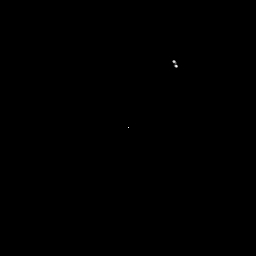}

\caption{The triple system ABC, composed of Gaia DR3 4493544185516358656 (G = 12.86), Gaia DR3 4493544082437142400 (G = 13.26), and Gaia DR3 4493544082437142784 (G = 14.13). (a) Power spectrum of subsystem BC. (b) High-resolution autocorrelation of subsystem BC. (c) High-resolution speckle holography image. It was derived from Speckle Interferometry \citep{Labeyrie1970}, the main difference being that speckle holography works so as to remove atmospheric aberrations from each short exposure image and the phase information is kept. The separation between components B and C is 0.1\arcsec  (8 au).  Panels (a), (b) and (c) are from left to right.}
\label{fig:PWS_ACF}
\end{figure}

For most wide binaries, hidden stellar components were not detected (Figure \ref{fig:SingleACF}). However, during data processing we identified several interesting systems — for example, Gaia DR3 4493544185516358656, which turned out to be a triple star system (Figure~\ref{fig:PWS_ACF}). To investigate such cases in more detail, we applied speckle holography to the entire dataset. The method can be summarized as follows.

If one has a point source $O^r(\vec{x})=\delta(\vec{x}=\vec{x^r})$ inside the area, Equation~(\ref{eq:Furr}) for this star becomes:
\begin{equation}
I_n^r (\vec{\omega})= OTF_n (\vec{\omega}).  
 \label{eq:OTFR}
\end{equation}

Substitution of Equation~(\ref{eq:OTFR}) into Equation~(\ref{eq:Furr}) gives:
\begin{equation}
    I_n (\vec{\omega})=O(\vec{\omega}) \times I_n^r (\vec{\omega}).  
 \label{eq:Eq12p}
\end{equation}
The Fourier transform of the object intensity distributions can be estimated by : 
\begin{equation}
     O(\vec{\omega})  = \frac{\langle I_n (\vec{\omega})I_n^{r*} (\vec{\omega}) \rangle}{\langle |I_n^r (\vec{\omega})|^2 \rangle},
\end{equation}
where $\langle ... \rangle $ means averaging over all images and $^*$ is the complex conjugation. Finally, the object intensity distributions can be obtained by the inverse Fourier transform.

\subsection{detection limit}

The detection limit describes how faint a companion star could be and still be identifiable in the data. It is defined by the minimum signal-to-noise ratio needed to separate a real signal from random noise. In practice, astronomers usually adopt a threshold of  $5\sigma$, which makes the chance of mistaking noise for a real star well below one percent. Sometimes a lower threshold of $3\sigma$ is used, but this is risky. At this level, it is rare, but possible, to detect spurious companions caused by noise \citep{Tokovinin_2010}. 

\begin{figure}[h!]
    \centering

    \label{fig:speck}
    \includegraphics[width=0.45\columnwidth]{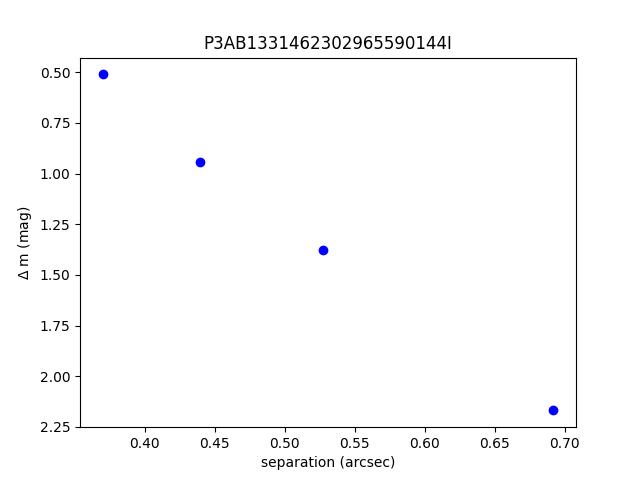}
    \label{fig:med}
    \includegraphics[width=0.45\columnwidth]{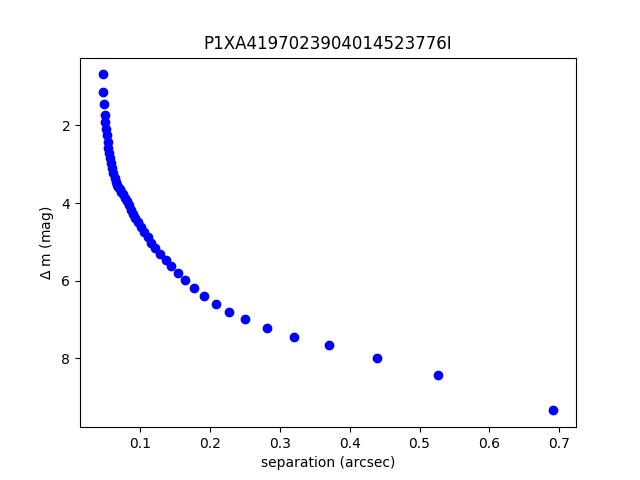}   
\caption{The $5\sigma$ Contrast curves. The contrast curve for Gaia DR3 1331462302965590144 ("Too weak") with a magnitude of 14.3 (left) and the curve for Gaia DR3 4197023904014523776 with a magnitude of 8.57 (right). }
\label{fig:ContrCurv}
\end{figure}

At large separations from the primary star (more than about three times the seeing disk) the noise reaches a steady level, and the detection limit becomes roughly constant. 
At smaller separations, however, it strongly depends on angular distance from the primary star. This behavior is summarized with a contrast curve (Figure \ref{fig:ContrCurv}). The contrast curve plots the faintest detectable companion (in terms of brightness difference, $\Delta m$) as a function of separation from the primary star. In other words, it tells: If there had been a companion this bright at this distance, we would have seen it.

Since no companions were detected in the vast majority of cases, for each star we present only two representative points of the contrast curve, corresponding $\Delta m = 3$ and $\Delta m = 4$.

In total, we obtained 1019 measurements for 390 selected wide binaries. Many targets were observed two or more times, allowing us to retain only the highest-quality measurements for each system. Table~\ref{Tab:ltapp} lists the results for 780 stars.

%\onecolumn

\startlongtable
\begin{deluxetable}{lllll}
\tablecaption{\textbf{Summary of Speckle Observations}\label{Tab:ltapp}}
\centerwidetable
\tabletypesize{\footnotesize}
\startdata
Comment& Identifier & Julian Year & Dm 4 & Dm 3  \\
	&	Gaia DR3 1003223584897948160	&	2024.865	&	0.140	&	0.100	\\
	&	Gaia DR3 1003223614961194752	&	2024.865	&	0.190	&	0.135	\\
	&	Gaia DR3 1020666649676726784	&	2025.2123	&	0.130	&	0.100	\\
	&	Gaia DR3 1020667714828594304	&	2025.2123	&	0.185	&	0.130	\\
	&	Gaia DR3 1023887978228642176	&	2025.2176	&	0.155	&	0.110	\\
	&	Gaia DR3 1023888012588380032	&	2024.868	&	0.125	&	0.090	\\
	&	Gaia DR3 1028432805245868544	&	2024.868	&	0.120	&	0.085	\\
\enddata
\end{deluxetable}

\end{document}